\documentclass[twocolumn,superscriptaddress,nofootinbib,floatfix,aps,prb,citeautoscript,longbibliography]{revtex4-2}
\usepackage[utf8]{inputenc}
\usepackage[T1]{fontenc}
\usepackage{lineno}

\usepackage{hyperref}
\usepackage{graphicx}
\usepackage{color}
\usepackage{array}
\usepackage{dcolumn}
\usepackage{bm}
\usepackage{multirow}
\usepackage[version=4]{mhchem}
\usepackage{cleveref}
\usepackage{amsfonts}

\usepackage{orcidlink}

\usepackage{tikz}

\newcommand{\ehull}{$E_\text{hull}$}

\usepackage{xtab,afterpage}

\begin{document}
\newcommand{\bochum}{Research Center Future Energy Materials and Systems of the University Alliance Ruhr and Interdisciplinary Centre for Advanced Materials Simulation, Ruhr University Bochum, Universitätsstraße 150, D-44801 Bochum, Germany}

\title{Generative AI for Crystal Structures: A Review}

\author{Pierre-Paul De Breuck  \orcidlink{0000-0002-3173-2058}}
\affiliation{\bochum}

\author{Hai-Chen Wang \orcidlink{0000-0002-2892-5879}}
\affiliation{\bochum}

\author{Gian-Marco Rignanese \orcidlink{0000-0002-1422-1205}}
\affiliation{UCLouvain, Institute of Condensed Matter and Nanosciences (IMCN), Chemin des Étoiles 8, Louvain-la-Neuve 1348, Belgium}
\affiliation{WEL Research Institute, Avenue Pasteur, 6, 1300 Wavre, Belgium}

\author{Silvana Botti \orcidlink{0000-0002-4920-2370}}
\affiliation{\bochum}
\author{Miguel A. L. Marques \orcidlink{0000-0003-0170-8222}}
\affiliation{\bochum}

\date{\today}

\begin{abstract}
As in many other fields, the rapid rise of generative artificial intelligence is reshaping materials discovery by offering new ways to propose crystal structures and, in some cases, even predict desired properties. This review provides a comprehensive survey of recent advancements in generative models specifically for inorganic crystalline materials. We begin by introducing the fundamentals of generative modeling and invertible material descriptors. We then propose a taxonomy based on architecture, representation, conditioning, and materials domain to categorize the diverse range of current generative AI models. We discuss data sources and address challenges related to performance metrics, emphasizing the need for standardized benchmarks. Specific examples and applications of novel generated structures are presented. Finally, we examine current limitations and future directions in this rapidly evolving field, highlighting its potential to accelerate the discovery of new inorganic materials.

\end{abstract}


\maketitle




\section{Introduction}
The rapid advancement of technology is closely tied to the discovery and development of new materials~\cite{lewisCostEffectiveSolarEnergy2007, mageeQuantificationRoleMaterials2012, snyderComplexThermoelectricMaterials2010}. From energy storage and conversion to electronics and catalysis, progress in these areas relies on materials with well-designed properties. A key step in this process is the computational design and characterization of stable or metastable crystal structures, which can then be prioritized for experimental synthesis and testing. This principle, called materials screening, is central in today's materials discovery.

One of the earliest problems in this context is enumerating plausible crystal structures for a given chemical formula. Traditionally, this task --- known as crystal structure prediction (CSP) --- used methods to explore the potential energy surface to identify global low-energy configurations. Techniques such as genetic algorithms (USPEX~\cite{oganovCrystalStructurePrediction2006,podryabinkinAcceleratingCrystalStructure2019}, random search~\cite{pickardInitioRandomStructure2011}), particle swarm optimization (CALYPSO~\cite{wangCrystalStructurePrediction2010}), and minima hopping~\cite{goedeckerMinimaHoppingEfficient2004} have been widely used to navigate this complex landscape, iteratively refining candidate structures by evaluating their energies. These approaches mimic natural processes and have successfully enabled the design of novel materials~\cite{yamashitaCrySPYCrystalStructure2021a,fallsXtalOptEvolutionaryAlgorithm2021}.

However, CSP methods should be distinguished from fully generative approaches.
The former traditionally start from a predefined chemical composition and often a specified number of atoms per unit cell. Moreover, they are typically based on an iterative screening process where each candidate structure must undergo an explicit energy calculation. This step is computationally expensive, particularly with high-accuracy methods. Furthermore, the search space in CSP grows exponentially with system size, posing a significant combinatorial challenge.

In contrast, generative methods from artificial intelligence learn the underlying data distribution and chemical rules from large crystal structure databases. Consequently, they can directly suggest novel and plausible crystal structures without \textit{a priori} constraints on chemistry or stoichiometry. This bypasses the computationally intensive search and initial energy evaluation steps inherent to CSP. Moreover, many generative models can be conditioned on specific target properties, enabling a more direct and efficient path to discovering materials with desired characteristics.



This represents a paradigm shift for materials discovery and design. Historically, discovery was driven by empirical trial-and-error methods, as illustrated by Edison’s search for light bulb filaments. Over time, the field has evolved from heuristic approaches to theory-guided synthesis and high-throughput computational screening. Today, generative AI enables proactive material generation, prioritizing candidates \textit{in silico} before experimental validation.

This review aims to provide a comprehensive overview of recent developments in generative models for inorganic crystalline materials. We focus on the design and architecture of these models, their representations, and their ability to incorporate constraints through conditioning. We emphasize that this review is limited to \textit{in silico} approaches for generating crystal structures, such as in the form of crystallographic information files (CIFs), and does not address synthesis, processing, or experimental characterization.

The remainder of this review is structured as follows: \Cref{sec:preliminaries} introduces the foundational concepts of generative modeling, including the mathematical formulation of probability distributions, common algorithms, and material representations. \Cref{sec:data} discusses the importance of training data and currently available datasets. \Cref{sec:taxonomy} surveys various generative models applied to inorganic materials, organizing them into a taxonomy based on representation, architecture, conditioning, and materials domain. \Cref{sec:metrics} covers benchmarking efforts and current gaps in evaluation metrics. Finally, \cref{sec:applications} presents practical use cases, \cref{sec:discussion} outlines future directions, and \cref{sec:conclusion} provides our conclusions.

\section{Preliminaries}
\label{sec:preliminaries}

\subsection{Problem statement}
\begin{figure}
    \centering
    \includegraphics[width=0.5\textwidth]{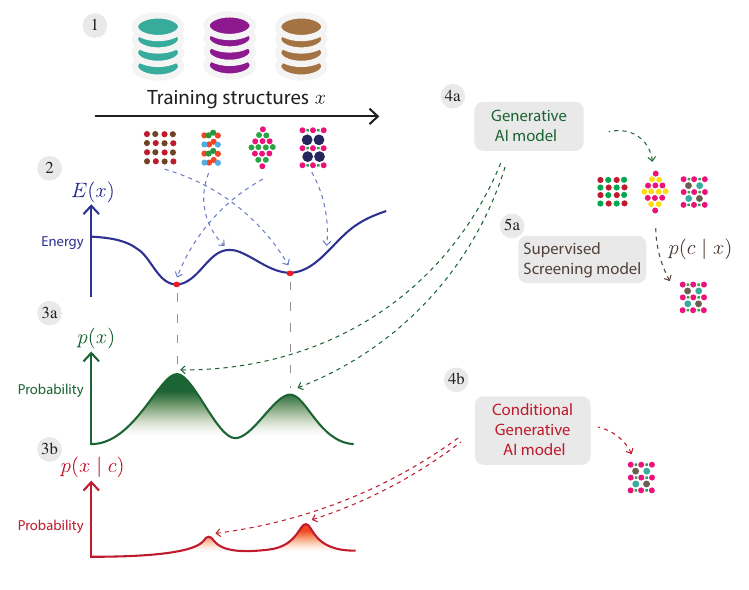}
    \caption{Schematic overview of generative AI for crystal structure generation. (1)~Crystal structure training data is collected from various databases. (2)~The data follows a distribution $\mathbf{x}$ in materials space and that can converted into an energy distribution $E(\mathbf{x})$. (3)~This distribution is implicitly learned by the generative model, providing the basis for the sampling probability $p(\mathbf{x})$. Two approaches are possible: (4a)~Unconditional sampling, optionally followed by post-processing such as (5a)~property screening; or (3b)~Conditional learning of $p(\mathbf{x} \mid c)$, enabling (4b)~direct conditional sampling.}

    \label{fig:genai_overview}
\end{figure}
Generative modeling constitutes a foundational paradigm within artificial intelligence and machine learning, primarily focused on learning the underlying probability distribution, $p(\mathbf{x})$, of a given dataset. Unlike supervised models, which learn a mapping from inputs to outputs to predict a label $y$ given an input $\mathbf{x}$ (i.e., learning conditional probability $p(y|\mathbf{x})$), generative models capture the inherent $p(\mathbf{x})$ of the data itself. 

Their core objective is to train a model parameterized by $\theta$ such that $p_\theta(\mathbf{x}) \approx p(\mathbf{x})$, allowing to generate from $p_\theta(\mathbf{x})$ novel samples $\mathbf{x'}$ that are statistically indistinguishable to samples drawn from the true data distribution $p(\mathbf{x})$.

This framework naturally extends to solid crystalline materials, where $\mathbf{x}$ represents an atomic configuration and $p(\mathbf{x})$ denotes the distribution over all possible material structures. In practice, some configurations are significantly more likely than others, effectively reducing the space of all possible (random) atomic arrangements to those near global or local energy minima. For example, assuming a Boltzmann distribution, $p(\mathbf{x}) \propto \exp(-E(\mathbf{x})/k_B T)$. As a result, low-energy configurations, corresponding to (meta-)stable materials, form the high-probability modes of this distribution and are the primary targets for generative sampling. This is schematically illustrated in \Cref{fig:genai_overview}.

Rather than relying on explicit chemical rules or direct energy evaluation, generative models implicitly learn the complex, multimodal probability distribution $p(\mathbf{x})$ from datasets of known structures. In doing so, they capture the essential structural motifs and bonding mechanisms of different materials classes.

\subsection{Conditioned generation}
Many applications in materials discovery require sampling from conditional distributions of the form $p(\mathbf{x}|c)$, where $c$ represents a specific constraint or target attribute, such as a desired chemical composition, space group symmetry, or functional properties such as the electronic band gap, thermal expansion coefficient, superconductivity transition temperature, etc. Conditional models, therefore, enable the targeted generation of materials that are not only structurally valid but also optimized for specific applications. When trained appropriately, such models capture which structures are most likely to exhibit the desired property, as summarized schematically in \Cref{fig:genai_overview}~(3b) and (4b).


\subsection{Generative architectures}

\begin{figure*}[ht]
\centering
\includegraphics[width=0.8\textwidth]{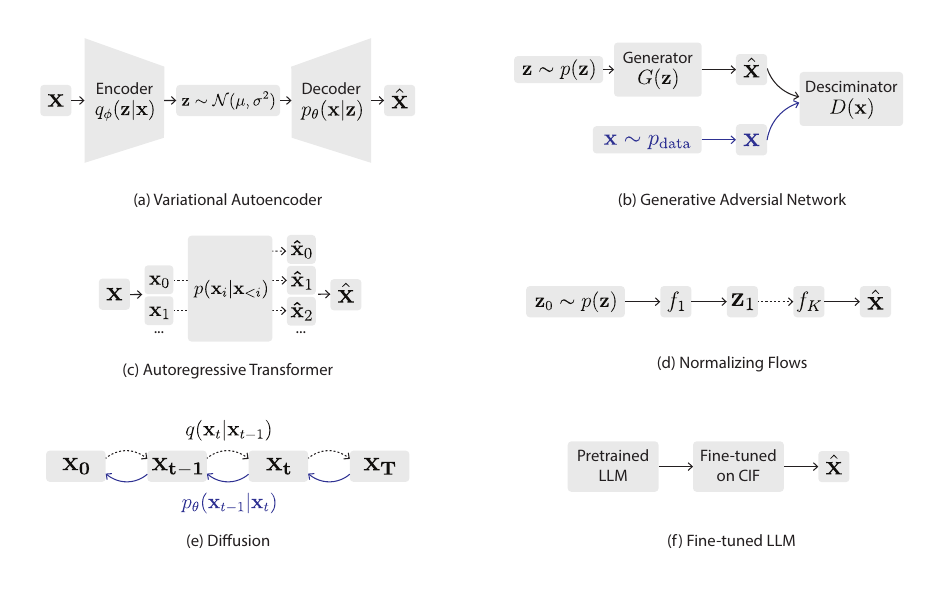}
\caption{Schematic overviews of the main generative model architectures. (a) variational autoencoder (VAE), (b) generative adversarial network (GAN), (c) autoregressive transformer (MLPs), (d) normalizing flow, (e) diffusion model, and (f) fine-tuned large language model (LLM). For detailed mathematical formulations and discussions, see the main text.}

\label{fig:architectures}
\end{figure*}

In this section, we outline the theoretical foundations underlying the main generative architectures used for materials generation. Here, architecture refers to the probabilistic framework that defines how a model samples from $p(\mathbf{x})$. We discuss variational autoencoders (VAEs), generative adversarial networks (GANs), transformers, normalizing flows, diffusion models, and fine-tuned large language models (LLMs). A schematic illustration of these different architectures is provided in \Cref{fig:architectures}.

\subsubsection{Variational Autoencoders (VAEs)}
Variational autoencoders (VAEs), introduced by Kingma and Welling ~\cite{kingmaAutoEncodingVariationalBayes2022}, combine dimensionality reduction with probabilistic generative modeling. They extend the classical autoencoder architecture, i.e., an encoder network that maps a high-dimensional input $\mathbf{x}$ to a lower-dimensional latent vector $\mathbf{z}$ and a decoder network that reconstructs $\mathbf{x}$ from $\mathbf{z}$, by introducing a probabilistic treatment of the latent space.

The VAEs define an approximate posterior distribution $q_{\phi}(\mathbf{z}|\mathbf{x})$, typically chosen as a multivariate Gaussian $\mathcal{N}(\boldsymbol\mu_{\phi}(\mathbf{x}), \mathrm{diag}(\boldsymbol\sigma^2_{\phi}(\mathbf{x})))$. From this distribution, one can draw a sample $\mathbf{z} \sim q_{\phi}(\mathbf{z}|\mathbf{x})$ and pass it to the decoder, which models the likelihood $p_{\theta}(\mathbf{x}|\mathbf{z})$. This stochastic encoding makes the decoder more robust to variations in the latent space and enables generation of novel samples by sampling directly from the prior.

The prior distribution $p(\mathbf{z})$, usually set to a standard normal $\mathcal{N}(0, I)$, imposes a smooth, continuous structure on the latent manifold. VAE training consists in maximizing the evidence lower bound (ELBO) on the marginal log-likelihood $\log p(\mathbf{x})$:

\begin{align}
\mathcal{L}_{\mathrm{ELBO}}(\mathbf{x}; \theta, \phi)
& = \mathbb{E}_{q_{\phi}(\mathbf{z}|\mathbf{x})}[\log p_{\theta}(\mathbf{x}|\mathbf{z})] \notag\\
\quad & - D_{\mathrm{KL}}\bigl(q_{\phi}(\mathbf{z}|\mathbf{x}) \,\|\, p(\mathbf{z})\bigr).
\end{align}

The first term, the \emph{reconstruction} loss, encourages accurate recovery of $\mathbf{x}$ from $\mathbf{z}$, while the second one, the Kullback–Leibler (KL) \emph{divergence}, regularizes the approximate posterior $q_{\phi}(\mathbf{z}|\mathbf{x})$ toward the prior $p(\mathbf{z})$. 
This ensures that the latent space is distributed in a way that permits smooth sampling.

Once the VAE is trained, the generation of new samples $\mathbf{x}'$ is straightforward: sample $\mathbf{z}' \sim p(\mathbf{z})$ and compute $\mathbf{x}' \sim p_{\theta}(\mathbf{x}|\mathbf{z}')$. This two-step process transforms the complex problem of modeling the data distribution $p(\mathbf{x})$ into learning and sampling from a well-behaved latent space, enabling the generation of new, plausible data points.

\subsubsection{Generative Adversarial Networks (GANs)}

Instead of explicitly learning the probability distribution $p(\mathbf{x})$, generative adversarial networks (GANs), introduced by Goodfellow \textit{et al.}~\cite{goodfellowGenerativeAdversarialNets2014, radfordUnsupervisedRepresentationLearning2016}, are trained to produce samples through an antagonistic process involving two distinct neural networks: a generator ($G$) and a discriminator ($D$).
The generator tries to fool the discriminator into believing its outputs are real, while the discriminator attempts to distinguish between real samples and synthetic ones produced by the generator.

In this two-players-game scenario, the generator learns to map the latent vector $\mathbf{z}$, typically drawn from a simple prior distribution such as a standard normal $\mathcal{N}(0, I)$, to a data-like output $\mathbf{\tilde{x}} = G(\mathbf{z})$. Meanwhile, the discriminator aims to maximize the probability of correctly classifying generated data versus real samples from the true data distribution $p_{\text{data}}(\mathbf{x})$. The adversarial training objective is formally expressed as:

\begin{align}
\mathcal{L} = \min_G \max_D \quad& \mathbb{E}_{\mathbf{x} \sim p_{\text{data}}(\mathbf{x})} [\log D(\mathbf{x})]
\notag\\
+& \mathbb{E}_{\mathbf{z} \sim p(\mathbf{z})} [\log(1 - D(G(\mathbf{z})))]
\end{align}

With this loss, the generator iteratively improves its ability to produce samples that are indistinguishable from the true data, while the discriminator becomes more adept at identifying subtle artifacts. Ideally, training converges to a Nash equilibrium where the discriminator cannot reliably distinguish real from generated samples, i.e., $D(\mathbf{x}) = 0.5$ for all inputs, and the generator captures the underlying data distribution $p(\mathbf{x})$.

Unlike VAEs, which explicitly optimize a likelihood-based objective and impose structure on the latent space through regularization, GANs do not require an explicit form for $p(\mathbf{x})$ or a reconstruction loss. 
However, GANs often suffer from instabilities during training and mode collapse, where the generator only produces a limited diversity of samples.


\subsubsection{Transformers}

The transformer, introduced by Vaswani \textit{et al.}~\cite{vaswaniAttentionAllYou2017}, is a foundational deep learning model renowned for its scalability, parallelizability, and ability to capture long-range dependencies. It dispenses with recurrence and convolution, relying instead on self-attention mechanisms to process input sequences.

The model operates on a sequence of input embeddings, $\{\mathbf{x}_1, \mathbf{x}_2, \dots, \mathbf{x}_n\}$, which are enhanced with positional encodings to preserve token order. In materials science, structural information like atoms, bonds, or symmetry descriptors can be represented as sequences, making the transformer well-suited for learning from material structure data.

Decoder-only transformers are used for generative tasks by learning the joint probability distribution of sequences. This is achieved by factorizing it using the chain rule:
\begin{equation}
    p(\mathbf{x}) = \prod_{i=1}^{N} p(\mathbf{x}_i | \mathbf{x}_{<i})
\end{equation}
Here, $x_i$ represents the $i$-th token in the sequence. This autoregressive formulation allows for the generation of new material structures by sequentially sampling tokens from the learned distribution.

To maintain this autoregressive property, the decoder employs a masked self-attention layer, which prevents access to future positions during training. The self-attention mechanism computes a weighted sum of all token representations in the sequence, allowing it to capture contextual dependencies regardless of their distance. For each attention head, the input is projected into queries ($\mathbf{Q}$), keys ($\mathbf{K}$), and values ($\mathbf{V}$), and the output is calculated as:
\begin{equation}
    \mathrm{Attention}(\mathbf{Q}, \mathbf{K}, \mathbf{V}) = 
    \mathrm{softmax}\left(\frac{\mathbf{Q}\mathbf{K}^\top}{\sqrt{d_k}}\right)\mathbf{V}
\end{equation}

where $d_k$ is the dimensionality of the key vectors. This is extended by multi-head attention, which performs several such computations in parallel to capture diverse relationships across different sub-spaces of representations.

The efficient and flexible design of the transformer has led to state-of-the-art performance in numerous sequence modeling tasks, driving its adoption in fields ranging from natural language processing to protein folding and materials discovery.

\subsubsection{Normalizing Flows}
Normalizing flows are a class of generative models that transform a simple probability distribution into a complex one through a sequence of invertible and differentiable mappings. Unlike VAEs or GANs, flows provide exact log-likelihood estimation and tractable sampling by explicitly modeling the data distribution~\cite{rezendeVariationalInferenceNormalizing2015,dinhDensityEstimationUsing2017}.

Given a base distribution $p_{\mathbf{z}}(\mathbf{z})$, typically a standard normal $\mathcal{N}(0, I)$, a normalizing flow defines a bijective transformation $f_\theta: \mathbf{z} \leftrightarrow \mathbf{x}$ such that the output $\mathbf{x}$ follows the target data distribution. The inverse $f_\theta^{-1}: \mathbf{x} \leftrightarrow \mathbf{z}$ allows for the computation of the exact density using the change of variables formula:

\begin{equation}
\log p_{\mathbf{x}}(\mathbf{x}) = \log p_{\mathbf{z}}(f_\theta^{-1}(\mathbf{x})) + \log \left| \det \left( \dfrac{\partial f_\theta^{-1}(\mathbf{x})}{\partial \mathbf{x}} \right) \right|.
\end{equation}

By composing multiple simple transformations $\mathbf{z}_0 \leftrightarrow \mathbf{z}_1 \leftrightarrow \dots \leftrightarrow \mathbf{z}_K = \mathbf{x}$, complex distributions can be modeled while maintaining tractability. Each transformation must be invertible and have a Jacobian determinant that is easy to compute.

Beyond discrete flows, continuous normalizing flows (CNFs)~\cite{dinhDensityEstimationUsing2017} replace the sequence of mappings with an ordinary differential equation parameterized by a neural network, enabling more flexible transformations via continuous-time dynamics. More recently, flow matching (FM)~\cite{lipmanFlowMatchingGenerative2023} provides a simplified training objective for learning such continuous flows by directly matching vector fields between distributions, leading to stable and scalable generative models. By learning smooth mappings between structured latent spaces and high-dimensional material representations, these models can learn to generate new materials.



\subsubsection{Diffusion}
Diffusion models have emerged as a highly successful class of generative models, synthesizing data by systematically reversing a gradual noising process, with early application to images~\cite{dhariwalDiffusionModelsBeat2021}, point clouds~\cite{caiLearningGradientFields2020} and molecular conformations~\cite{shiLearningGradientFields2021}, often exceeding GANs in quality. Originally inspired by non-equilibrium thermodynamics (see the work of Sohl-Dickstein \textit{et al.}~\cite{sohl-dicksteinDeepUnsupervisedLearning2015}), these models consist of two main parts: a fixed forward process that incrementally adds noise to data until it resembles pure noise, and a learned reverse process that reconstructs realistic samples by denoising at each step.

We explain here the denoising diffusion probabilistic models (DDPMs)~\cite{hoDenoisingDiffusionProbabilistic2020a}. The forward process is defined as a Markov chain that adds Gaussian noise to a data sample $\mathbf{x}_0$ over $T$ discrete time steps according to a fixed variance schedule $\{\beta_t\}_{t=1}^T$:
\begin{equation}
 q(\mathbf{x}_t|\mathbf{x}_{t-1}) = \mathcal{N}(\mathbf{x}_t; \sqrt{1 - \beta_t} \mathbf{x}_{t-1}, \beta_t \mathbf{I}).
\end{equation}
As $t \to T$, the distribution $q(\mathbf{x}_T)$ converges to an isotropic Gaussian distribution $\mathcal{N}(0, \mathbf{I})$. The core task is to learn the reverse process, $p_\theta(\mathbf{x}_{t-1} | \mathbf{x}_t)$, which approximates the true (but intractable) posterior $q(\mathbf{x}_{t-1} | \mathbf{x}_t)$. This is typically achieved by training a neural network, $\boldsymbol{\epsilon}_\theta$, to predict the noise that was added at a given step $t$. The training objective simplifies to the following loss:
\begin{equation}
\mathcal{L} = \mathbb{E}_{t, \mathbf{x}_0, \boldsymbol{\epsilon}} \left[ \left\| \boldsymbol{\epsilon} - \boldsymbol{\epsilon}_\theta(\mathbf{x}_t, t) \right\|^2 \right],
\end{equation}
where $\boldsymbol{\epsilon}$ is the sampled noise and $\mathbf{x}_t$ is the noised version of $\mathbf{x}_0$ at step $t$.

To generate a new sample, one starts with pure noise $\mathbf{x}_T \sim \mathcal{N}(0, \mathbf{I})$ and iteratively applies the learned denoising function to step backward in time, eventually yielding a clean sample $\mathbf{x}_0$. Key variants of this are the denoising diffusion implicit models (DDIMs), which introduce a deterministic sampling path that allows for much faster generation with fewer steps. Another variant includes modelling the gradient~\cite{songGenerativeModelingEstimating2019} and a further generalization casts the process in continuous time, leading to score-based models that solve stochastic differential equations (SDEs) to transform noise into data~\cite{songScoreBasedGenerativeModeling2021}.

\subsubsection{Large Language Models (LLMs)}
A final category involves the large language models (LLMs)~\cite{radford2018improving,devlinBERTPretrainingDeep2019,brownLanguageModelsAre2020}, often fine-tuned on crystallographic data. Although these models are fundamentally based on the transformer architecture, and are therefore not a different architecture on their own, they represent a different paradigm from models training point of view. Here, a massive model pre-trained on a general text corpus is repurposed (fine-tuned) for a specialized scientific domain.

The core idea is to treat structural information as a form of language. Crystallographic information files (CIFs), simplified molecular-input line-entry systems (SMILES), or other textual representations of atomic structures are used as training data. The LLM is then fine-tuned on this dataset, adapting its learned linguistic capabilities to understand the ``grammar'' and ``vocabulary'' of crystal structures, i.e., the rules governing symmetry, atomic arrangements, and bonding. By re-training the model with the specific objectives of generating valid text-based structural descriptions, it learns to produce novel CIFs or other representations that can be further interpreted (by machine or human) into physically plausible materials.

\subsection{Representations}
Modeling the probability distribution $p(\mathbf{x})$ of materials requires an appropriate representation for $\mathbf{x}$. 
Unlike supervised learning, where simple descriptors (e.g., average mass or volume) may suffice, generative modeling demands an invertible representation---one that preserves all structural/chemical information and allows exact reconstruction of materials from their representations.

Ideally, representations should be invariant to translation, rotation, and permutation. However, these invariances are not strict requirements, as data augmentation can help models learn such symmetries~\cite{quirogaRevisitingDataAugmentation2020,mazitovPETMADLightweightUniversal2025}. Below, we summarize the main invertible representations used in crystal generative AI models.

\subsubsection{Point Cloud or AXL}
One of the most straightforward ways to describe a crystal structure is as a set of points in $\mathbb{R}^3$, where each point corresponds to an actual atom. The set of total $N$ atoms in a crystal cell can be expressed as a set $\{s_i, (x_i, y_i, z_i)\}_{i=1}^N$, where $s_i$ and $(x_i, y_i, z_i)$ are respectively the chemical species and Cartesian coordinate of the $i$-th atom. Next, the periodicity of the cell is captured by including the lattice $\mathbf{L}$, which is typically defined by lengths of the lattice vectors $(a, b, c)$ and the angles between the lattice vectors $(\alpha, \beta, \gamma)$. Thus, the complete representation becomes $\mathbf{P} = \{s_i, (x_i, y_i, z_i)\}_{i=1}^N, (a, b, c, \alpha, \beta, \gamma)$. In the literature, this is also often referred to as the $(\mathbf{A}, \mathbf{X}, \mathbf{L})$ representation, where $\mathbf{A}$ corresponds to the species, $\mathbf{X}$ to the coordinates, and $\mathbf{L}$ to the lattice. This representation is invertible as required, since with any given $(\mathbf{A}, \mathbf{X}, \mathbf{L})$ the corresponding crystal structure can be reconstructed.

\subsubsection{Voxel grid}

The voxel grid representation discretizes the 3D space of the crystal into a regular grid of voxels, similar to a 3D image. Each voxel in the grid holds a value that reflects the presence or density of atoms within that region. Therefore, atoms are represented as continuous functions, typically Gaussians, centered at their fractional (reduced) coordinates within the unit cell. This provides a smooth representation of atomic positions. For materials with multiple chemical species, the voxel grid can have multiple channels, with each channel dedicated to a specific element. The resolution of the grid is a crucial hyperparameter, dictating the level of detail the representation captures and the computational cost. The crystal lattice itself can be encoded by applying transformations to a canonical unit cell. For instance, a unit-centered Gaussian within the unit cell can be generated and then transformed according to the lattice parameters $(a, b, c, \alpha, \beta, \gamma)$ into a cubic grid. This representation allows the model to operate on image-like 3D data and leverage convolutional operations. As it naturally drew inspiration from the image processing field, voxel was one of the earliest representations used in crystal generative AI models.

\subsubsection{Graph}
The graph representation encodes a crystal structure as a multi-graph $G = (V, E)$, where each node $v_i \in V$ corresponds to an atom and each edge $e_{ij} \in E$ represents an interaction or bond between atoms $i$ and $j$. Nodes are typically annotated with elemental features (e.g., atomic number, electronegativity, etc.), while edges can carry information about distances (bond length), bond types, or bond angles. Optionally, secondary and further graphs can be added, incorporate higher-order interactions. Unlike point cloud or voxel representations, graphs are not inherently invertible and lose exact geometric information, making them unsuitable as stand-alone representations for generative models. Instead, they are often used jointly with point cloud representations to iteratively update atomic positions during generation.

\subsubsection{Reciprocal Space}  
The reciprocal space representation leverages the Fourier transform of the crystal atomic positions to express periodicity and long-range order. A central quantity in this framework is the \textit{structure factor} $F(h,k,l)$, defined as:
\begin{equation}
  F(h, k, l) = \sum_j f_j \, e^{-2\pi i (h x_j + k y_j + l z_j)},  
\end{equation}
where $f_j$ is the atomic scattering factor (or form factor) of atom $j$, and $(x_j, y_j, z_j)$ are the corresponding fractional coordinates. In practical implementations, scattering factors may be replaced or augmented by learned elemental embeddings, and the sum can be decomposed into multiple channels.

To ensure invertibility, the direct lattice $L$, defined by its parameters $(a, b, c, \alpha, \beta, \gamma)$, must be encoded alongside the set of structure factors. Given the information of $L$, the inverse Fourier transform can uniquely reconstruct the periodic positions and crystal structure from the reciprocal space representation.

\subsubsection{Wyckoff Positions}  

This representation encodes a structure based on its crystallographic information. Each atom in a crystal is assigned to a \textit{Wyckoff position}, i.e., a set of symmetry-equivalent points defined within the space group of the crystal. In three-dimensional crystallography, there are 230 space groups comprising 1\,771 distinct Wyckoff positions.

A Wyckoff position is specified by its multiplicity and label (e.g., $4a$, $4b$, $8c$, etc.), the associated space group, and any free parameters. These parameters define the coordinates of a representative point within the asymmetric unit, while the symmetry operations of the space group generate the complete set of symmetrically equivalent atomic positions. Given information about the atomic occupancy (i.e., which atoms occupy which Wyckoff positions) along with the lattice parameters, one can uniquely reconstruct the complete crystal structure.


As this representation essentially reduces the crystal structure to the asymmetric unit, it eliminates redundancy and naturally enforces symmetry constraints.

\section{Data Sources}
\label{sec:data}


\begin{table*}[ht!]
\centering
\caption{Common inorganic crystal datasets used for training generative models.}
\label{tab:datasources}
\begin{tabular}{lp{6cm}@{\hspace{10pt}}l@{\hspace{10pt}}l@{\hspace{10pt}}p{2cm}}
\hline\hline
\textbf{Dataset} & \textbf{Description} & \textbf{Size} & \textbf{URL} & \textbf{License} \\
\hline
\\
Materials Project~\cite{jainMaterialsProjectMaterials2013} & DFT-optimized inorganic crystals (PBE GGA/GGA+U), includes experimental and hypothetical structures & 155K & \url{https://materialsproject.org} & CC BY 4.0 \\\\
Alexandria~\cite{schmidtMachineLearningAssistedDeterminationGlobal2023,Alex2d} & Large-scale DFT-optimized inorganic crystals, compatible with MP settings & 4.5M & \url{https://alexandria.icams.rub.de} & CC BY 4.0\\\\
ICSD~\cite{belskyNewDevelopmentsInorganic2002} & Experimentally known inorganic crystal structures; subset included in MP & 300K & \url{https://icsd.fiz-karlsruhe.de} & Commercial (restricted) \\\\
AFLOWLIB~\cite{curtaroloAFLOWLIBORGDistributedMaterials2012} & Automated high-throughput DFT database of materials properties & 3.5M & \url{http://aflowlib.org} & Academic use only\\\\
OQMD~\cite{saalMaterialsDesignDiscovery2013} & DFT database focused on thermodynamic stability and phase diagrams & 1.3M & \url{http://oqmd.org} & CC BY 4.0\\
JARVIS~\cite{jarvis} & DFT database including properties calculated using PBEsol, OptB88vdW, and TBmBJ  & 80K & \url{https://jarvis.nist.gov} & Open access (MIT for tools)\\
\hline\hline
\end{tabular}
\end{table*}

Beyond the choice of architecture (e.g., VAE, GAN) and structure representation, it is crucial to access to high-quality training data which typically include low-energy, thermodynamically stable crystal structures therefore sample the underlying distribution $p(\mathbf{x})$.

In practice, most datasets rely on density functional theory (DFT) calculations at the generalized gradient approximation (GGA) level~\cite{lehtolaRecentDevelopmentsLibxc2018}, typically using the Perdew, Burke, and Ernzerhof (PBE) functional~\cite{perdewGeneralizedGradientApproximation1996}. The fidelity of the generated crystal structures and their predicted properties is fundamentally constrained by the accuracy of the underlying DFT training data, as models will naturally reproduce both the strengths and limitations of the density functional approximations used in their training datasets. \Cref{tab:datasources} summarizes key resources in this area.

A widely-used source is the Materials Project (MP)~\cite{jainMaterialsProjectMaterials2013}, which contains a diverse set of DFT-relaxed inorganic structures. Many of these entries are derived from the Inorganic Crystal Structure Database (ICSD)~\cite{belskyNewDevelopmentsInorganic2002}, providing a link to experimentally observed compounds. Filtering MP entries by ICSD provenance or by applying an threshold on the energy above the hull (e.g., $< 80$ meV/atom) can help ensure thermodynamic stability.

The Alexandria database~\cite{schmidtMachineLearningAssistedDeterminationGlobal2023,Alex2d} significantly expands the available data, offering over 4.5 million 3D structures computed under MP-compatible settings using VASP~\cite{kresseEfficientIterativeSchemes1996}. This massive size enables significant improvement in quality in generative frameworks such as MatterGen~\cite{zeniGenerativeModelInorganic2025} and Matra-Genoa~\cite{breuckGenerativeMaterialTransformer2025} that were trained on its contents.

Additional valuable resources include AFLOWLIB~\cite{curtaroloAFLOWLIBORGDistributedMaterials2012},  OQMD~\cite{saalMaterialsDesignDiscovery2013}, and JARVIS~\cite{choudharyJointAutomatedRepository2020} all emphasizing high-throughput DFT calculations.
In this framework, it is worth mentioning the OPTIMADE API, which provides users with an easy common access to many of these world leading materials databases~\cite{Andersen2021,Evans2024}. A full list of the databases implementing the API is available on the \href{https://www.optimade.org/providers-dashboard/}{OPTIMADE providers dashboard}.

Finally, Xie \textit{et al.}~\cite{xieCrystalDiffusionVariational2022} introduced task-specific datasets for generative models, including Perov-5, a curated set of $\sim$19,000 perovskites designed for water splitting~\cite{castelliComputationalScreeningPerovskite2012, castelliNewCubicPerovskites2012}, and Carbon-24, which contains over 100,000 carbon allotropes generated via \textit{ab initio} random structure searching~\cite{pickardInitioRandomStructure2011}.

\section{Taxonomy}
\label{sec:taxonomy}

\begin{table*}[ht!]
\caption{\label{tab:taxonomydef}Taxonomy for generative models applied to crystal structures. Generative models can be distinguished by four key aspects: the choice of data representation for the crystal, the core model architecture, the conditioning information used to guide material generation, and the target material domain (limited by the architecture or training data).}
\centering
\begin{tabular}{l@{\hspace{1em}} l@{\hspace{1em}} l}
\hline\hline
\textbf{Category} & \textbf{Description} & \textbf{Possible Values} \\
\hline
\\
\textbf{Representation} &
How the crystal is encoded as input/output &
\begin{tabular}[t]{l}
Voxel grid \\
Graph \\
Wyckoff positions \\
Point cloud \\
Reciprocal space\\
\textit{etc.}\\
\end{tabular} \\[8pt]
\\
\textbf{Architecture} &
Core generative model framework &
\begin{tabular}[t]{l}
Variational Autoencoder (VAE)\\
Generative Adversarial Network (GAN)\\
Normalizing Flow \\
Diffusion Model \\
Transformer \\
Fine-tuned LLM\\
\textit{etc.}\\
\end{tabular} \\[8pt]
\\
\textbf{Conditioning} &
Extra information used to guide generation &
\begin{tabular}[t]{l}
Unconditional (no external constraints) \\
Composition (elemental fractions) \\
Space group or crystal system \\
Band gap (numeric) \\
Formation energy \\
Target density or porosity \\
\textit{etc.}\\
\end{tabular} \\[8pt]
\\
\textbf{Material Domain} &
Subset of materials modeled or generated &
\begin{tabular}[t]{l}
Oxides, Halides or any limited chemistry \\
Perovskites, cubic materials or any structural prototype\\
Alloys \\
All structures (unconstrained) \\
\textit{etc.}\\
\end{tabular} \\

\hline\hline
\end{tabular}
\end{table*}
\begin{figure*}[ht!]
    \centering
    \includegraphics[width=0.8\textwidth]{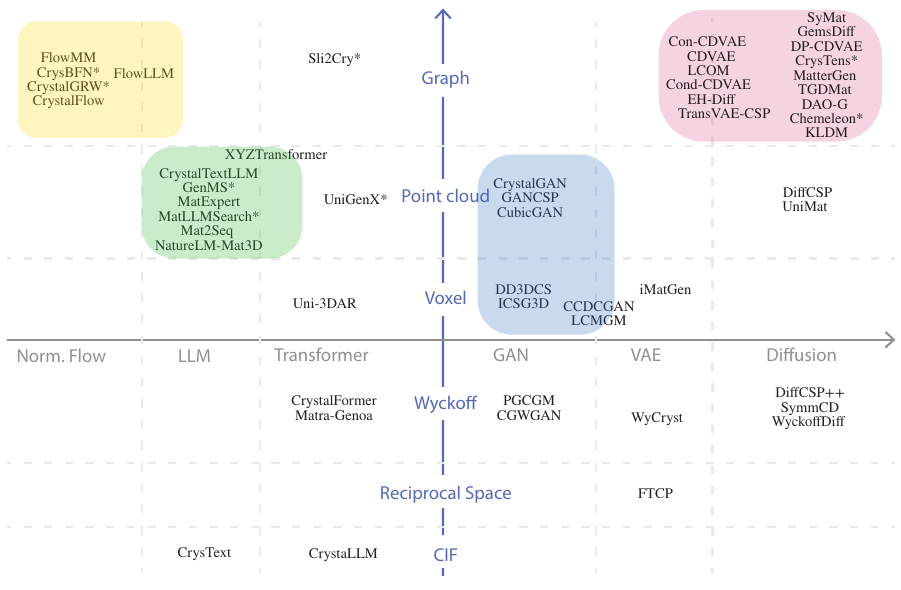}
    \caption{Schematic representation of the taxonomy of models in a two-dimensional architecture–representation space. The space is discrete, indicated by dashed lines. Some models lie at intersections, combining multiple architectures. The main categories are highlighted with colored bubbles, reflecting historically established groupings with significant contributions: GANs on point clouds or voxels (blue), graph diffusion (red), LLMs on point clouds (green), and normalizing flows on point clouds (yellow). Models with architectural variations relative to the indicated category are denoted by (*), see~\Cref{tab:generative_models_summary} for additional details.}
    \label{fig:taxonomy}
\end{figure*}

\begin{table*}[ht!]
\caption{Summary of generative models found in literature following the presented taxonomy. Refer to Table~\ref{tab:taxonomydef} and main text for definition of "representation", "architecture" and "conditioning". The latter is marked as "yes" only when the authors explicitly demonstrate conditioning on any functional property.}
\label{tab:generative_models_summary}
\centering
\begin{tabular}{l@{\hskip 10pt}>{\raggedright\arraybackslash}p{4cm}@{\hskip 10pt}>{\raggedright\arraybackslash}p{3cm}@{\hskip 10pt}c@{\hskip 10pt}>{\raggedright\arraybackslash}p{2cm}@{\hskip 12pt}cl}
\hline\hline
\textbf{Acronym} & \textbf{Representation} & \textbf{Architecture} & \textbf{Condit.} & \textbf{Domain} & \textbf{Ref.} & \textbf{Year} \\
\hline
\\
    \textbf{CrystalGAN} & Point Cloud & GAN & No & Hydrides & \cite{nouiraCrystalGANLearningDiscover2019} & 2019 \\
    \textbf{DD3DCS} & Voxel & GAN & No & All Structures & \cite{hoffmannDataDrivenApproachEncoding2019} & 2019 \\
    \textbf{CondGAN} & Bag of Atoms & GAN & Yes & Compositions & \cite{sawadaStudyDeepGenerative2019} & 2019 \\
    \textbf{iMatGen} & Voxel Grid & VAE & No & Vanadium Oxides & \cite{nohInverseDesignSolidState2019} & 2019 \\
    \textbf{MatGAN} & One-Hot Elements & GAN & Yes & Compositions & \cite{danGenerativeAdversarialNetworks2020} & 2020 \\
    \textbf{GANCSP} & Point Cloud & GAN & Yes & All Structures & \cite{kimGenerativeAdversarialNetworks2020} & 2020 \\
    \textbf{ICSG3D} & Voxel & GAN & Yes & Cubic Alloys, Perovskites, Heusler Compounds & \cite{court3DInorganicCrystal2020} & 2020 \\
    \textbf{CubicGAN} & Point Cloud & GAN & Yes & Cubic Structures & \cite{zhaoHighThroughputDiscoveryNovel2021b} & 2021 \\
    \textbf{CCDCGAN} & Voxel Grid & VAE, GAN & Yes & Bi-Se Materials & \cite{longConstrainedCrystalsDeep2021} & 2021 \\
    \textbf{FTCP} & Reciprocal Space, Point Cloud & VAE & No & All Structures & \cite{renInvertibleCrystallographicRepresentation2022} & 2022 \\
    \textbf{CDVAE} & Point Cloud, Graph & VAE, Diffusion & Yes & All Structures & \cite{xieCrystalDiffusionVariational2022} & 2022 \\
    \textbf{CCDCGAN} & Voxel Grid & VAE, GAN & Yes & All Structures & \cite{longInverseDesignCrystal2022} & 2022 \\
    \textbf{PGCGM} & Wyckoff & GAN & Yes & Ternaries & \cite{zhaoPhysicsGuidedDeep2023} & 2023 \\
    \textbf{XYZTransformer} & Point Cloud & LLM & No & All Structures & \cite{flam-shepherdLanguageModelsCan2023} & 2023 \\
    \textbf{PCVAE} & Composition, Prototype & VAE & No & Prototypes & \cite{liuPCVAEPhysicsinformedNeural2023} & 2023 \\
    \textbf{LCOM} & Graph, Point Cloud & VAE, Diffusion & Yes & All Structures & \cite{qiLatentConservativeObjective2023} & 2023 \\
    \textbf{CHGlownet} & Graph & GFlowNet & No & All Structures & \cite{NeurIPSHierarchicalGFlowNet2023} & 2023 \\
    \textbf{SLI2Cry} & Graph (SLICES) & RNN & No & All Structures & \cite{xiaoInvertibleInvariantCrystal2023} & 2023 \\
    \textbf{SyMat} & Graph, Point Cloud & Diffusion & No & All Structures & \cite{luoSymmetryAwareGenerationPeriodic2023} & 2023 \\
    \textbf{Crystal-GFN} & Prototype (Space Group) & GFlowNet & Yes & Structural Prototypes & \cite{ai4scienceCrystalGFNSamplingCrystals2023} & 2023 \\
    \textbf{GemsDiff} & Graph, Point Cloud & Diffusion & No & All Structures & \cite{klipfelVectorFieldOriented2023} & 2023 \\
    \textbf{CGMD} & Point Cloud & Diffusion, VAE, Flow Matching & No & All Structures & \cite{novitskiyUnleashingPowerNovel2024} & 2024 \\
    \textbf{DP-CDVAE} & Point Cloud, Graph & Diffusion & No & All Structures & \cite{pakornchoteDiffusionProbabilisticModels2024} & 2024 \\
    \textbf{CrysTens} & Point Cloud (Pairwise Distance) & GAN, Diffusion & No & All Structures & \cite{alversonGenerativeAdversarialNetworks2024} & 2024 \\
    \textbf{CrystalTextLLM} & Point Cloud & LLM & Yes & All Structures & \cite{gruverFineTunedLanguageModels2024} & 2024 \\
    \textbf{DiffCSP} & Point Cloud & Diffusion & Yes & All Structures & \cite{jiaoCrystalStructurePrediction2024} & 2024 \\
    \textbf{DiffCSP++} & Wyckoff & Diffusion & No & All Structures & \cite{jiaoSpaceGroupConstrained2024} & 2024 \\
    \textbf{Con-CDVAE} & Point Cloud, Graph & VAE, Diffusion & Yes & All Structures & \cite{yeConCDVAEMethodConditional2024} & 2024 \\
    \textbf{NSGAN} & Composition Vector & GAN, GA & No & Alloys & \cite{liNSGANNondominantSorting2024} & 2024 \\
    \textbf{UniMat} & Point Cloud & Diffusion & Yes & All Structures & \cite{yangScalableDiffusionMaterials2024} & 2024 \\
    \textbf{FlowMM} & Point Cloud (Flat Manifold) & Flow Matching & No & All Structures & \cite{millerFlowMMGeneratingMaterials2024} & 2024 \\
    \textbf{StructRepDiff} & Embedded Atom Density & Diffusion & No & All Structures & \cite{sinhaRepresentationspaceDiffusionModels2024} & 2024 \\
    \textbf{CrystalFormer} & Wyckoff & Transformer & Yes & All Structures & \cite{caoSpaceGroupInformed2024} & 2024 \\
    \textbf{VGD-CG} & Composition One-Hot & GAN, VAE, Diffusion & Yes & Compositions & \cite{qinInverseDesignSemiconductor2024} & 2024 \\
    \textbf{LCMGM} & Reciprocal Space & VAE, GAN & No & Perovskites & \cite{chenebuahDeepGenerativeModeling2024} & 2024 \\
    \textbf{GenMS} & Point Cloud & LLM, Diffusion & Yes & All Structures & \cite{yangGenerativeHierarchicalMaterials2024} & 2024 \\
    \textbf{WyCryst} & Wyckoff & VAE & Yes & All Structures & \cite{zhuWyCrystWyckoffInorganic2024} & 2024 \\
    \textbf{MatExpert} & Point Cloud (Conversational) & LLM & Yes & All Structures & \cite{dingMatExpertDecomposingMaterials2024} & 2024 \\
    \textbf{FlowLLM} & Point Cloud (Flat Manifold) & LLM, Flow Matching & Yes & All Structures & \cite{sriramFlowLLMFlowMatching2024} & 2024 \\
    \textbf{Cond-CDVAE} & Point Cloud, Graph & VAE, Diffusion & Yes & All Structures & \cite{luoDeepLearningGenerative2024} & 2024 \\
    \textbf{CrystaLLM} & CIF File & LLM, Transformer & Yes & All Structures & \cite{antunesCrystalStructureGeneration2024} & 2024 \\
    \textbf{CrysText} & CIF File & LLM & Yes & All Structures & \cite{mohantyCrysTextGenerativeAI2024} & 2024 \\
    \textbf{CGWGAN} & Wyckoff & GAN & No & All Structures & \cite{suCGWGANCrystalGenerative2024} & 2024 \\
    \textbf{Matra-Genoa} & Wyckoff & Transformer & Yes & All Structures & \cite{breuckGenerativeMaterialTransformer2025} & 2025 \\
    \textbf{EH-Diff} & Hypergraph, Point Cloud & Diffusion & No & All Structures & \cite{liuEquivariantHypergraphDiffusion2025} & 2025 \\
\end{tabular}
\end{table*}

\begin{table*}[ht!]
\label{tab:generative_models_summary_part2}
\centering
\begin{tabular}{l@{\hskip 10pt}>{\raggedright\arraybackslash}p{4cm}@{\hskip 10pt}>{\raggedright\arraybackslash}p{3cm}@{\hskip 10pt}c@{\hskip 10pt}>{\raggedright\arraybackslash}p{2cm}@{\hskip 12pt}cl}
    \textbf{CrysBFN} & Point Cloud (Hyper-Torus) & Bayesian & No & All Structures & \cite{wuPeriodicBayesianFlow2025} & 2025 \\
    \textbf{TransVAE-CSP} & Graph, Point Cloud, RBF & VAE, Diffusion & No & All Structures & \cite{chenTransformerEnhancedVariationalAutoencoder2025} & 2025 \\
    \textbf{CrystalFlow} & Graph, Point Cloud & Continuous Normalizing Flow & Yes & All Structures & \cite{luoCrystalFlowFlowBasedGenerative2025} & 2025 \\
    \textbf{MatLLMSearch} & Text (JSON) & LLM, Evolutionary Search & Yes & All Structures & \cite{ganLargeLanguageModels2025} & 2025 \\
    \textbf{Mat2Seq} & Invariant Sequence & LLM & Yes & All Structures & \cite{yanInvariantTokenizationCrystalline2025} & 2025 \\
    \textbf{MatterGen} & Point Cloud, Graph & Diffusion & Yes & All Structures & \cite{zeniGenerativeModelInorganic2025} & 2025 \\
    \textbf{TGDMat} & Point Cloud, Contextual & Diffusion & Yes & All Structures & \cite{dasPeriodicMaterialsGeneration2025} & 2025 \\
    \textbf{NatureLM-Mat3D} & Point Cloud & LLM & Yes & All Structures & \cite{xiaNatureLanguageModel2025} & 2025 \\
    \textbf{CrystalGRW} & Graph, Manifold & Geodesic Random Walk & Yes & All Structures & \cite{tangsongcharoenCrystalGRWGenerativeModeling2025} & 2025 \\
    \textbf{UniGenX} & Point Cloud & Transformer, Diffusion & Yes & All Structures & \cite{zhangUniGenXUnifiedGeneration2025} & 2025 \\
    \textbf{DAO-G} & Graph, Point Cloud & Diffusion (EBM) & No & All Structures & \cite{wuSiameseFoundationModels2025} & 2025 \\
    \textbf{Uni-3DAR} & Voxel (Compressed) & Transformer & No & All Structures & \cite{luUni3DARUnified3D2025} & 2025 \\
    \textbf{Chemeleon} & Graph, Point Cloud, Text & Diffusion, Contrastive Learning & Yes & All Structures & \cite{parkExplorationCrystalChemical2025} & 2025 \\
    \textbf{SymmCD} & Wyckoff (Binary Matrix), Graph & Diffusion & No & All Structures & \cite{levySymmCDSymmetryPreservingCrystal2025} & 2025 \\
    \textbf{WyckoffDiff} & Wyckoff, Graph & Diffusion & No & Protostructures & \cite{kelviniusWyckoffDiffGenerativeDiffusion2025} & 2025 \\
    \textbf{KLDM} & Manifold, Graph & Diffusion & No & All Structures & \cite{cornetKineticLangevinDiffusion2025} & 2025 \\
\hline
\end{tabular}
\end{table*}

The methodology and intrinsic workings of generative models for crystal structures can be systematically understood through four key pillars: \textit{representation}, \textit{architecture}, \textit{conditioning}, and \textit{materials domain}. Together, these elements define the design space of existing approaches and form the basis of the taxonomy presented in~\Cref{tab:taxonomydef,tab:generative_models_summary}.

The \textit{representation} specifies how crystal structures are encoded, ranging from point clouds and voxel grids to graphs, reciprocal-space descriptors, and Wyckoff positions, as explained above. This choice determines how symmetry, periodicity, and atomic details are captured. In practice, many combinations are possible, and models often employ multiple representations simultaneously. For example, FTCP~\cite{renInvertibleCrystallographicRepresentation2022} integrates reciprocal- and real-space features into a hybrid representation, while graph-based models are commonly paired with point clouds to iteratively refine atomic positions.

The \textit{architecture}, as defined here, refers to the foundational generative framework, including GANs, VAEs, transformers, normalizing flows, diffusion models, and fine-tuned large language models. These architectures learn the underlying distribution of crystal structures based on the chosen representation.

The \textit{conditioning} extends generative modeling from learning an unconditional distribution $p(\mathbf{x})$ to a conditional one $p(\mathbf{x}|c)$, where $c$ represents a target property. While formation energy is often implicitly modeled, some approaches explicitly condition on additional properties such as magnetism, electrical conductivity, or thermal expansion to guide generation. For simplicity, we label models as “conditioned” (i.e., marked as ``yes'' in~\Cref{tab:generative_models_summary}) only when the authors explicitly demonstrate conditioning on any functional property. Otherwise ``no'' is stated.

The \textit{materials domain} defines the scope of data used for training and thus generation of new compounds. This can be either limited by chemistry (e.g., oxides, halides), structural prototypes (e.g., perovskites, cubic materials) or others subdomains such as alloys or 2D materials.

\Cref{fig:taxonomy} illustrates how the first two dimensions of the taxonomy---architecture and representation---creates a discrete space for organizing current models, thereby visually summarizing the current landscape.

The following sections review these models in detail, grouping them by their common architectural and representational strategies into common subsections.

\subsection{GANs and VAEs}

Inspired by successes in image generation, early generative models in materials science adapted similar techniques. GANs and VAEs operating on voxel representations emerged as the first methods for crystal structure generation. These models treat 3D atomic arrangements as image-like density grids, compressing them into continuous latent spaces for sampling new structures.

Hoffman \textit{et al.}~\cite{hoffmannDataDrivenApproachEncoding2019} represented structures within a 10~\AA\ cube, discretized into a $30\times30\times30$ grid. Each voxel held a density value derived from Gaussian distributions centered on atoms, scaled by atomic number. A U-Net architecture was then used to segment this density grid to determine atomic positions. Despite data augmentation and high accuracy on the MP database, the model struggled to generate physically stable structures.

iMatGen~\cite{nohInverseDesignSolidState2019} extended the voxel approach using fractional coordinates and separate grids for unit cell representation (using a transformed Gaussian). A hierarchical, two-step VAE encoded the materials into a latent space. When tested on vanadium oxides, it rediscovered 25 of 31 known vanadium oxide structures, even when held out from training. The channel-based design in principle allowed this model to be extended to other chemical systems.

ICSG3D~\cite{court3DInorganicCrystal2020} represented crystals using voxelized electron density grids with summed Gaussian densities. It used a 3D U-Net for segmentation to recover atomic coordinates. A one hot encoding of the energy was appended to the latent space of the VAE to condition on formation energy. Trained on cubic binary alloys, ternary perovskites, and Heusler compounds, ICSG3D demonstrated the ability to interpolate between known structures—for instance, varying the A-site element in a perovskite across a periodic row—by traversing its latent space.

CCDCGAN~\cite{longConstrainedCrystalsDeep2021} introduced a two-stage approach in which a VAE was first trained to create a structured latent space, which subsequently served as input for a GAN. This framework incorporated a formation energy predictor directly into the loss function, compelling the generator to find low-energy structures by seeking minima in the latent space. This method successfully discovered previously unreported, low-energy crystal structures in the Bi--Se system. The work was later extended to multi-component systems~\cite{longInverseDesignCrystal2022}.

While voxel-based methods are prominent, GANs using other representations are also found in the literature. For instance, CrystalGAN~\cite{nouiraCrystalGANLearningDiscover2019} and GANCSP~\cite{kimGenerativeAdversarialNetworks2020} operated directly on point clouds of atomic coordinates, with the latter also enabling conditioning on the composition. More recent approach evolved the voxel concept for greater efficiency, as an example, Uni-3DAR~\cite{luUni3DARUnified3D2025} used an octree-compressed voxel representation to convert a 3D structure into a compact token sequence. A transformer was then autoregressively used on these tokens to rapid sample of new crystals.

\subsection{Models based on diffusion}

In 2021, Xie \textit{et al.} \cite{xieCrystalDiffusionVariational2022} introduced the Crystal Diffusion Variational Autoencoder (CDVAE) for periodic material generation. This model has since served as a foundation for many subsequent iterations of diffusion-based approaches. The original generation pipeline began with an encoder, i.e., a periodic graph neural network (PGNN) that mapped materials into a latent space. From this latent representation, the model predicted composition, lattice parameters, and number of atoms, which served as the initial structure. A diffusion decoder subsequently refined this initial structure by performing Langevin dynamics, which simultaneously denoised atomic coordinates and updated atom types. A noise-conditioned score network guided the denoising process. This network used another PGNN that estimated the gradient of the probability density at various noise levels while explicitly incorporating interactions across periodic boundaries. CDVAE maintained permutation, translation and rotation invariances through SE(3) equivariant graph neural networks, however failed to achieve translational invariance in denoising. Nevertheless, benchmark evaluations demonstrated that CDVAE achieved relatively high validity and coverage for materials generation.

Several extensions were built on the CDVAE architecture. Qi \textit{et al.} proposed LCOM \cite{qiLatentConservativeObjective2023}, which extended CDVAE by introducing a surrogate conservative model aimed at minimizing formation energy in the latent space. Similarly, the SyMat \cite{luoSymmetryAwareGenerationPeriodic2023} framework followed the CDVAE architecture but further strengthened invariance with respect to translation. The GemmsDiff model developed by Klipfell \textit{et al.}~\cite{klipfelVectorFieldOriented2023}, also employed an equivariant GNN, further extended CDVAE. It allowed lattice parameters to evolve within the diffusion process, in contrast to the original CDVAE that fixed them after the VAE.

Another variation, DP-CDVAE \cite{pakornchoteDiffusionProbabilisticModels2024}, modified CDVAE by replacing its score network with a denoising diffusion probabilistic model (DDPM). This modification enabled the model to generate structures that were closer to their ground states compared to the original CDVAE. Jiao \textit{et al.} introduced DiffCSP \cite{jiaoCrystalStructurePrediction2024}, which performed joint diffusion on both lattice parameters and reduced coordinates using a periodic E(3) equivariant denoising model. Unlike CDVAE, DiffCSP started from randomly initialized structures drawn from prior distributions, eliminating the need for a VAE. The model demonstrated excellent validity and coverage. Its successor, DiffCSP++ \cite{jiaoSpaceGroupConstrained2024}, further incorporated crystal symmetry by conditioning the diffusion process on a specific space group.
Luo \textit{et al.} extended CDVAE to Cond-CDVAE~\cite{luoDeepLearningGenerative2024}, enabling conditional generation based on both chemical composition and external pressure.  Trained on a combination of the Materials Project dataset and high-pressure CALYPSO CSP data, Cond-CDVAE could generate crystal structures under user-defined conditions via injecting a conditional vector into both the latent space and Langevin dynamics.

MatterGen~\cite{zeniGenerativeModelInorganic2025} further evolved the CDVAE architecture, by removing the VAE entirely (as in DiffCSP) and employing a fully joint diffusion process over atom types, coordinates, and lattice parameters. Thus, MatterGen allowed the lattice to change during generation. Instead of applying Gaussian noise in Cartesian coordinates, MatterGen diffused in fractional coordinates using a wrapped normal distribution, which inherently respected periodicity and accommodates varying unit cells. MatterGen also used larger training datasets compared to previous diffusion models and implemented conditional generation through inserting lightweight adapter modules into the score network. This enabled property control without retraining the full model.

Recent developments explored various architectural enhancements. EH-Diff~\cite{liuEquivariantHypergraphDiffusion2025} replaced the graph representation by a hypergraph, linking more than two atoms by an edge, capturing better multi-atomic interactions. TransVAE-CSP~\cite{chenTransformerEnhancedVariationalAutoencoder2025} used a methodology similar to CDVAE but enhanced the encoder by using a transformer with equivariant dot product attention. TGDMat~\cite{dasPeriodicMaterialsGeneration2025} used jointly diffussion on atom types, fractional coordinates, and lattice structure using a periodic E(3) equivariant denoising model, and further integrated textual information to guide generation with user defined prompts. In a similar fashion, Chemeleon~\cite{parkExplorationCrystalChemical2025} also integrated textual descriptors. The embeddings for textual and structural data were trained using contrastive learning. Then these embeddings were used in diffusion, enabling material sampling with simple prompts such as ``\ce{LiMnO4} with orthorhombic structure''. DAO~\cite{wuSiameseFoundationModels2025} used two models for diffusion, by incorporating energy considerations into the diffusion paths. The approach was also pretrained on non-stable structures for enhancing model capability.

Other geometric approaches offered alternative perspectives. CrystalGRW~\cite{tangsongcharoenCrystalGRWGenerativeModeling2025} proposed diffusion defined on manifolds, i.e., torus for coordinates and simplices for atom types, using geodesic random walks instead of Euclidean Gaussian noise. This method respected better the natural geometry of crystal representations. Drawing inspiration from physical Langevin dynamics, KLDM~\cite{cornetKineticLangevinDiffusion2025} used manifolds for structure representation but performed diffusion on auxiliary velocity variables in associated Euclidean space. This avoided complex diffusion directly on curved manifolds while maintaining geometric constraints. 

In another direction, UniMat~\cite{yangScalableDiffusionMaterials2024} proposed a universal representation in which materials were encoded as a point cloud shaped like the periodic table, with DDPM-based diffusion applied on top. This approach generated structures with lower decomposition energies compared to CDVAE. It is also worth noting that diffusion has also been integrated with transformer architectures. For instance, UniGenX~\cite{zhangUniGenXUnifiedGeneration2025} generated element types (i.e., chemical formulas) autoregressively via a transformer decoder, while refining atomic coordinates using a diffusion head applied to the next-token embeddings of transformer. This hybrid design offered controllability under both compositional and structural constraints.

\subsection{Models based on Wyckoff descriptors}

One of the earliest works to incorporate symmetry information into generative models was PGCGM by Zhao \textit{et al.}~\cite{zhaoPhysicsGuidedDeep2023}. This approach represented symmetry operations using affine matrices and encoded a single fractional coordinate for each set of equivalent sites, thereby implicitly preserving symmetry throughout the generation process. A GAN architecture was used, combined with physics-inspired loss functions: specifically, inter- and intra-atomic distance-based and symmetry-compliant based losses, as well as average full-coordinate losses. Using this framework, the authors successfully generated 2,000 materials, with 5\% of them lying within 0.25~eV/atom of the Materials Project hull. However, the method was limited to ternary systems, though it could likely be extended to more complex compositions.

Built on the use of Wyckoff positions, the WyCryst~\cite{zhuWyCrystWyckoffInorganic2024} model encoded them, as one-hot matrices—referred to, as Wyckoff genes, which were then embedded into a latent space via a VAE. This latent space was guided by a property-prediction branch, ensured that the space was structured along property gradients for targeted sampling. While this enabled symmetry-aware and property-driven generation, the workflow required additional steps, including using PyXtal~\cite{fredericksPyXtalPythonLibrary2021} to generate atomic coordinates and performing DFT calculations for final structure refinement and validation.

A related approach is CGWGAN~\cite{suCGWGANCrystalGenerative2024} proposed by Su \textit{et al.}, which also leveraged Wyckoff descriptors but adopted a two-step generation process. In the first step, a GAN was used to generate crystal templates defined by the asymmetric unit, space group, and lattice vectors. In the second step, atoms were filled into these templates, guided by a supervised force field model (specifically, M3GNet~\cite{chenUniversalGraphDeep2022}) to assess thermodynamic and mechanical stabilities. Promising candidates were then validated through DFT calculations, which led to the discovery of seven stable structures in the ternary Ba--Ru--O system.

More recently, Cao \textit{et al.} introduced CrystalFormer~\cite{caoSpaceGroupInformed2024}, a decoder-only transformer model that generated crystal structures by sampling Wyckoff descriptors conditioned on the space group. The model iteratively predicted Wyckoff sites, fractional coordinates, and lattice parameters, achieving high validity and diversity while outperforming earlier methods in maintaining symmetry consistency. A parallel work by De Breuck \textit{et al.}~\cite{breuckGenerativeMaterialTransformer2025} introduced Matra-Genoa, an autoregressive transformer that encoded crystals in a fully invertible sequence representation based on Wyckoff positions, including both discrete sites and continuous free coordinates. Trained on over two million structures, Matra-Genoa could also be conditioned on thermodynamic stability by explicitly incorporating the distance to the convex hull during training using a stability token. Together with CrystalFormer, Matra-Genoa combined for the first time discrete Wyckoff information with continuous positional data.

Finally, Wyckoff descriptors can also be fully utilized using other types of architectures, such as diffusion. SymmCD~\cite{levySymmCDSymmetryPreservingCrystal2025} learned the joint distribution of the asymmetric unit cell and corresponding symmetry operations using diffusion. It encoded crystals as a reduced set of crystallographic orbits with a binary symmetry representation and allows conditioning on space group and number of orbits, producing complete 3D structures with atomic coordinates. WyckoffDiff~\cite{kelviniusWyckoffDiffGenerativeDiffusion2025} also employed diffusion but operated purely on Wyckoff positions, starting from all positions and diffusing element assignments to produce proto-structures. Unlike SymmCD, it did not output exact coordinates, and it required DFT relaxation for stability assessment.

\subsection{Models based on LLMs}

One of the earliest explorations of applying large language models (LLMs) to structural data files, such as XYZ, CIF, and PDB formats, was carried out by Flam-Shepherd and Aspuru-Guzik~\cite{flam-shepherdLanguageModelsCan2023}. They trained a GPT-like transformer architecture from scratch using next-token prediction, rather than fine-tuning an existing language model. While this means that their work could also be categorized under “transformer” approaches, the focus was on demonstrating that an LLMs could directly generate novel and valid structures without relying on advanced, domain-specific representations. This pioneering result sparked further research into leveraging more advanced LLMs for modeling crystal structures directly from text-based formats. Gruver \textit{et al.} introduced CrystalTextLLM~\cite{gruverFineTunedLanguageModels2024}, a model which fine-tuned Llama 2-70B~\cite{touvronLlama2Open2023} on textual point cloud descriptions of crystals. The model tokenized numerical values digit-by-digit, enabling the language model to process them. This setup supported flexible text-conditioned generation, allowed users to prompt the model with specifications like ``Below is a description of a bulk material. [The chemical formula is Pm2ZnRh]. Generate a description of the lengths and angles of the lattice vectors and then the elements and coordinates for each atom within the lattice.'' The model also performed infilling by masking parts of the input and proposing replacements, giving a flexible framework for the user.

Antunes \textit{et al.} explored CIF generation by training a LLaMA-based~\cite{touvronLLaMAOpenEfficient2023} language model from scratch \cite{antunesCrystalStructureGeneration2024}. Given a chemical formula and the number of formula units, their model predicted complete CIFs token-by-token, successfully generating a wide range of structure types including perovskites, spinels, zeolites, and metal–organic frameworks. Around 53\% of the generated structures were symmetry-valid, and they obtained a 17\% match to known prototypes identified by crystal classification tools, demonstrating the ability of LLMs to produce realistic and diverse crystal structures.

Mohanty~\textit{et al.} introduced CrysText~\cite{mohantyCrysTextGenerativeAI2024}, which combined the CIFs representation from CrystaLLM with the pre-trained LLM approach of CrystalTextLLM. Using a Llama 3.1–8B~\cite{grattafioriLlama3Herd2024} model fine-tuned with QLoRA~\cite{dettmersQLoRAEfficientFinetuning2023}, CrysText learned the relationships between crystallographic parameters while retained the flexibility of LLM-based conditioning on natural language prompts.

Moving beyond direct CIFs generation, GenMS~\cite{yangGenerativeHierarchicalMaterials2024} proposed a hierarchical approach in which an LLM first predicted plausible intermediate chemical formulas from natural language descriptions. A diffusion model then generated detailed 3D atomic structures conditioned on these formulas. This two-step process outperformed direct CIF generation via prompting, produced structures with higher validity, lower formation energies, and better adherence to user specifications.

Other works focused on making LLM-powered crystal generation more agentic. Ding \textit{et al.} developed MatExpert~\cite{dingMatExpertDecomposingMaterials2024}, a conversational multi-step agent that started with a set of user-defined property requirements. The system then retrieved similar materials from a pre-embedded database using contrastive learning, after which a fine-tuned LLM proposed modifications to improve the target properties. For example, it might suggest: ``Mg would be replaced by Na, which coordinates differently with sulfur, forming NaS\textsubscript{6} octahedra with a mix of edge- and corner-sharing geometries.'' This design not only generated final crystal structures but also explained the reasoning behind changes, making the process more transparent to human.

Other recent approaches continued to demonstrate the versatility of LLM-based crystal generation. MatLLMSearch \cite{ganLargeLanguageModels2025} showed that large language models could propose novel structures without fine-tuning. They combined a LLM with evolutionary search: the LLM received two parent structures and was asked to propose new structures, therefore performing implicitly crossovers and mutations. The proposed crystal structures was then optimized, filtered, and fed back into the LLM for another round of evolution. Another work, Mat2Seq~\cite{yanInvariantTokenizationCrystalline2025} introduced a systematic way to encode crystals as unique, invariant 1D sequences to enable language-model-based crystal generation. NatureLM-Mat3D~\cite{xiaNatureLanguageModel2025} used a fine-tuned Llama 3-8B model that represented crystals as lists of elements, space groups, and coordinates. It demonstrated strong performance in crystal generation and had the advantage of being trained on multiple scientific domains by converting them all to sequences, providing a larger science foundation model.


\subsection{Models using Flow Matching}

Miller \textit{et al.} introduced FlowMM~\cite{millerFlowMMGeneratingMaterials2024}, the first application of flow matching to crystal structure generation. By learning symmetry-aware vector fields on manifolds that reflect the geometries and periodicity of crystals, FlowMM performed Riemannian flow matching to efficiently model both CSP and \textit{de novo} generation. It outperformed prior methods like CDVAE and DiffCSP in accuracy and stability while requiring significantly fewer integration steps, achieved up to 3 times faster inference with comparable or better results. The authors later extended this work by using an LLM for the base distribution~\cite{sriramFlowLLMFlowMatching2024}. This approach added a fine-tuned LLM at the start to sample a noisy material description that was subsequently refined using Riemannian flow matching. This yielded structures with a better stability rate than FlowMM.

Although not directly related to flow matching and closer to diffusion approaches, CrysBFN~\cite{wuPeriodicBayesianFlow2025} used Bayesian flow networks, a class of models in which iterative Bayesian updates are used to update parameters of the target distribution. Another work in this category includes CrystalFlow~\cite{wuPeriodicBayesianFlow2025}, which introduced continuous flow matching in combination with a graph-based equivariant neural network. It also demonstrated the possibility of conditioning on external variables such as pressure.

\subsection{Other models}

In this section, we briefly mention other works that do not fall into the previous main categories, as they combine multiple or different representations and architectures. Generative Flow Networks (GFlowNets~\cite{bengioFlowNetworkBased}) are a general generative modeling framework that sequentially sample actions using a policy model trained on a reward signal. By enabling sampling proportional to a reward distribution, GFlowNets efficiently explored multiple modes in the search space, making them a promising approach for crystal generation. Within this line of work, the Crystal-GFN~\cite{ai4scienceCrystalGFNSamplingCrystals2023} model represented one of the first attempts, although it did not generate full structures. CHGlowNet~\cite{NeurIPSHierarchicalGFlowNet2023} adopted a hierarchical GFlowNet-based design for crystal generation.

A different approach focused on reversible graph representations: SlI2Cry~\cite{xiaoInvertibleInvariantCrystal2023} introduced an invertible SLICES-based representation using RNNs, achieving successful crystal generation. Additional efforts include CGMD~\cite{novitskiyUnleashingPowerNovel2024}, which combined point cloud inputs with diffusion, VAE, and flow-matching models. CrysTens~\cite{alversonGenerativeAdversarialNetworks2024} converted atomic coordinates into an image-like representation based on pairwise distances, applying both GANs and diffusion models, with diffusion showing superior results. NSGAN~\cite{liNSGANNondominantSorting2024} integrated GANs with a genetic algorithm operating in the latent space of aluminum alloys. StructRepDiff~\cite{liNSGANNondominantSorting2024} employed a diffusion model in an invariant descriptor space derived from the embedded atom method (EAM~\cite{zhangEmbeddedAtomNeural2019}), reconstructing atomic positions via gradient optimization to preserve symmetries.

Finally, Qin \textit{et al.} proposed VGD-CG~\cite{qinInverseDesignSemiconductor2024}, a composition generator that combined VAE, GAN, and diffusion, conditioned on properties such as decomposition enthalpy, synthesizability, and band gap. The predicted compositions were then converted into structures using a template-based predictor, led to the discovery of stable and metastable semiconductors, including promising photocatalysts such as VBi$_3$O$_7$.

\section{Evaluation Metrics and Benchmarks}
\label{sec:metrics}

Assessing whether a generative model works well requires a comprehensive multi-metric evaluation. A truly effective generative model should demonstrate not only the ability to discover novel materials with desired properties that are stable and synthesizable, but also the computational efficiency and scalability to do so in practical applications.

To achieve this comprehensive assessment, the evaluation of model performance involves both domain-knowledge-based criteria and data statistical metrics. Early models applied the \textit{validity metric}, where the generated structure is considered valid when certain chemical or physical constraints are satisfied. For example, the charges (oxidation states) must be balanced, the shortest distances between atoms must be reasonable, and the space groups assigned to the structures by the model must match those from symmetry analysis. For successful models, the validity rates, i.e., the fraction of generated structures that are physically meaningful, are often $>90\%$. However, these domain knowledge-based validity checks do not guarantee the stability of the generated structures, which are more stringently evaluated later through DFT calculations.

The \textit{coverage metric} evaluates how well the generated materials match that of a reference test set by computing precision (the fraction of generated structures that are valid and match the reference) and recall (the fraction of reference structures successfully predicted). However, this metric strongly depends on the choice and completeness of the test set. Missing structures in the test set are not inherently problematic---novelty and creativity are desirable in generative models---so the metric has clear limitations and, in our view, is not always well-suited for comparing models.

The \textit{realism metric} measures the distance between realistic references and the generated structures. To quantify this distance, one can use the root-mean-square displacement (RMSD) between reference and generated structures. Another approach is to check the difference in terms of symmetries using structure matching algorithms. Distances (e.g., Wasserstein distance, WD) in distributions of properties (energies, volumes, or targets for conditional models) has also been used. However, such distribution-based distances make comparisons challenging, as smaller improvements do not necessarily indicate a better model but only a closer match to the reference set. Metrics include the number of optimization steps required to relax a generated geometry to a local minimum, which can serve as a distance measure. Shorter optimization paths between generated and realistic structures indicate that the samples are largely physically plausible. In practice, DFT-optimized counterparts or algorithm-matched experimental structures are typically used as realistic references. Another related measure is the synthesizability metric, which estimates how feasible it would be to synthesize a generated structure under current laboratory conditions. However, synthesizability is difficult to standardize due to the scarcity of large-scale, high-throughput synthesis validation. As a result, energy-based metrics, described hereafter, often provide a more robust alternative.

The \textit{stability metrics} only appeared much later, making comparison with prior works difficult. These metrics rely on the evaluation of the thermodynamic and/or dynamic stabilities of the generated structures. Commonly, thermodynamic stability (with respect to competing phases) is represented by the distance to the convex hull. Usually, a structure is considered to be (meta-)stable when it lies within a threshold (e.g., 100~meV/atom) above the hull. For structures to be dynamically stable, the harmonic phonons across the Brillouin zone must exhibit real frequencies. Due to the computational expense of phonon calculations via DFT, dynamic stability checks are applied only in select cases, e.g., for benchmarking of the WyCryst model~\cite{zhuWyCrystWyckoffInorganic2024}. Recently, with the improving accuracy of universal machine learning interatomic potentials (uMLIPs)~\cite{matterbench,Loew2025}, some researchers have explored the possibility of replacing DFT with uMLIPs in stability checks~\cite{zeniGenerativeModelInorganic2025}. In model performance benchmarking, the fraction of generated samples remaining stable after DFT or uMLIPs relaxation is reported. We note that to the best of the authors' knowledge, there is neither a universal threshold nor a standardized choice of reference convex hull across all benchmarking studies of generative models. Therefore, comparisons between different models based on this metric are often not straightforward (or even available) or standardized and should be interpreted with caution.

The \textit{uniqueness metric} measures the ability of the model to generate diverse candidates. However, diversity is difficult to quantify. In practice, the uniqueness of the samples is used instead. A generated structure is considered unique if it does not duplicate any other structure generated in the \textit{same batch}. Another metric, the \textit{novelty metric}, measures how many structures are new compared to the training set. Unlike uniqueness, which refers to whether generated samples duplicate among themselves, this metric refers to whether a generated structure is different from a fixed reference (usually the training) dataset. A similar metric is the \textit{rediscovery rate}, which measures the percentage of the test set being recovered by the model. For these metrics, the algorithm used to judge the match between structures plays a key role. For example, some algorithms treat different ordered configurations for a disordered alloy as unique structures. Therefore, it is crucial to account for differences in algorithms when comparing model performance. Moreover, these three metrics are functions of the generated batch size and the reference datasets. With increasing batch size, the novelty and uniqueness ratios usually drop while the recovery rate increases. Nevertheless, within the same batch, a model might generate each novel structure multiple times, resulting in high novelty but low uniqueness. A successful model should be able to generate stable, unique, and novel structures. Therefore, these three metrics are often combined as the \textit{S.U.N. metric}~\cite{zeniGenerativeModelInorganic2025}.

The \textit{efficiency metric} represents a critical practical consideration for real-world deployment of generative models. Efficiency encompasses multiple dimensions including computational training cost, data efficiency, scalability, etc. Recent studies reveal significant disparities in resource consumption across different models, with training duration varying dramatically depending on the model architecture and dataset size. For instance, TGDMat required only 500 epochs to train on Perov-5 and Carbon-24 datasets, compared to more than $3000$ epochs for CDVAE and DiffCSP~\cite{jiaoCrystalStructurePrediction2024}. Some approaches obviate the computational costs for curating data that in some cases are several orders of magnitude larger than the training costs. The scaling behavior becomes particularly important when models are applied to larger chemical spaces or high-throughput screening scenarios, where thousands or millions of candidate structures need to be generated. However, most benchmarks focus exclusively on generation quality while leaving efficiency metrics overlooked.

We must emphasize the lack of unified standards in evaluating generative materials models. This issue manifests across multiple interconnected dimensions that collectively undermine meaningful cross-model comparisons. Therefore, we do not attempt to compare or benchmark models in this review, as the absence of standardized metrics currently makes such evaluations unreliable. Additional limitations of the current benchmarking landscape are discussed hereafter.

Firstly, early models were constrained by limited training datasets available at the time, such as Perov-5 or Carbon-24, which were restricted to specific structure motifs or chemical compositions. While recent models leverage substantially larger datasets as shown in~\Cref{tab:datasources}, their performance metrics are typically benchmarked on in-house random-split test sets. This approach prevents meaningful cross-model comparisons and necessitates the development of universal standardized benchmarking test sets. 

Secondly, additional factors such as matching algorithms and reference convex hulls vary significantly across the literature, as previously stated. Even when studies employ identical reference datasets and algorithms, they frequently apply different evaluation criteria and thresholds, severely hampering cross-study comparisons. A particularly concerning issue is that many models rely exclusively on formation energies for stability metrics, either lacking proper convex hull analysis for competing phases or disregarding it entirely. Additionally, matching algorithms might erroneously discriminate ordered configurations from their disordered prototype structure. This issue has been discussed across the literature~\cite{Cheetham2024,Leeman2024,Juelsholt2025}. 

Thirdly, while validity metrics and recovery rates are commonly treated as key performance indicators, simplistic metrics such as ``machine readability of generated structure files'' or ``successful termination of DFT calculations'' are insufficient for meaningful evaluation. The fundamental balance between validity (which can reflect memorization) and creativity (which enables exploration of chemical space) deserves greater attention in benchmarking frameworks. Current approaches often prioritize structural correctness without adequately assessing genuine materials discovery potential.

Last but not the least, the realism metrics, particularly synthesizability assessments, often suffer from bias present in reference datasets---a limitation that remains poorly discussed in the literature. These biases can systematically skew evaluation results and misrepresent model capabilities. All these interconnected aspects highlight the critical need for comprehensive, multi-criteria benchmarking standards.

\section{Applications}
\label{sec:applications}

\begin{figure*}
    \centering
    \begin{tabular}{c c c c c}
    \includegraphics[height=3cm]{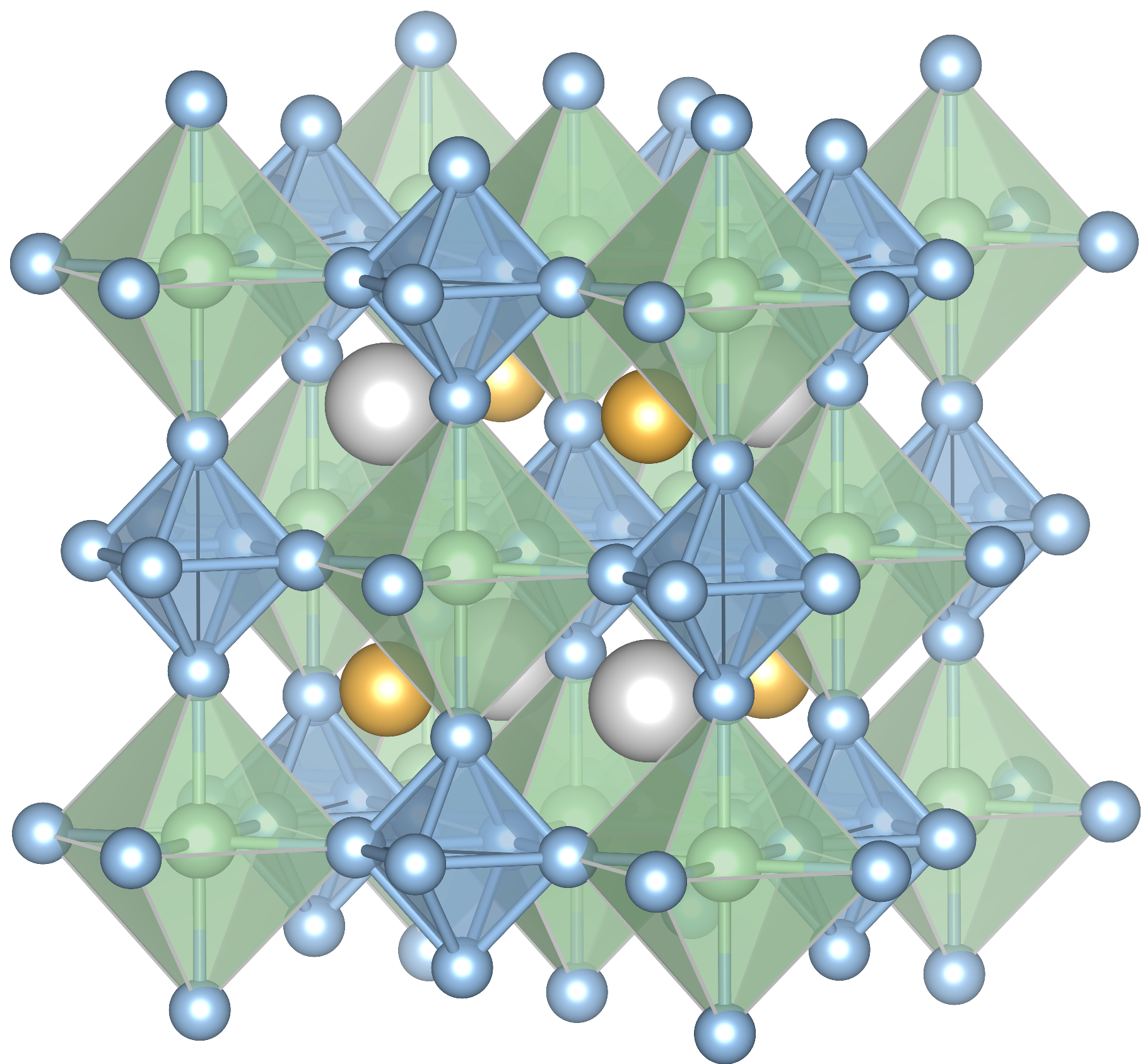}   &
    \includegraphics[height=3cm]{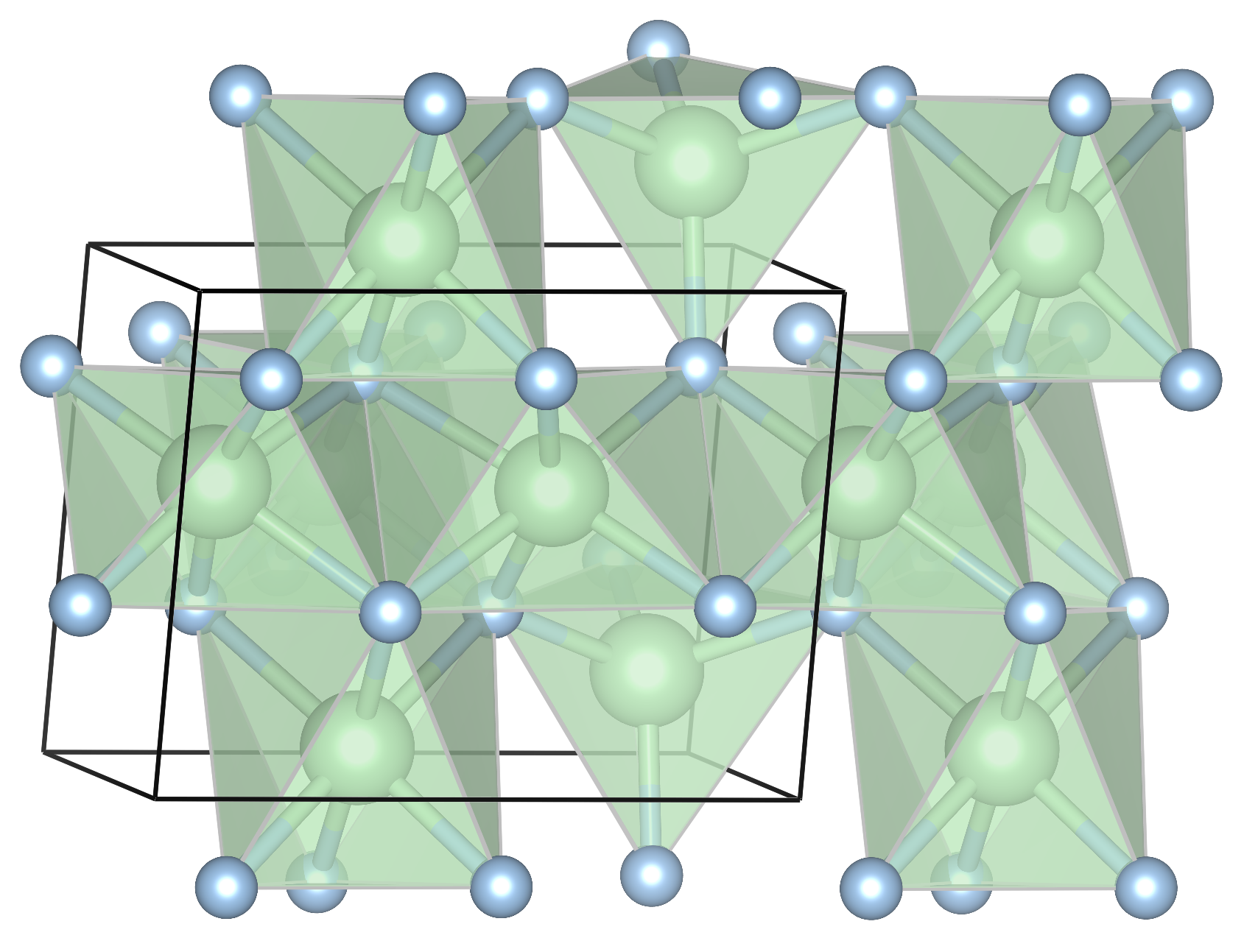}       &
    \includegraphics[height=3cm]{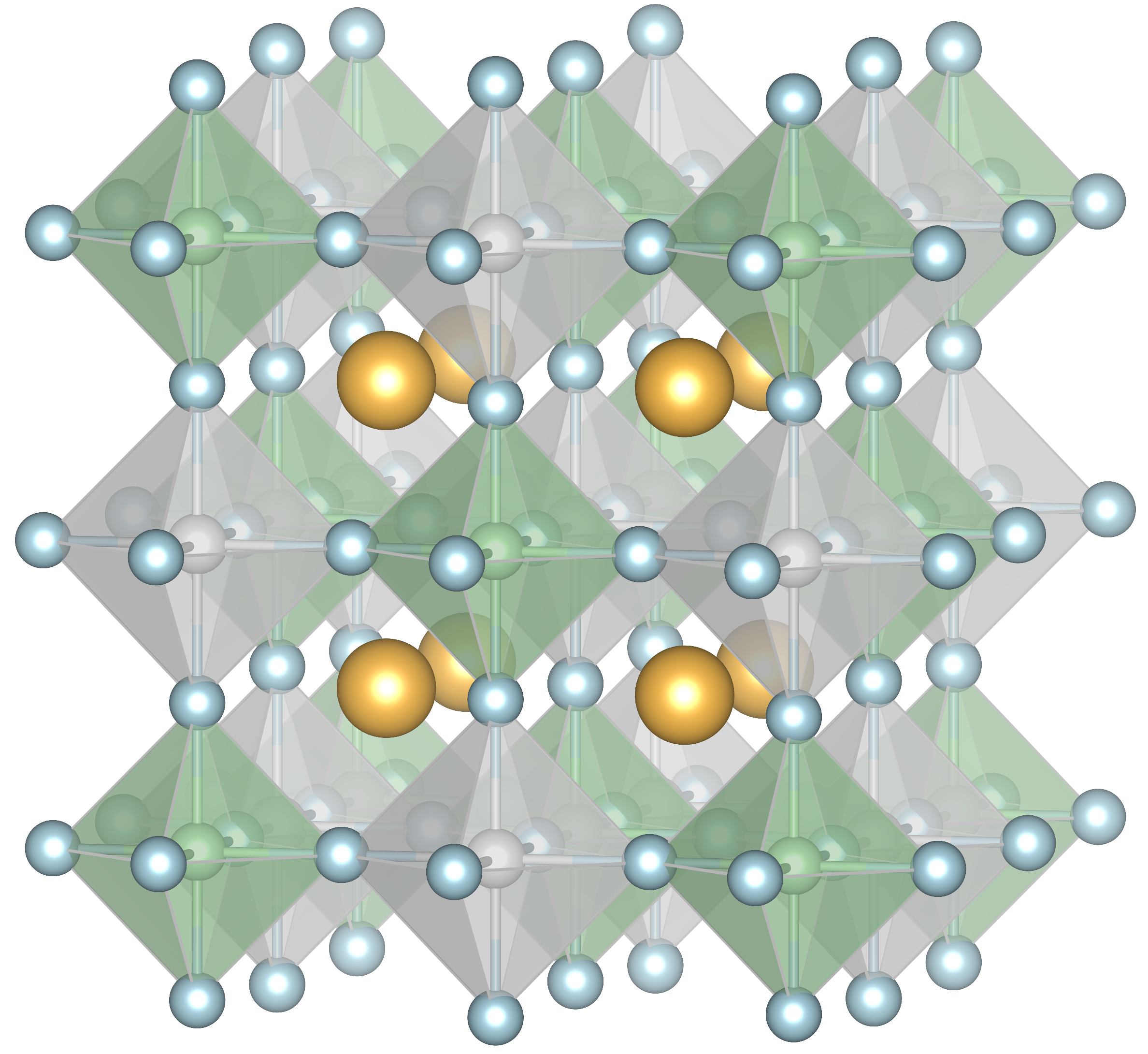} &
    \includegraphics[height=3cm]{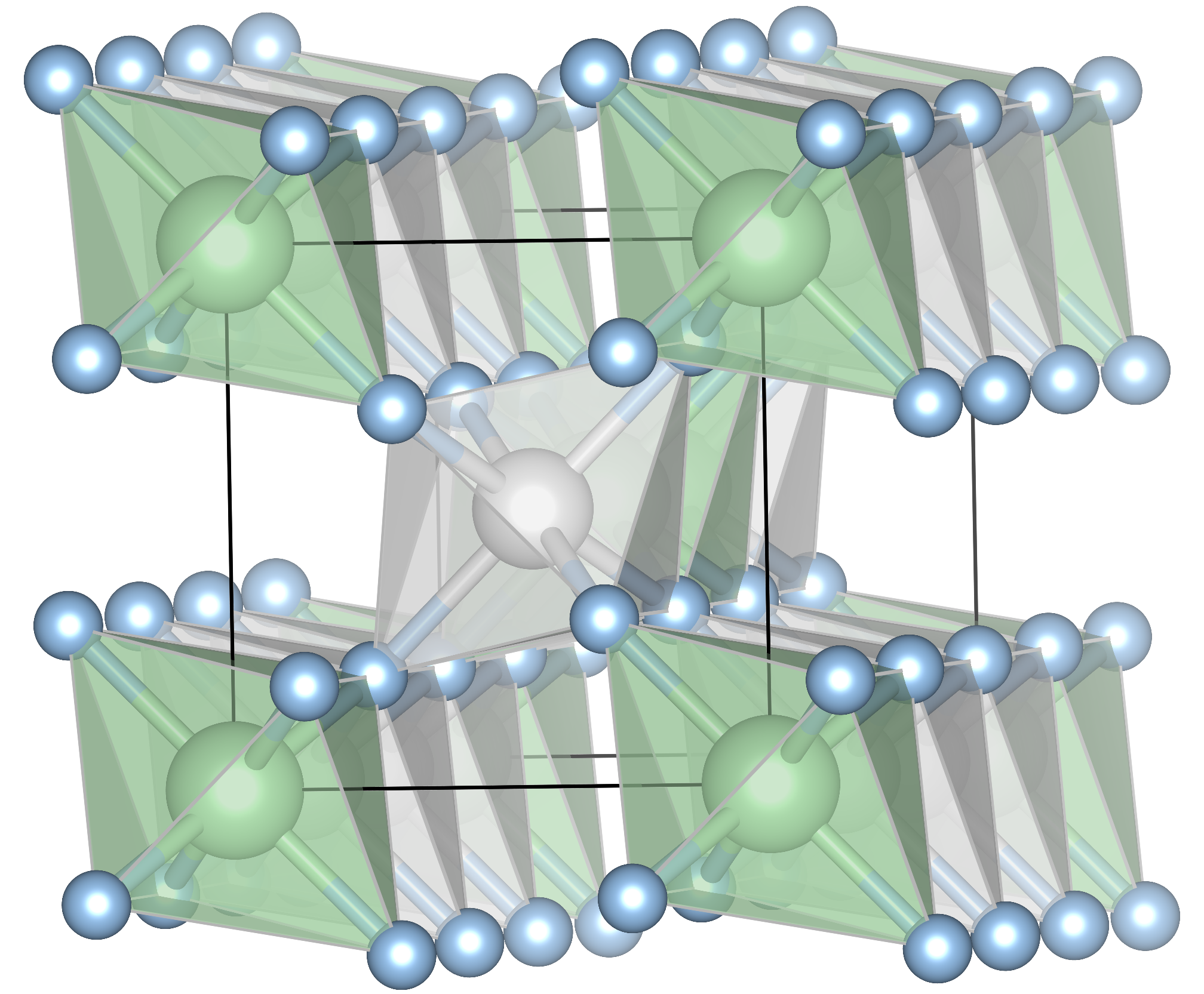}   &
    \includegraphics[height=3cm]{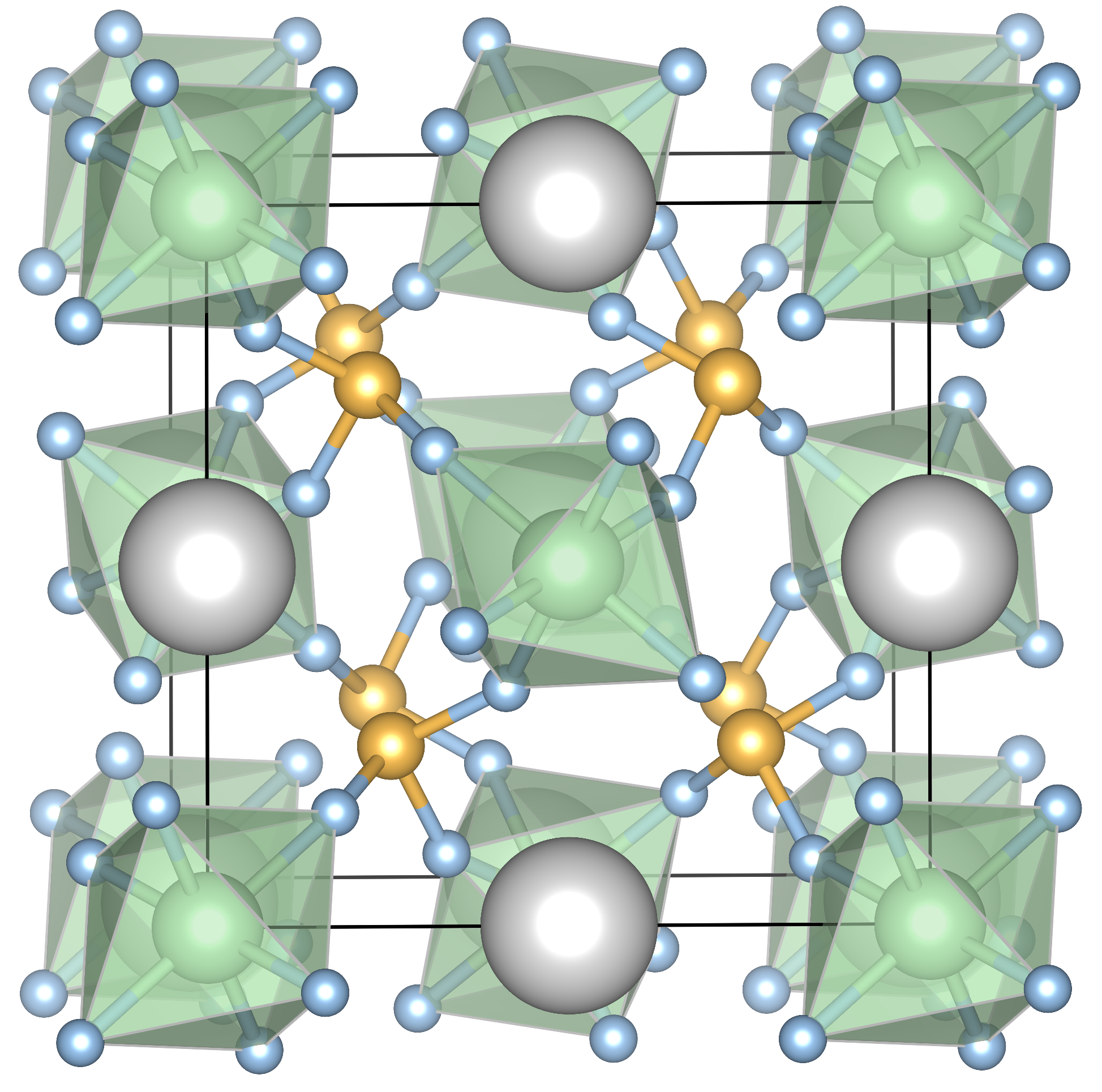} \\
    (a) \ce{KYNbSi6} & (b) \ce{V5O8} & (c) \ce{K2AgMoI6} & (d) \ce{TaCr2O6} & (e) \ce{CsLuSe2O6} \\
    \end{tabular}
    \caption{Examples of generated structures by a) cubicGAN, b) iMatGen, c) CrystaLLM, d) MatterGen, and e) Matra-Genoa models, replotted from the CIF files in Refs.~\cite{zhaoHighThroughputDiscoveryNovel2021b}, \cite{nohInverseDesignSolidState2019}, ~\cite{antunesCrystalStructureGeneration2024}, \cite{zeniGenerativeModelInorganic2025}, and ~\cite{breuckGenerativeMaterialTransformer2025}, respectively using VESTA~\cite{VESTA}.}
    \label{fig:structs}
\end{figure*}

Beyond benchmarking, the real value of generative models in materials science lies in their ability to discover new, stable compounds. Beyond simply reproducing known data, these models can explore novel chemistries, prototypes, and topologies, as shown by the generated structure for several models in \Cref{fig:structs}. This section reviews applications in two stages. First, \emph{large-scale, unconstrained screening}, where models generate vast libraries of candidate structures across broad structural classes. Second, \emph{targeted, property-driven generation}, where models are steered toward specific functional objectives, such as battery performance, superconductivity, or optoelectronic properties.

\subsection{Large-scale screening of crystal structures}

The CubicGAN model~\cite{zhaoHighThroughputDiscoveryNovel2021b} is an early example of large-scale generative screening in materials science. Trained to produce structures in three cubic space groups (\#216, \#225, and \#221), it generated 10 million candidates—recovering most known cubic crystals from MP and ICSD and identified 24 novel prototypes. Filtered by (i) CIF validity, (ii) composition uniqueness, (iii) charge neutrality, and (iv) negative formation energy, the model produced 17,303 ternary and 91,594 quaternary materials. DFT optimization succeeded for 83.4\% (14,433) of ternaries and 24.5\% (22,414) of quaternaries. Mechanical and dynamical stability screening yielded 506 stable structures, including four new prototypes. The method generalized beyond cubic symmetry~\cite{zhaoPhysicsGuidedDeep2023}, produced 1,869 DFT-optimized non-cubic structures, $\sim5.3$\% of which lied within 250~meV/atom of the Materials Project convex hull.

Variants of the CDVAE model~\cite{xieCrystalDiffusionVariational2022} have been applied to low-dimensional materials generation. Lyngby \textit{et al.}~\cite{Lyngby2022} trained on 2,615 2D structures from the C2DB database~\cite{Gjerding2021} and generated 10,000 candidates, 89\% of which passed geometric and charge neutrality checks. After deduplication and elemental substitution augmentation, systematic DFT calculations showed that the generated set has similar formation energies to the training set but with distinct compositions and structures. The approach was later extended to 1D materials~\cite{Moustafa2023}, produced 1,895 candidates, of which 1,008 were thermodynamically stable ($E_\text{hull}<200$~meV/atom) and included completely new prototypes.

MatterGen, trained on 607,683 structures from MP and Alexandria, further showed the ability of generative models to find S.U.N. structures. Interestingly, the uniqueness declined from 100\% for 1,000 samples to 52\% for 10 million, while novelty stayed near 60\% up to 1 million samples before dropping to $\sim$47\%. It recovered over 2,000 experimental ICSD structures absent from the training set. DFT validation of 1,000 samples showed RMSE $<$0.076~\AA\ between generated and optimized structures, with 75\% within 100~meV/atom of the convex hull.

\subsection{Generation of target functional materials}

Generative models can be fine-tuned or conditioned to produce candidates meeting specific functional targets. MatterGen, for example, supports conditioning on composition, space group, and mechanical, electronic, or magnetic properties. Parida \textit{et al}.~\cite{paridaMiningChemicalSpace2025} targeted Li-ion battery materials with $E_\text{hull}^\text{DFT}\le30$~meV/atom, and generated 32,600 candidates. Using the MatterSim interatomic potential~\cite{MatterSim} for screening, they identified 12,550 unique novel structures within 100~meV/atom, including 817 Li-compounds without heavy elements, $\sim$11\% of which met the target stability threshold.

MatterGen was also employed to generate metal oxides for thermochemical water-splitting (TCH) applications. Conditioned on \ehull{} $\le10$~meV/atom and 13 desired chemical systems, the model generated around 15\,000 samples. Further, the samples were pre-filtered using S.U.N checks (for stability the \ehull{} values were also predicted by the MatterSim potential). The most notable candidate is a novel thermodynamically stable quinary oxide \ce{Ba2SrInFeO6} that exhibits oxygen defect formation energies appropriate for TCH.

CDVAE was also used to design new conventional superconductors~\cite{10.1021/acs.jpclett.3c01260}. Trained on 1,058 JARVIS structures~\cite{jarvis}, it generated 3\,000 candidates, which were property-screened using a pretrained ALIGNN model~\cite{Choudhary2021}. While element abundance was similar to the training set, the generated set included novel stoichiometries and structures, though all belonged to space group $P1$, showing a limitation of the CDVAE model in generating high symmetry crystal structures.

The GANCSP model was applied to generate Mg--Mn--O ternaries for photoanode applications, and identified 23 new compounds with promising aqueous stability and band gaps. Two structures (MgMn$_4$O$_8$ and Mg$_2$MnO$_4$) were at or near the convex hull, representing stable new phases within DFT accuracy.

Recently, Okabe \textit{et al.} showed that structural constraints can be integrated into any diffusion-based generative model by strategic masking of the denoising step to steer the generation~\cite{Li2024SCIGEN}. Using DiffCSP trained on MP-20 dataset as the diffusion model, the SCIGEN model generated 7.87 million compounds in Archimedean lattices. Pre-filtering DFT calculations were performed on 26\,000 samples and 13\,800 materials were found to be converged to a maximum force smaller than 0.01~eV/\AA. 

Another example of targeting specific structural conditions can be found in Ref.~\cite{Gao2025Deep}. Via conditioning the CDVAE model, the ConditionCDVAE+ model was trained on a dataset of 19\,926 two-dimensional Janus III–VI van der Waals heterostructures and could generate novel heterostructures for specific chemical compositions. 

\section{Discussion and Future Directions}
\label{sec:discussion}

Generative AI for inorganic crystal structures has evolved from early feasibility demonstrations on narrow material classes to high-quality models that can produce symmetry-aware, stable, and sometimes property-targeted candidates. Despite this progress, much remains to be done. Below, we outline and discuss a research agenda organized around methodology, physics, control, data, and deployment.

As shown in~\Cref{fig:taxonomy}, several regions of the methodological design space remain unexplored. While some regions make less sense (e.g., diffusion models or LLMs on voxel), some remain underexplored (e.g., diffusion in reciprocal space or normalizing flows on Wyckoff positions). Other directions include masked denoising, manifold-constrained diffusion, and flow-matching on crystallographic manifolds (e.g., fixed space group/Wyckoff sets), which can enforce space-group consistency, charge neutrality, and minimal interatomic distances by design, thereby reducing the fraction of invalid samples and post-hoc filtering. We also expect more hybrid approaches combining complementary architectures and representations.

On the physics side, most current models assume fully ordered crystals at $T=0$~K with full site occupancies. Extending representations to include partial occupancies, i.e., alloying, vacancies, and point defects will be critical for applications in catalysis, solid electrolytes, and thermoelectrics. Addressing these non-idealities may require stepping beyond the unit cell and modeling larger-scale structures. Additional auxiliary information, such as magnetic moments and charge states, can also be integrated into structure representations. Moreover, conditioning generative models on thermodynamic variables (temperature, pressure, and chemical potentials) would allow targeting of metastable phases and entropically stabilized structures.

Beyond standard likelihood-based learning of $p(\mathbf{x})$ or $p(\mathbf{x}\mid\mathbf{c})$, coupling to energy surrogates $E_\phi(\mathbf{x})$ (e.g., uMLIPs or differentiable property predictors) enables Boltzmann-like likelihoods of the form $p(\mathbf{x}) \propto \exp(-\beta E_\phi(\mathbf{x}))$. This can be realized through energy-based models (EBMs), which directly parameterize $E_\theta(x)$ and rely on Markov chain Monte Carlo sampling~\cite{lecunTutorialEnergybasedLearning2006,duImplicitGenerationModeling2019}. Related approaches use energy as a reward signal: reinforcement learning can exploit $f(E(\mathbf{x}))$ for reward shaping, while Generative Flow Networks (GFlowNets~\cite{bengioFlowNetworkBased2021}) learn policies that sample proportional to a reward function.

Score-based methods, which learn $\nabla_x \log p(\mathbf{x})$ rather than $p(\mathbf{x})$ itself, can also be connected to energy-based perspectives since the score can be derived from the energy function. Diffusion models, for example, can be interpreted as score-based generative models~\cite{songScoreBasedGenerativeModeling2021}. These generative methods are therefore conceptually related to global optimization approaches---such as simulated annealing~\cite{kirkpatrickOptimizationSimulatedAnnealing1983} and minima hopping~\cite{goedeckerMinimaHoppingEfficient2004}---but are, in principle, more powerful, as they operate over high-dimensional probability distributions that encompass composition and other relevant aspects.


Further developments will also include better conditional control. Models with hard conditional guarantees (e.g., composition, charge balance, space group, prototype class) will increase success rates. Many applications also require balancing trade-offs, such as $\Delta E_\mathrm{hull}$, band gap, ionic conductivity, critical temperature, and abundance/toxicity. Multi-objective approaches such as GFlowNets and reinforcement learning with vector-valued rewards can explicitly learn to sample across Pareto fronts rather than optimize a single scalar. Natural-language conditioning is also promising: as demonstrated by Chemeleon, LLM-conditioned generators can parse free-text prompts (e.g., ``wide band gap, nontoxic, layered $p$-doped perovskite with large polaron mobility'') into structured constraints, supporting interactive dialogue.

On the data side, models are limited by the data they are trained on. Existing datasets are biased toward certain chemistries, prototypes, small unit cells, and GGA-level energetics. Better curation is needed to align training data with generation goals. This includes careful dataset design, clear licenses, transparent recipes, and provenance tracking for reproducibility. Online training, where new data are generated and used to fine-tune models, offers an attractive direction. Where available, attaching synthesis metadata (e.g., temperature, atmosphere, precursors, success/failure) can help bridge the gap between generative proposals and laboratory realization by enabling models to suggest not just what to make, but also how to make it.

On the evaluation side, as noted earlier, there is a lack of rigorous benchmarks. There is a need for improvement from simple sanity checks (validity, coverage, precision-recall rates) to real comparative benchmarks that include stability, novelty, and time complexity. Community-maintained suites should include frozen convex hulls to assess stability, fixed training sets with standard matching algorithms for novelty, and explicit reporting of computational cost for scalability. Benchmarks should span conditional tasks across chemistries and properties, not only unconditional generation. Following the example of property prediction~\cite{debreuckRobustModelBenchmarking2021}, incorporating calibrated uncertainties for structures and properties would enable more reliable screening and prioritization.

At scale, generative models will serve as seeds for new candidates in autonomous materials discovery. Crucially, generative AI should not be viewed as a one-shot predictor of ``the best'' material, but rather as a component in iterative pipelines. Fast, lightweight models are needed to propose candidates for downstream filtering by uMLIPs, high-throughput DFT, and experimental validation. Active learning loops, where models propose, labs validate, and data are recycled, will be central. Sustainability constraints---earth abundance, toxicity, embodied energy, and end-of-life---must also be built into discovery pipelines. Finally, models with interpretable outputs will allow experts to audit and steer generation, improving trust and accelerating insight.

\section{Conclusion}
\label{sec:conclusion}
Generative AI is beginning to reshape how we explore crystalline materials. By learning probability distributions from large databases, models such as VAEs, GANs, flows, diffusion, transformers, and LLMs can propose novel structures and, in some cases, target desired properties. In this review, we outlined the core architectures, representations, as well as data sources, and organized them through a taxonomy of representations, conditioning strategies, and application domains. More than 50 models have been discussed, with applications spanning large-scale screening, discovery of novel phases, and early advances in energy and catalysis.

Despite these advances, generative models for crystal structures are still in an early stage of development. Benchmarks are still fragmented, stability assessments vary, and synthesizability is often overlooked. Disorder, finite-temperature effects, and multi-objective trade-offs are not yet fully addressed. Fortunately, many promising paths exist to improve them---through advances in methodology, physics, control, data, and integration into discovery pipelines. With continued development, generative models could become indispensable tools for computational materials scientists, accelerating the discovery and understanding of the whole materials landscape.

\clearpage
\bibliography{com}

\begin{thebibliography}{133}%
\makeatletter
\providecommand \@ifxundefined [1]{%
 \@ifx{#1\undefined}
}%
\providecommand \@ifnum [1]{%
 \ifnum #1\expandafter \@firstoftwo
 \else \expandafter \@secondoftwo
 \fi
}%
\providecommand \@ifx [1]{%
 \ifx #1\expandafter \@firstoftwo
 \else \expandafter \@secondoftwo
 \fi
}%
\providecommand \natexlab [1]{#1}%
\providecommand \enquote  [1]{``#1''}%
\providecommand \bibnamefont  [1]{#1}%
\providecommand \bibfnamefont [1]{#1}%
\providecommand \citenamefont [1]{#1}%
\providecommand \href@noop [0]{\@secondoftwo}%
\providecommand \href [0]{\begingroup \@sanitize@url \@href}%
\providecommand \@href[1]{\@@startlink{#1}\@@href}%
\providecommand \@@href[1]{\endgroup#1\@@endlink}%
\providecommand \@sanitize@url [0]{\catcode `\\12\catcode `\$12\catcode `\&12\catcode `\#12\catcode `\^12\catcode `\_12\catcode `\%12\relax}%
\providecommand \@@startlink[1]{}%
\providecommand \@@endlink[0]{}%
\providecommand \url  [0]{\begingroup\@sanitize@url \@url }%
\providecommand \@url [1]{\endgroup\@href {#1}{\urlprefix }}%
\providecommand \urlprefix  [0]{URL }%
\providecommand \Eprint [0]{\href }%
\providecommand \doibase [0]{https://doi.org/}%
\providecommand \selectlanguage [0]{\@gobble}%
\providecommand \bibinfo  [0]{\@secondoftwo}%
\providecommand \bibfield  [0]{\@secondoftwo}%
\providecommand \translation [1]{[#1]}%
\providecommand \BibitemOpen [0]{}%
\providecommand \bibitemStop [0]{}%
\providecommand \bibitemNoStop [0]{.\EOS\space}%
\providecommand \EOS [0]{\spacefactor3000\relax}%
\providecommand \BibitemShut  [1]{\csname bibitem#1\endcsname}%
\let\auto@bib@innerbib\@empty
\bibitem [{\citenamefont {Lewis}(2007)}]{lewisCostEffectiveSolarEnergy2007}%
  \BibitemOpen
  \bibfield  {author} {\bibinfo {author} {\bibfnamefont {N.~S.}\ \bibnamefont {Lewis}},\ }\bibfield  {title} {\bibinfo {title} {Toward cost-effective solar energy use},\ }\href {https://doi.org/10.1126/science.1137014} {\bibfield  {journal} {\bibinfo  {journal} {Science}\ }\textbf {\bibinfo {volume} {315}},\ \bibinfo {pages} {798} (\bibinfo {year} {2007})}\BibitemShut {NoStop}%
\bibitem [{\citenamefont {Magee}(2012)}]{mageeQuantificationRoleMaterials2012}%
  \BibitemOpen
  \bibfield  {author} {\bibinfo {author} {\bibfnamefont {C.~L.}\ \bibnamefont {Magee}},\ }\bibfield  {title} {\bibinfo {title} {Towards quantification of the role of materials innovation in overall technological development},\ }\href {https://doi.org/10.1002/cplx.20309} {\bibfield  {journal} {\bibinfo  {journal} {Complexity}\ }\textbf {\bibinfo {volume} {18}},\ \bibinfo {pages} {10} (\bibinfo {year} {2012})}\BibitemShut {NoStop}%
\bibitem [{\citenamefont {Snyder}\ and\ \citenamefont {Toberer}(2010)}]{snyderComplexThermoelectricMaterials2010}%
  \BibitemOpen
  \bibfield  {author} {\bibinfo {author} {\bibfnamefont {G.~J.}\ \bibnamefont {Snyder}}\ and\ \bibinfo {author} {\bibfnamefont {E.~S.}\ \bibnamefont {Toberer}},\ }\bibfield  {title} {\bibinfo {title} {Complex thermoelectric materials},\ }in\ \href {https://doi.org/10.1142/9789814317665_0016} {\emph {\bibinfo {booktitle} {Materials for {{Sustainable Energy}}}}}\ (\bibinfo  {publisher} {Co-Published with Macmillan Publishers Ltd, UK},\ \bibinfo {year} {2010})\ pp.\ \bibinfo {pages} {101--110}\BibitemShut {NoStop}%
\bibitem [{\citenamefont {Oganov}\ and\ \citenamefont {Glass}(2006)}]{oganovCrystalStructurePrediction2006}%
  \BibitemOpen
  \bibfield  {author} {\bibinfo {author} {\bibfnamefont {A.~R.}\ \bibnamefont {Oganov}}\ and\ \bibinfo {author} {\bibfnamefont {C.~W.}\ \bibnamefont {Glass}},\ }\bibfield  {title} {\bibinfo {title} {Crystal structure prediction using ab initio evolutionary techniques: {Principles} and applications},\ }\href {https://doi.org/10.1063/1.2210932} {\bibfield  {journal} {\bibinfo  {journal} {J. Chem. Phys.}\ }\textbf {\bibinfo {volume} {124}},\ \bibinfo {pages} {244704} (\bibinfo {year} {2006})}\BibitemShut {NoStop}%
\bibitem [{\citenamefont {Podryabinkin}\ \emph {et~al.}(2019)\citenamefont {Podryabinkin}, \citenamefont {Tikhonov}, \citenamefont {Shapeev},\ and\ \citenamefont {Oganov}}]{podryabinkinAcceleratingCrystalStructure2019}%
  \BibitemOpen
  \bibfield  {author} {\bibinfo {author} {\bibfnamefont {E.~V.}\ \bibnamefont {Podryabinkin}}, \bibinfo {author} {\bibfnamefont {E.~V.}\ \bibnamefont {Tikhonov}}, \bibinfo {author} {\bibfnamefont {A.~V.}\ \bibnamefont {Shapeev}},\ and\ \bibinfo {author} {\bibfnamefont {A.~R.}\ \bibnamefont {Oganov}},\ }\bibfield  {title} {\bibinfo {title} {Accelerating crystal structure prediction by machine-learning interatomic potentials with active learning},\ }\href {https://doi.org/10.1103/PhysRevB.99.064114} {\bibfield  {journal} {\bibinfo  {journal} {Phys. Rev. B}\ }\textbf {\bibinfo {volume} {99}},\ \bibinfo {pages} {064114} (\bibinfo {year} {2019})}\BibitemShut {NoStop}%
\bibitem [{\citenamefont {Pickard}\ and\ \citenamefont {Needs}(2011)}]{pickardInitioRandomStructure2011}%
  \BibitemOpen
  \bibfield  {author} {\bibinfo {author} {\bibfnamefont {C.~J.}\ \bibnamefont {Pickard}}\ and\ \bibinfo {author} {\bibfnamefont {R.~J.}\ \bibnamefont {Needs}},\ }\bibfield  {title} {\bibinfo {title} {Ab initio random structure searching},\ }\href {https://doi.org/10.1088/0953-8984/23/5/053201} {\bibfield  {journal} {\bibinfo  {journal} {J. Phys.: Condens. Matter}\ }\textbf {\bibinfo {volume} {23}},\ \bibinfo {pages} {053201} (\bibinfo {year} {2011})}\BibitemShut {NoStop}%
\bibitem [{\citenamefont {Wang}\ \emph {et~al.}(2010)\citenamefont {Wang}, \citenamefont {Lv}, \citenamefont {Zhu},\ and\ \citenamefont {Ma}}]{wangCrystalStructurePrediction2010}%
  \BibitemOpen
  \bibfield  {author} {\bibinfo {author} {\bibfnamefont {Y.}~\bibnamefont {Wang}}, \bibinfo {author} {\bibfnamefont {J.}~\bibnamefont {Lv}}, \bibinfo {author} {\bibfnamefont {L.}~\bibnamefont {Zhu}},\ and\ \bibinfo {author} {\bibfnamefont {Y.}~\bibnamefont {Ma}},\ }\bibfield  {title} {\bibinfo {title} {Crystal structure prediction via particle-swarm optimization},\ }\href {https://doi.org/10.1103/PhysRevB.82.094116} {\bibfield  {journal} {\bibinfo  {journal} {Phys. Rev. B}\ }\textbf {\bibinfo {volume} {82}},\ \bibinfo {pages} {094116} (\bibinfo {year} {2010})}\BibitemShut {NoStop}%
\bibitem [{\citenamefont {Goedecker}(2004)}]{goedeckerMinimaHoppingEfficient2004}%
  \BibitemOpen
  \bibfield  {author} {\bibinfo {author} {\bibfnamefont {S.}~\bibnamefont {Goedecker}},\ }\bibfield  {title} {\bibinfo {title} {Minima hopping: {{An}} efficient search method for the global minimum of the potential energy surface of complex molecular systems},\ }\href {https://doi.org/10.1063/1.1724816} {\bibfield  {journal} {\bibinfo  {journal} {J. Chem. Phys.}\ }\textbf {\bibinfo {volume} {120}},\ \bibinfo {pages} {9911} (\bibinfo {year} {2004})}\BibitemShut {NoStop}%
\bibitem [{\citenamefont {Yamashita}\ \emph {et~al.}(2021)\citenamefont {Yamashita}, \citenamefont {~}, \citenamefont {~}, \citenamefont {~}, \citenamefont {~}, \citenamefont {~}, \citenamefont {~}, \citenamefont {~}, \citenamefont {~}, \citenamefont {~},\ and\ \citenamefont {{and Oguchi}}}]{yamashitaCrySPYCrystalStructure2021a}%
  \BibitemOpen
  \bibfield  {author} {\bibinfo {author} {\bibfnamefont {T.}~\bibnamefont {Yamashita}}, \bibinfo {author} {\bibfnamefont {K.}~\bibnamefont {~}, \bibfnamefont {Shinichi}}, \bibinfo {author} {\bibfnamefont {S.}~\bibnamefont {~}, \bibfnamefont {Nobuya}}, \bibinfo {author} {\bibfnamefont {K.}~\bibnamefont {~}, \bibfnamefont {Hiori}}, \bibinfo {author} {\bibfnamefont {T.}~\bibnamefont {~}, \bibfnamefont {Kei}}, \bibinfo {author} {\bibfnamefont {S.}~\bibnamefont {~}, \bibfnamefont {Hikaru}}, \bibinfo {author} {\bibfnamefont {S.}~\bibnamefont {~}, \bibfnamefont {Takumi}}, \bibinfo {author} {\bibfnamefont {U.}~\bibnamefont {~}, \bibfnamefont {Futoshi}}, \bibinfo {author} {\bibfnamefont {T.}~\bibnamefont {~}, \bibfnamefont {Koji}}, \bibinfo {author} {\bibfnamefont {M.}~\bibnamefont {~}, \bibfnamefont {Takashi}},\ and\ \bibinfo {author} {\bibfnamefont {T.}~\bibnamefont {{and Oguchi}}},\ }\bibfield  {title} {\bibinfo {title} {{{CrySPY}}: {A} crystal structure prediction tool accelerated by machine learning},\ }\href {https://doi.org/10.1080/27660400.2021.1943171} {\bibfield  {journal} {\bibinfo  {journal} {Sci. Technol. Adv. Mater.: Methods}\ }\textbf {\bibinfo {volume} {1}},\ \bibinfo {pages} {87} (\bibinfo {year} {2021})}\BibitemShut {NoStop}%
\bibitem [{\citenamefont {Falls}\ \emph {et~al.}(2021)\citenamefont {Falls}, \citenamefont {Avery}, \citenamefont {Wang}, \citenamefont {Hilleke},\ and\ \citenamefont {Zurek}}]{fallsXtalOptEvolutionaryAlgorithm2021}%
  \BibitemOpen
  \bibfield  {author} {\bibinfo {author} {\bibfnamefont {Z.}~\bibnamefont {Falls}}, \bibinfo {author} {\bibfnamefont {P.}~\bibnamefont {Avery}}, \bibinfo {author} {\bibfnamefont {X.}~\bibnamefont {Wang}}, \bibinfo {author} {\bibfnamefont {K.~P.}\ \bibnamefont {Hilleke}},\ and\ \bibinfo {author} {\bibfnamefont {E.}~\bibnamefont {Zurek}},\ }\bibfield  {title} {\bibinfo {title} {The {XtalOpt} evolutionary algorithm for crystal structure prediction},\ }\href {https://doi.org/10.1021/acs.jpcc.0c09531} {\bibfield  {journal} {\bibinfo  {journal} {J. Phys. Chem. C}\ }\textbf {\bibinfo {volume} {125}},\ \bibinfo {pages} {1601} (\bibinfo {year} {2021})}\BibitemShut {NoStop}%
\bibitem [{\citenamefont {Kingma}\ and\ \citenamefont {Welling}(2022)}]{kingmaAutoEncodingVariationalBayes2022}%
  \BibitemOpen
  \bibfield  {author} {\bibinfo {author} {\bibfnamefont {D.~P.}\ \bibnamefont {Kingma}}\ and\ \bibinfo {author} {\bibfnamefont {M.}~\bibnamefont {Welling}},\ }\href {https://doi.org/10.48550/arXiv.1312.6114} {\bibinfo {title} {Auto-encoding variational bayes}} (\bibinfo {year} {2022}),\ \Eprint {https://arxiv.org/abs/1312.6114} {arXiv:1312.6114 [stat]} \BibitemShut {NoStop}%
\bibitem [{\citenamefont {Goodfellow}\ \emph {et~al.}(2014)\citenamefont {Goodfellow}, \citenamefont {{Pouget-Abadie}}, \citenamefont {Mirza}, \citenamefont {Xu}, \citenamefont {{Warde-Farley}}, \citenamefont {Ozair}, \citenamefont {Courville},\ and\ \citenamefont {Bengio}}]{goodfellowGenerativeAdversarialNets2014}%
  \BibitemOpen
  \bibfield  {author} {\bibinfo {author} {\bibfnamefont {I.~J.}\ \bibnamefont {Goodfellow}}, \bibinfo {author} {\bibfnamefont {J.}~\bibnamefont {{Pouget-Abadie}}}, \bibinfo {author} {\bibfnamefont {M.}~\bibnamefont {Mirza}}, \bibinfo {author} {\bibfnamefont {B.}~\bibnamefont {Xu}}, \bibinfo {author} {\bibfnamefont {D.}~\bibnamefont {{Warde-Farley}}}, \bibinfo {author} {\bibfnamefont {S.}~\bibnamefont {Ozair}}, \bibinfo {author} {\bibfnamefont {A.}~\bibnamefont {Courville}},\ and\ \bibinfo {author} {\bibfnamefont {Y.}~\bibnamefont {Bengio}},\ }\bibfield  {title} {\bibinfo {title} {Generative adversarial nets},\ }in\ \href@noop {} {\emph {\bibinfo {booktitle} {Advances in Neural Information Processing Systems}}},\ Vol.~\bibinfo {volume} {27},\ \bibinfo {editor} {edited by\ \bibinfo {editor} {\bibfnamefont {Z.}~\bibnamefont {Ghahramani}}, \bibinfo {editor} {\bibfnamefont {M.}~\bibnamefont {Welling}}, \bibinfo {editor} {\bibfnamefont {C.}~\bibnamefont {Cortes}}, \bibinfo {editor} {\bibfnamefont {N.}~\bibnamefont {Lawrence}},\ and\ \bibinfo {editor} {\bibfnamefont {K.}~\bibnamefont {Weinberger}}}\ (\bibinfo  {publisher} {Curran Associates, Inc.},\ \bibinfo {year} {2014})\BibitemShut {NoStop}%
\bibitem [{\citenamefont {Radford}\ \emph {et~al.}(2016)\citenamefont {Radford}, \citenamefont {Metz},\ and\ \citenamefont {Chintala}}]{radfordUnsupervisedRepresentationLearning2016}%
  \BibitemOpen
  \bibfield  {author} {\bibinfo {author} {\bibfnamefont {A.}~\bibnamefont {Radford}}, \bibinfo {author} {\bibfnamefont {L.}~\bibnamefont {Metz}},\ and\ \bibinfo {author} {\bibfnamefont {S.}~\bibnamefont {Chintala}},\ }\href {https://doi.org/10.48550/arXiv.1511.06434} {\bibinfo {title} {Unsupervised representation learning with deep convolutional generative adversarial networks}} (\bibinfo {year} {2016}),\ \Eprint {https://arxiv.org/abs/1511.06434} {arXiv:1511.06434 [cs]} \BibitemShut {NoStop}%
\bibitem [{\citenamefont {Vaswani}\ \emph {et~al.}(2017)\citenamefont {Vaswani}, \citenamefont {Shazeer}, \citenamefont {Parmar}, \citenamefont {Uszkoreit}, \citenamefont {Jones}, \citenamefont {Gomez}, \citenamefont {Kaiser},\ and\ \citenamefont {Polosukhin}}]{vaswaniAttentionAllYou2017}%
  \BibitemOpen
  \bibfield  {author} {\bibinfo {author} {\bibfnamefont {A.}~\bibnamefont {Vaswani}}, \bibinfo {author} {\bibfnamefont {N.}~\bibnamefont {Shazeer}}, \bibinfo {author} {\bibfnamefont {N.}~\bibnamefont {Parmar}}, \bibinfo {author} {\bibfnamefont {J.}~\bibnamefont {Uszkoreit}}, \bibinfo {author} {\bibfnamefont {L.}~\bibnamefont {Jones}}, \bibinfo {author} {\bibfnamefont {A.~N.}\ \bibnamefont {Gomez}}, \bibinfo {author} {\bibfnamefont {{\L}.}~\bibnamefont {Kaiser}},\ and\ \bibinfo {author} {\bibfnamefont {I.}~\bibnamefont {Polosukhin}},\ }\bibfield  {title} {\bibinfo {title} {Attention is all you need},\ }in\ \href@noop {} {\emph {\bibinfo {booktitle} {Advances in Neural Information Processing Systems}}},\ Vol.~\bibinfo {volume} {30},\ \bibinfo {editor} {edited by\ \bibinfo {editor} {\bibfnamefont {I.}~\bibnamefont {Guyon}}, \bibinfo {editor} {\bibfnamefont {U.~V.}\ \bibnamefont {Luxburg}}, \bibinfo {editor} {\bibfnamefont {S.}~\bibnamefont {Bengio}}, \bibinfo {editor} {\bibfnamefont {H.}~\bibnamefont {Wallach}}, \bibinfo {editor} {\bibfnamefont {R.}~\bibnamefont {Fergus}}, \bibinfo {editor} {\bibfnamefont {S.}~\bibnamefont {Vishwanathan}},\ and\ \bibinfo {editor} {\bibfnamefont {R.}~\bibnamefont {Garnett}}}\ (\bibinfo  {publisher} {Curran Associates, Inc.},\ \bibinfo {year} {2017})\BibitemShut {NoStop}%
\bibitem [{\citenamefont {Rezende}\ and\ \citenamefont {Mohamed}(2015)}]{rezendeVariationalInferenceNormalizing2015}%
  \BibitemOpen
  \bibfield  {author} {\bibinfo {author} {\bibfnamefont {D.}~\bibnamefont {Rezende}}\ and\ \bibinfo {author} {\bibfnamefont {S.}~\bibnamefont {Mohamed}},\ }\bibfield  {title} {\bibinfo {title} {Variational inference with normalizing flows},\ }in\ \href@noop {} {\emph {\bibinfo {booktitle} {Proceedings of the 32nd International Conference on Machine Learning}}},\ \bibinfo {series} {Proceedings of Machine Learning Research}, Vol.~\bibinfo {volume} {37},\ \bibinfo {editor} {edited by\ \bibinfo {editor} {\bibfnamefont {F.}~\bibnamefont {Bach}}\ and\ \bibinfo {editor} {\bibfnamefont {D.}~\bibnamefont {Blei}}}\ (\bibinfo  {publisher} {PMLR},\ \bibinfo {address} {Lille, France},\ \bibinfo {year} {2015})\ pp.\ \bibinfo {pages} {1530--1538}\BibitemShut {NoStop}%
\bibitem [{\citenamefont {Dinh}\ \emph {et~al.}(2017)\citenamefont {Dinh}, \citenamefont {{Sohl-Dickstein}},\ and\ \citenamefont {Bengio}}]{dinhDensityEstimationUsing2017}%
  \BibitemOpen
  \bibfield  {author} {\bibinfo {author} {\bibfnamefont {L.}~\bibnamefont {Dinh}}, \bibinfo {author} {\bibfnamefont {J.}~\bibnamefont {{Sohl-Dickstein}}},\ and\ \bibinfo {author} {\bibfnamefont {S.}~\bibnamefont {Bengio}},\ }\bibfield  {title} {\bibinfo {title} {Density estimation using real {{NVP}}.},\ }in\ \href@noop {} {\emph {\bibinfo {booktitle} {{{ICLR}} (Poster)}}}\ (\bibinfo  {publisher} {OpenReview.net},\ \bibinfo {year} {2017})\BibitemShut {NoStop}%
\bibitem [{\citenamefont {Lipman}\ \emph {et~al.}(2023)\citenamefont {Lipman}, \citenamefont {Chen}, \citenamefont {{Ben-Hamu}}, \citenamefont {Nickel},\ and\ \citenamefont {Le}}]{lipmanFlowMatchingGenerative2023}%
  \BibitemOpen
  \bibfield  {author} {\bibinfo {author} {\bibfnamefont {Y.}~\bibnamefont {Lipman}}, \bibinfo {author} {\bibfnamefont {R.~T.~Q.}\ \bibnamefont {Chen}}, \bibinfo {author} {\bibfnamefont {H.}~\bibnamefont {{Ben-Hamu}}}, \bibinfo {author} {\bibfnamefont {M.}~\bibnamefont {Nickel}},\ and\ \bibinfo {author} {\bibfnamefont {M.}~\bibnamefont {Le}},\ }\href {https://doi.org/10.48550/arXiv.2210.02747} {\bibinfo {title} {Flow matching for generative modeling}} (\bibinfo {year} {2023}),\ \Eprint {https://arxiv.org/abs/2210.02747} {arXiv:2210.02747 [cs]} \BibitemShut {NoStop}%
\bibitem [{\citenamefont {Dhariwal}\ and\ \citenamefont {Nichol}(2021)}]{dhariwalDiffusionModelsBeat2021}%
  \BibitemOpen
  \bibfield  {author} {\bibinfo {author} {\bibfnamefont {P.}~\bibnamefont {Dhariwal}}\ and\ \bibinfo {author} {\bibfnamefont {A.}~\bibnamefont {Nichol}},\ }\bibfield  {title} {\bibinfo {title} {Diffusion models beat gans on image synthesis},\ }in\ \href@noop {} {\emph {\bibinfo {booktitle} {Advances in Neural Information Processing Systems}}},\ Vol.~\bibinfo {volume} {34},\ \bibinfo {editor} {edited by\ \bibinfo {editor} {\bibfnamefont {M.}~\bibnamefont {Ranzato}}, \bibinfo {editor} {\bibfnamefont {A.}~\bibnamefont {Beygelzimer}}, \bibinfo {editor} {\bibfnamefont {Y.}~\bibnamefont {Dauphin}}, \bibinfo {editor} {\bibfnamefont {P.}~\bibnamefont {Liang}},\ and\ \bibinfo {editor} {\bibfnamefont {J.~W.}\ \bibnamefont {Vaughan}}}\ (\bibinfo  {publisher} {Curran Associates, Inc.},\ \bibinfo {year} {2021})\ pp.\ \bibinfo {pages} {8780--8794}\BibitemShut {NoStop}%
\bibitem [{\citenamefont {Cai}\ \emph {et~al.}(2020)\citenamefont {Cai}, \citenamefont {Yang}, \citenamefont {{Averbuch-Elor}}, \citenamefont {Hao}, \citenamefont {Belongie}, \citenamefont {Snavely},\ and\ \citenamefont {Hariharan}}]{caiLearningGradientFields2020}%
  \BibitemOpen
  \bibfield  {author} {\bibinfo {author} {\bibfnamefont {R.}~\bibnamefont {Cai}}, \bibinfo {author} {\bibfnamefont {G.}~\bibnamefont {Yang}}, \bibinfo {author} {\bibfnamefont {H.}~\bibnamefont {{Averbuch-Elor}}}, \bibinfo {author} {\bibfnamefont {Z.}~\bibnamefont {Hao}}, \bibinfo {author} {\bibfnamefont {S.}~\bibnamefont {Belongie}}, \bibinfo {author} {\bibfnamefont {N.}~\bibnamefont {Snavely}},\ and\ \bibinfo {author} {\bibfnamefont {B.}~\bibnamefont {Hariharan}},\ }\href {https://doi.org/10.48550/arXiv.2008.06520} {\bibinfo {title} {Learning {{Gradient Fields}} for {{Shape Generation}}}} (\bibinfo {year} {2020}),\ \Eprint {https://arxiv.org/abs/2008.06520} {arXiv:2008.06520 [cs]} \BibitemShut {NoStop}%
\bibitem [{\citenamefont {Shi}\ \emph {et~al.}(2021)\citenamefont {Shi}, \citenamefont {Luo}, \citenamefont {Xu},\ and\ \citenamefont {Tang}}]{shiLearningGradientFields2021}%
  \BibitemOpen
  \bibfield  {author} {\bibinfo {author} {\bibfnamefont {C.}~\bibnamefont {Shi}}, \bibinfo {author} {\bibfnamefont {S.}~\bibnamefont {Luo}}, \bibinfo {author} {\bibfnamefont {M.}~\bibnamefont {Xu}},\ and\ \bibinfo {author} {\bibfnamefont {J.}~\bibnamefont {Tang}},\ }\bibfield  {title} {\bibinfo {title} {Learning gradient fields for molecular conformation generation},\ }in\ \href@noop {} {\emph {\bibinfo {booktitle} {Proceedings of the 38th {{International Conference}} on {{Machine Learning}}}}}\ (\bibinfo  {publisher} {PMLR},\ \bibinfo {year} {2021})\ pp.\ \bibinfo {pages} {9558--9568}\BibitemShut {NoStop}%
\bibitem [{\citenamefont {{Sohl-Dickstein}}\ \emph {et~al.}(2015)\citenamefont {{Sohl-Dickstein}}, \citenamefont {Weiss}, \citenamefont {Maheswaranathan},\ and\ \citenamefont {Ganguli}}]{sohl-dicksteinDeepUnsupervisedLearning2015}%
  \BibitemOpen
  \bibfield  {author} {\bibinfo {author} {\bibfnamefont {J.}~\bibnamefont {{Sohl-Dickstein}}}, \bibinfo {author} {\bibfnamefont {E.}~\bibnamefont {Weiss}}, \bibinfo {author} {\bibfnamefont {N.}~\bibnamefont {Maheswaranathan}},\ and\ \bibinfo {author} {\bibfnamefont {S.}~\bibnamefont {Ganguli}},\ }\bibfield  {title} {\bibinfo {title} {Deep unsupervised learning using nonequilibrium thermodynamics},\ }in\ \href@noop {} {\emph {\bibinfo {booktitle} {Proceedings of the 32nd International Conference on Machine Learning}}},\ \bibinfo {series} {Proceedings of Machine Learning Research}, Vol.~\bibinfo {volume} {37},\ \bibinfo {editor} {edited by\ \bibinfo {editor} {\bibfnamefont {F.}~\bibnamefont {Bach}}\ and\ \bibinfo {editor} {\bibfnamefont {D.}~\bibnamefont {Blei}}}\ (\bibinfo  {publisher} {PMLR},\ \bibinfo {address} {Lille, France},\ \bibinfo {year} {2015})\ pp.\ \bibinfo {pages} {2256--2265}\BibitemShut {NoStop}%
\bibitem [{\citenamefont {Ho}\ \emph {et~al.}(2020)\citenamefont {Ho}, \citenamefont {Jain},\ and\ \citenamefont {Abbeel}}]{hoDenoisingDiffusionProbabilistic2020a}%
  \BibitemOpen
  \bibfield  {author} {\bibinfo {author} {\bibfnamefont {J.}~\bibnamefont {Ho}}, \bibinfo {author} {\bibfnamefont {A.}~\bibnamefont {Jain}},\ and\ \bibinfo {author} {\bibfnamefont {P.}~\bibnamefont {Abbeel}},\ }\bibfield  {title} {\bibinfo {title} {Denoising diffusion probabilistic models},\ }in\ \href@noop {} {\emph {\bibinfo {booktitle} {Advances in {{Neural Information Processing Systems}}}}},\ Vol.~\bibinfo {volume} {33}\ (\bibinfo  {publisher} {Curran Associates, Inc.},\ \bibinfo {year} {2020})\ pp.\ \bibinfo {pages} {6840--6851}\BibitemShut {NoStop}%
\bibitem [{\citenamefont {Song}\ and\ \citenamefont {Ermon}(2019)}]{songGenerativeModelingEstimating2019}%
  \BibitemOpen
  \bibfield  {author} {\bibinfo {author} {\bibfnamefont {Y.}~\bibnamefont {Song}}\ and\ \bibinfo {author} {\bibfnamefont {S.}~\bibnamefont {Ermon}},\ }\bibfield  {title} {\bibinfo {title} {Generative modeling by estimating gradients of the data distribution},\ }in\ \href@noop {} {\emph {\bibinfo {booktitle} {Advances in {{Neural Information Processing Systems}}}}},\ Vol.~\bibinfo {volume} {32}\ (\bibinfo  {publisher} {Curran Associates, Inc.},\ \bibinfo {year} {2019})\BibitemShut {NoStop}%
\bibitem [{\citenamefont {Song}\ \emph {et~al.}(2021)\citenamefont {Song}, \citenamefont {{Sohl-Dickstein}}, \citenamefont {Kingma}, \citenamefont {Kumar}, \citenamefont {Ermon},\ and\ \citenamefont {Poole}}]{songScoreBasedGenerativeModeling2021}%
  \BibitemOpen
  \bibfield  {author} {\bibinfo {author} {\bibfnamefont {Y.}~\bibnamefont {Song}}, \bibinfo {author} {\bibfnamefont {J.}~\bibnamefont {{Sohl-Dickstein}}}, \bibinfo {author} {\bibfnamefont {D.~P.}\ \bibnamefont {Kingma}}, \bibinfo {author} {\bibfnamefont {A.}~\bibnamefont {Kumar}}, \bibinfo {author} {\bibfnamefont {S.}~\bibnamefont {Ermon}},\ and\ \bibinfo {author} {\bibfnamefont {B.}~\bibnamefont {Poole}},\ }\href {https://doi.org/10.48550/arXiv.2011.13456} {\bibinfo {title} {Score-based generative modeling through stochastic differential equations}} (\bibinfo {year} {2021}),\ \Eprint {https://arxiv.org/abs/2011.13456} {arXiv:2011.13456 [cs]} \BibitemShut {NoStop}%
\bibitem [{\citenamefont {Radford}\ \emph {et~al.}(2018)\citenamefont {Radford}, \citenamefont {Narasimhan}, \citenamefont {Salimans},\ and\ \citenamefont {Sutskever}}]{radford2018improving}%
  \BibitemOpen
  \bibfield  {author} {\bibinfo {author} {\bibfnamefont {A.}~\bibnamefont {Radford}}, \bibinfo {author} {\bibfnamefont {K.}~\bibnamefont {Narasimhan}}, \bibinfo {author} {\bibfnamefont {T.}~\bibnamefont {Salimans}},\ and\ \bibinfo {author} {\bibfnamefont {I.}~\bibnamefont {Sutskever}},\ }\bibfield  {title} {\bibinfo {title} {Improving language understanding by generative pre-training},\ }\href@noop {} {\bibfield  {journal} {\bibinfo  {journal} {Preprint}\ } (\bibinfo {year} {2018})}\BibitemShut {NoStop}%
\bibitem [{\citenamefont {Devlin}\ \emph {et~al.}(2019)\citenamefont {Devlin}, \citenamefont {Chang}, \citenamefont {Lee},\ and\ \citenamefont {Toutanova}}]{devlinBERTPretrainingDeep2019}%
  \BibitemOpen
  \bibfield  {author} {\bibinfo {author} {\bibfnamefont {J.}~\bibnamefont {Devlin}}, \bibinfo {author} {\bibfnamefont {M.-W.}\ \bibnamefont {Chang}}, \bibinfo {author} {\bibfnamefont {K.}~\bibnamefont {Lee}},\ and\ \bibinfo {author} {\bibfnamefont {K.}~\bibnamefont {Toutanova}},\ }\href {https://doi.org/10.48550/arXiv.1810.04805} {\bibinfo {title} {{{BERT}}: {{Pre-training}} of deep bidirectional transformers for language understanding}} (\bibinfo {year} {2019}),\ \Eprint {https://arxiv.org/abs/1810.04805} {arXiv:1810.04805 [cs]} \BibitemShut {NoStop}%
\bibitem [{\citenamefont {Brown}\ \emph {et~al.}(2020)\citenamefont {Brown}, \citenamefont {Mann}, \citenamefont {Ryder}, \citenamefont {Subbiah}, \citenamefont {Kaplan}, \citenamefont {Dhariwal}, \citenamefont {Neelakantan}, \citenamefont {Shyam}, \citenamefont {Sastry}, \citenamefont {Askell}, \citenamefont {Agarwal}, \citenamefont {{Herbert-Voss}}, \citenamefont {Krueger}, \citenamefont {Henighan}, \citenamefont {Child}, \citenamefont {Ramesh}, \citenamefont {Ziegler}, \citenamefont {Wu}, \citenamefont {Winter}, \citenamefont {Hesse}, \citenamefont {Chen}, \citenamefont {Sigler}, \citenamefont {Litwin}, \citenamefont {Gray}, \citenamefont {Chess}, \citenamefont {Clark}, \citenamefont {Berner}, \citenamefont {McCandlish}, \citenamefont {Radford}, \citenamefont {Sutskever},\ and\ \citenamefont {Amodei}}]{brownLanguageModelsAre2020}%
  \BibitemOpen
  \bibfield  {author} {\bibinfo {author} {\bibfnamefont {T.}~\bibnamefont {Brown}}, \bibinfo {author} {\bibfnamefont {B.}~\bibnamefont {Mann}}, \bibinfo {author} {\bibfnamefont {N.}~\bibnamefont {Ryder}}, \bibinfo {author} {\bibfnamefont {M.}~\bibnamefont {Subbiah}}, \bibinfo {author} {\bibfnamefont {J.~D.}\ \bibnamefont {Kaplan}}, \bibinfo {author} {\bibfnamefont {P.}~\bibnamefont {Dhariwal}}, \bibinfo {author} {\bibfnamefont {A.}~\bibnamefont {Neelakantan}}, \bibinfo {author} {\bibfnamefont {P.}~\bibnamefont {Shyam}}, \bibinfo {author} {\bibfnamefont {G.}~\bibnamefont {Sastry}}, \bibinfo {author} {\bibfnamefont {A.}~\bibnamefont {Askell}}, \bibinfo {author} {\bibfnamefont {S.}~\bibnamefont {Agarwal}}, \bibinfo {author} {\bibfnamefont {A.}~\bibnamefont {{Herbert-Voss}}}, \bibinfo {author} {\bibfnamefont {G.}~\bibnamefont {Krueger}}, \bibinfo {author} {\bibfnamefont {T.}~\bibnamefont {Henighan}}, \bibinfo {author} {\bibfnamefont {R.}~\bibnamefont {Child}}, \bibinfo {author} {\bibfnamefont {A.}~\bibnamefont {Ramesh}}, \bibinfo {author} {\bibfnamefont {D.}~\bibnamefont {Ziegler}}, \bibinfo {author} {\bibfnamefont {J.}~\bibnamefont {Wu}}, \bibinfo {author} {\bibfnamefont {C.}~\bibnamefont {Winter}}, \bibinfo {author} {\bibfnamefont {C.}~\bibnamefont {Hesse}}, \bibinfo {author} {\bibfnamefont {M.}~\bibnamefont {Chen}}, \bibinfo {author} {\bibfnamefont {E.}~\bibnamefont {Sigler}}, \bibinfo {author} {\bibfnamefont {M.}~\bibnamefont {Litwin}}, \bibinfo {author} {\bibfnamefont {S.}~\bibnamefont {Gray}}, \bibinfo {author} {\bibfnamefont {B.}~\bibnamefont {Chess}}, \bibinfo {author} {\bibfnamefont {J.}~\bibnamefont {Clark}}, \bibinfo {author} {\bibfnamefont {C.}~\bibnamefont {Berner}}, \bibinfo {author} {\bibfnamefont {S.}~\bibnamefont {McCandlish}}, \bibinfo {author} {\bibfnamefont {A.}~\bibnamefont {Radford}}, \bibinfo {author} {\bibfnamefont {I.}~\bibnamefont {Sutskever}},\ and\ \bibinfo {author} {\bibfnamefont {D.}~\bibnamefont {Amodei}},\ }\bibfield  {title} {\bibinfo {title} {Language models are few-shot learners},\ }in\ \href@noop {} {\emph {\bibinfo {booktitle} {Advances in Neural Information Processing Systems}}},\ Vol.~\bibinfo {volume} {33},\ \bibinfo {editor} {edited by\ \bibinfo {editor} {\bibfnamefont {H.}~\bibnamefont {Larochelle}}, \bibinfo {editor} {\bibfnamefont {M.}~\bibnamefont {Ranzato}}, \bibinfo {editor} {\bibfnamefont {R.}~\bibnamefont {Hadsell}}, \bibinfo {editor} {\bibfnamefont {M.}~\bibnamefont {Balcan}},\ and\ \bibinfo {editor} {\bibfnamefont {H.}~\bibnamefont {Lin}}}\ (\bibinfo  {publisher} {Curran Associates, Inc.},\ \bibinfo {year} {2020})\ pp.\ \bibinfo {pages} {1877--1901}\BibitemShut {NoStop}%
\bibitem [{\citenamefont {Quiroga}\ \emph {et~al.}(2020)\citenamefont {Quiroga}, \citenamefont {Ronchetti}, \citenamefont {Lanzarini},\ and\ \citenamefont {{Fernandez-Bariviera}}}]{quirogaRevisitingDataAugmentation2020}%
  \BibitemOpen
  \bibfield  {author} {\bibinfo {author} {\bibfnamefont {F.~M.}\ \bibnamefont {Quiroga}}, \bibinfo {author} {\bibfnamefont {F.}~\bibnamefont {Ronchetti}}, \bibinfo {author} {\bibfnamefont {L.}~\bibnamefont {Lanzarini}},\ and\ \bibinfo {author} {\bibfnamefont {A.}~\bibnamefont {{Fernandez-Bariviera}}},\ }\bibfield  {title} {\bibinfo {title} {Revisiting data augmentation for rotational invariance in convolutional neural networks}\ }(\bibinfo  {publisher} {arXiv},\ \bibinfo {year} {2020})\ pp.\ \bibinfo {pages} {127--141},\ \Eprint {https://arxiv.org/abs/2310.08429} {arXiv:2310.08429 [cs]} \BibitemShut {NoStop}%
\bibitem [{\citenamefont {Mazitov}\ \emph {et~al.}(2025)\citenamefont {Mazitov}, \citenamefont {Bigi}, \citenamefont {Kellner}, \citenamefont {Pegolo}, \citenamefont {Tisi}, \citenamefont {Fraux}, \citenamefont {Pozdnyakov}, \citenamefont {Loche},\ and\ \citenamefont {Ceriotti}}]{mazitovPETMADLightweightUniversal2025}%
  \BibitemOpen
  \bibfield  {author} {\bibinfo {author} {\bibfnamefont {A.}~\bibnamefont {Mazitov}}, \bibinfo {author} {\bibfnamefont {F.}~\bibnamefont {Bigi}}, \bibinfo {author} {\bibfnamefont {M.}~\bibnamefont {Kellner}}, \bibinfo {author} {\bibfnamefont {P.}~\bibnamefont {Pegolo}}, \bibinfo {author} {\bibfnamefont {D.}~\bibnamefont {Tisi}}, \bibinfo {author} {\bibfnamefont {G.}~\bibnamefont {Fraux}}, \bibinfo {author} {\bibfnamefont {S.}~\bibnamefont {Pozdnyakov}}, \bibinfo {author} {\bibfnamefont {P.}~\bibnamefont {Loche}},\ and\ \bibinfo {author} {\bibfnamefont {M.}~\bibnamefont {Ceriotti}},\ }\href {https://doi.org/10.48550/arXiv.2503.14118} {\bibinfo {title} {{{PET-MAD}}, a lightweight universal interatomic potential for advanced materials modeling}} (\bibinfo {year} {2025}),\ \Eprint {https://arxiv.org/abs/2503.14118} {arXiv:2503.14118 [cond-mat]} \BibitemShut {NoStop}%
\bibitem [{\citenamefont {Jain}\ \emph {et~al.}(2013)\citenamefont {Jain}, \citenamefont {Ong}, \citenamefont {Hautier}, \citenamefont {Chen}, \citenamefont {Richards}, \citenamefont {Dacek}, \citenamefont {Cholia}, \citenamefont {Gunter}, \citenamefont {Skinner}, \citenamefont {Ceder},\ and\ \citenamefont {Persson}}]{jainMaterialsProjectMaterials2013}%
  \BibitemOpen
  \bibfield  {author} {\bibinfo {author} {\bibfnamefont {A.}~\bibnamefont {Jain}}, \bibinfo {author} {\bibfnamefont {S.~P.}\ \bibnamefont {Ong}}, \bibinfo {author} {\bibfnamefont {G.}~\bibnamefont {Hautier}}, \bibinfo {author} {\bibfnamefont {W.}~\bibnamefont {Chen}}, \bibinfo {author} {\bibfnamefont {W.~D.}\ \bibnamefont {Richards}}, \bibinfo {author} {\bibfnamefont {S.}~\bibnamefont {Dacek}}, \bibinfo {author} {\bibfnamefont {S.}~\bibnamefont {Cholia}}, \bibinfo {author} {\bibfnamefont {D.}~\bibnamefont {Gunter}}, \bibinfo {author} {\bibfnamefont {D.}~\bibnamefont {Skinner}}, \bibinfo {author} {\bibfnamefont {G.}~\bibnamefont {Ceder}},\ and\ \bibinfo {author} {\bibfnamefont {K.~A.}\ \bibnamefont {Persson}},\ }\bibfield  {title} {\bibinfo {title} {The {Materials Project}: {{A}} materials genome approach to accelerating materials innovation},\ }\href {https://doi.org/10.1063/1.4812323} {\bibfield  {journal} {\bibinfo  {journal} {APL Mater.}\ }\textbf {\bibinfo {volume} {1}},\ \bibinfo {pages} {011002} (\bibinfo {year} {2013})}\BibitemShut {NoStop}%
\bibitem [{\citenamefont {Schmidt}\ \emph {et~al.}(2023)\citenamefont {Schmidt}, \citenamefont {Hoffmann}, \citenamefont {Wang}, \citenamefont {Borlido}, \citenamefont {Carri{\c c}o}, \citenamefont {Cerqueira}, \citenamefont {Botti},\ and\ \citenamefont {Marques}}]{schmidtMachineLearningAssistedDeterminationGlobal2023}%
  \BibitemOpen
  \bibfield  {author} {\bibinfo {author} {\bibfnamefont {J.}~\bibnamefont {Schmidt}}, \bibinfo {author} {\bibfnamefont {N.}~\bibnamefont {Hoffmann}}, \bibinfo {author} {\bibfnamefont {H.-C.}\ \bibnamefont {Wang}}, \bibinfo {author} {\bibfnamefont {P.}~\bibnamefont {Borlido}}, \bibinfo {author} {\bibfnamefont {P.~J. M.~A.}\ \bibnamefont {Carri{\c c}o}}, \bibinfo {author} {\bibfnamefont {T.~F.~T.}\ \bibnamefont {Cerqueira}}, \bibinfo {author} {\bibfnamefont {S.}~\bibnamefont {Botti}},\ and\ \bibinfo {author} {\bibfnamefont {M.~A.~L.}\ \bibnamefont {Marques}},\ }\bibfield  {title} {\bibinfo {title} {Machine-learning-assisted determination of the global zero-temperature phase diagram of materials},\ }\href {https://doi.org/10.1002/adma.202210788} {\bibfield  {journal} {\bibinfo  {journal} {Adv. Mater.}\ }\textbf {\bibinfo {volume} {35}},\ \bibinfo {pages} {2210788} (\bibinfo {year} {2023})}\BibitemShut {NoStop}%
\bibitem [{\citenamefont {Wang}\ \emph {et~al.}(2023)\citenamefont {Wang}, \citenamefont {Schmidt}, \citenamefont {Marques}, \citenamefont {Wirtz},\ and\ \citenamefont {Romero}}]{Alex2d}%
  \BibitemOpen
  \bibfield  {author} {\bibinfo {author} {\bibfnamefont {H.-C.}\ \bibnamefont {Wang}}, \bibinfo {author} {\bibfnamefont {J.}~\bibnamefont {Schmidt}}, \bibinfo {author} {\bibfnamefont {M.~A.~L.}\ \bibnamefont {Marques}}, \bibinfo {author} {\bibfnamefont {L.}~\bibnamefont {Wirtz}},\ and\ \bibinfo {author} {\bibfnamefont {A.~H.}\ \bibnamefont {Romero}},\ }\bibfield  {title} {\bibinfo {title} {Symmetry-based computational search for novel binary and ternary {2D} materials},\ }\href {https://doi.org/10.1088/2053-1583/accc43} {\bibfield  {journal} {\bibinfo  {journal} {2D Mater.}\ }\textbf {\bibinfo {volume} {10}},\ \bibinfo {pages} {035007} (\bibinfo {year} {2023})}\BibitemShut {NoStop}%
\bibitem [{\citenamefont {Belsky}\ \emph {et~al.}(2002)\citenamefont {Belsky}, \citenamefont {Hellenbrandt}, \citenamefont {Karen},\ and\ \citenamefont {Luksch}}]{belskyNewDevelopmentsInorganic2002}%
  \BibitemOpen
  \bibfield  {author} {\bibinfo {author} {\bibfnamefont {A.}~\bibnamefont {Belsky}}, \bibinfo {author} {\bibfnamefont {M.}~\bibnamefont {Hellenbrandt}}, \bibinfo {author} {\bibfnamefont {V.~L.}\ \bibnamefont {Karen}},\ and\ \bibinfo {author} {\bibfnamefont {P.}~\bibnamefont {Luksch}},\ }\bibfield  {title} {\bibinfo {title} {New developments in the {{Inorganic Crystal Structure Database}} ({{ICSD}}): Accessibility in support of materials research and design},\ }\bibfield  {journal} {\bibinfo  {journal} {Acta Crystallogr. Sect. B Struct. Sci.}\ }\textbf {\bibinfo {volume} {58}},\ \href {https://doi.org/10.1107/S0108768102006948} {10.1107/S0108768102006948} (\bibinfo {year} {2002})\BibitemShut {NoStop}%
\bibitem [{\citenamefont {Curtarolo}\ \emph {et~al.}(2012)\citenamefont {Curtarolo}, \citenamefont {Setyawan}, \citenamefont {Wang}, \citenamefont {Xue}, \citenamefont {Yang}, \citenamefont {Taylor}, \citenamefont {Nelson}, \citenamefont {Hart}, \citenamefont {Sanvito}, \citenamefont {{Buongiorno-Nardelli}}, \citenamefont {Mingo},\ and\ \citenamefont {Levy}}]{curtaroloAFLOWLIBORGDistributedMaterials2012}%
  \BibitemOpen
  \bibfield  {author} {\bibinfo {author} {\bibfnamefont {S.}~\bibnamefont {Curtarolo}}, \bibinfo {author} {\bibfnamefont {W.}~\bibnamefont {Setyawan}}, \bibinfo {author} {\bibfnamefont {S.}~\bibnamefont {Wang}}, \bibinfo {author} {\bibfnamefont {J.}~\bibnamefont {Xue}}, \bibinfo {author} {\bibfnamefont {K.}~\bibnamefont {Yang}}, \bibinfo {author} {\bibfnamefont {R.~H.}\ \bibnamefont {Taylor}}, \bibinfo {author} {\bibfnamefont {L.~J.}\ \bibnamefont {Nelson}}, \bibinfo {author} {\bibfnamefont {G.~L.~W.}\ \bibnamefont {Hart}}, \bibinfo {author} {\bibfnamefont {S.}~\bibnamefont {Sanvito}}, \bibinfo {author} {\bibfnamefont {M.}~\bibnamefont {{Buongiorno-Nardelli}}}, \bibinfo {author} {\bibfnamefont {N.}~\bibnamefont {Mingo}},\ and\ \bibinfo {author} {\bibfnamefont {O.}~\bibnamefont {Levy}},\ }\bibfield  {title} {\bibinfo {title} {{{AFLOWLIB}}.{{ORG}}: {{A}} distributed materials properties repository from high-throughput ab initio calculations},\ }\href {https://doi.org/10.1016/j.commatsci.2012.02.002} {\bibfield  {journal} {\bibinfo  {journal} {Comput. Mater. Sci.}\ }\textbf {\bibinfo {volume} {58}},\ \bibinfo {pages} {227} (\bibinfo {year} {2012})}\BibitemShut {NoStop}%
\bibitem [{\citenamefont {Saal}\ \emph {et~al.}(2013)\citenamefont {Saal}, \citenamefont {Kirklin}, \citenamefont {Aykol}, \citenamefont {Meredig},\ and\ \citenamefont {Wolverton}}]{saalMaterialsDesignDiscovery2013}%
  \BibitemOpen
  \bibfield  {author} {\bibinfo {author} {\bibfnamefont {J.~E.}\ \bibnamefont {Saal}}, \bibinfo {author} {\bibfnamefont {S.}~\bibnamefont {Kirklin}}, \bibinfo {author} {\bibfnamefont {M.}~\bibnamefont {Aykol}}, \bibinfo {author} {\bibfnamefont {B.}~\bibnamefont {Meredig}},\ and\ \bibinfo {author} {\bibfnamefont {C.}~\bibnamefont {Wolverton}},\ }\bibfield  {title} {\bibinfo {title} {Materials design and discovery with high-throughput density functional theory: {{The}} open quantum materials database ({{OQMD}})},\ }\href {https://doi.org/10.1007/s11837-013-0755-4} {\bibfield  {journal} {\bibinfo  {journal} {JOM}\ }\textbf {\bibinfo {volume} {65}},\ \bibinfo {pages} {1501} (\bibinfo {year} {2013})}\BibitemShut {NoStop}%
\bibitem [{\citenamefont {Choudhary}\ \emph {et~al.}(2020{\natexlab{a}})\citenamefont {Choudhary}, \citenamefont {Garrity}, \citenamefont {Reid}, \citenamefont {DeCost}, \citenamefont {Biacchi}, \citenamefont {Hight~Walker}, \citenamefont {Trautt}, \citenamefont {Hattrick-Simpers}, \citenamefont {Kusne}, \citenamefont {Centrone}, \citenamefont {Davydov}, \citenamefont {Jiang}, \citenamefont {Pachter}, \citenamefont {Cheon}, \citenamefont {Reed}, \citenamefont {Agrawal}, \citenamefont {Qian}, \citenamefont {Sharma}, \citenamefont {Zhuang}, \citenamefont {Kalinin}, \citenamefont {Sumpter}, \citenamefont {Pilania}, \citenamefont {Acar}, \citenamefont {Mandal}, \citenamefont {Haule}, \citenamefont {Vanderbilt}, \citenamefont {Rabe},\ and\ \citenamefont {Tavazza}}]{jarvis}%
  \BibitemOpen
  \bibfield  {author} {\bibinfo {author} {\bibfnamefont {K.}~\bibnamefont {Choudhary}}, \bibinfo {author} {\bibfnamefont {K.~F.}\ \bibnamefont {Garrity}}, \bibinfo {author} {\bibfnamefont {A.~C.~E.}\ \bibnamefont {Reid}}, \bibinfo {author} {\bibfnamefont {B.}~\bibnamefont {DeCost}}, \bibinfo {author} {\bibfnamefont {A.~J.}\ \bibnamefont {Biacchi}}, \bibinfo {author} {\bibfnamefont {A.~R.}\ \bibnamefont {Hight~Walker}}, \bibinfo {author} {\bibfnamefont {Z.}~\bibnamefont {Trautt}}, \bibinfo {author} {\bibfnamefont {J.}~\bibnamefont {Hattrick-Simpers}}, \bibinfo {author} {\bibfnamefont {A.~G.}\ \bibnamefont {Kusne}}, \bibinfo {author} {\bibfnamefont {A.}~\bibnamefont {Centrone}}, \bibinfo {author} {\bibfnamefont {A.}~\bibnamefont {Davydov}}, \bibinfo {author} {\bibfnamefont {J.}~\bibnamefont {Jiang}}, \bibinfo {author} {\bibfnamefont {R.}~\bibnamefont {Pachter}}, \bibinfo {author} {\bibfnamefont {G.}~\bibnamefont {Cheon}}, \bibinfo {author} {\bibfnamefont {E.}~\bibnamefont {Reed}}, \bibinfo {author} {\bibfnamefont {A.}~\bibnamefont {Agrawal}}, \bibinfo {author} {\bibfnamefont {X.}~\bibnamefont {Qian}}, \bibinfo {author} {\bibfnamefont {V.}~\bibnamefont {Sharma}}, \bibinfo {author} {\bibfnamefont {H.}~\bibnamefont {Zhuang}}, \bibinfo {author} {\bibfnamefont {S.~V.}\ \bibnamefont {Kalinin}}, \bibinfo {author} {\bibfnamefont {B.~G.}\ \bibnamefont {Sumpter}}, \bibinfo {author} {\bibfnamefont {G.}~\bibnamefont {Pilania}}, \bibinfo {author} {\bibfnamefont {P.}~\bibnamefont {Acar}}, \bibinfo {author} {\bibfnamefont {S.}~\bibnamefont {Mandal}}, \bibinfo {author} {\bibfnamefont {K.}~\bibnamefont {Haule}}, \bibinfo {author} {\bibfnamefont {D.}~\bibnamefont {Vanderbilt}}, \bibinfo {author} {\bibfnamefont {K.}~\bibnamefont {Rabe}},\ and\ \bibinfo {author} {\bibfnamefont {F.}~\bibnamefont {Tavazza}},\ }\bibfield  {title} {\bibinfo {title} {The joint automated repository for various integrated simulations (jarvis) for data-driven materials design},\ }\bibfield  {journal} {\bibinfo  {journal} {npj Comput. Mater.}\ }\textbf {\bibinfo {volume} {6}},\ \href {https://doi.org/10.1038/s41524-020-00440-1} {10.1038/s41524-020-00440-1} (\bibinfo {year} {2020}{\natexlab{a}})\BibitemShut {NoStop}%
\bibitem [{\citenamefont {Lehtola}\ \emph {et~al.}(2018)\citenamefont {Lehtola}, \citenamefont {Steigemann}, \citenamefont {Oliveira},\ and\ \citenamefont {Marques}}]{lehtolaRecentDevelopmentsLibxc2018}%
  \BibitemOpen
  \bibfield  {author} {\bibinfo {author} {\bibfnamefont {S.}~\bibnamefont {Lehtola}}, \bibinfo {author} {\bibfnamefont {C.}~\bibnamefont {Steigemann}}, \bibinfo {author} {\bibfnamefont {M.~J.}\ \bibnamefont {Oliveira}},\ and\ \bibinfo {author} {\bibfnamefont {M.~A.}\ \bibnamefont {Marques}},\ }\bibfield  {title} {\bibinfo {title} {Recent developments in libxc --- {{A}} comprehensive library of functionals for density functional theory},\ }\href {https://doi.org/10.1016/j.softx.2017.11.002} {\bibfield  {journal} {\bibinfo  {journal} {SoftwareX}\ }\textbf {\bibinfo {volume} {7}},\ \bibinfo {pages} {1} (\bibinfo {year} {2018})}\BibitemShut {NoStop}%
\bibitem [{\citenamefont {Perdew}\ \emph {et~al.}(1996)\citenamefont {Perdew}, \citenamefont {Burke},\ and\ \citenamefont {Ernzerhof}}]{perdewGeneralizedGradientApproximation1996}%
  \BibitemOpen
  \bibfield  {author} {\bibinfo {author} {\bibfnamefont {J.~P.}\ \bibnamefont {Perdew}}, \bibinfo {author} {\bibfnamefont {K.}~\bibnamefont {Burke}},\ and\ \bibinfo {author} {\bibfnamefont {M.}~\bibnamefont {Ernzerhof}},\ }\bibfield  {title} {\bibinfo {title} {Generalized gradient approximation made simple},\ }\href {https://doi.org/10.1103/PhysRevLett.77.3865} {\bibfield  {journal} {\bibinfo  {journal} {Phys. Rev. Lett.}\ }\textbf {\bibinfo {volume} {77}},\ \bibinfo {pages} {3865} (\bibinfo {year} {1996})}\BibitemShut {NoStop}%
\bibitem [{\citenamefont {Kresse}\ and\ \citenamefont {Furthm{\"u}ller}(1996)}]{kresseEfficientIterativeSchemes1996}%
  \BibitemOpen
  \bibfield  {author} {\bibinfo {author} {\bibfnamefont {G.}~\bibnamefont {Kresse}}\ and\ \bibinfo {author} {\bibfnamefont {J.}~\bibnamefont {Furthm{\"u}ller}},\ }\bibfield  {title} {\bibinfo {title} {Efficient iterative schemes for ab initio total-energy calculations using a plane-wave basis set},\ }\href {https://doi.org/10.1103/PhysRevB.54.11169} {\bibfield  {journal} {\bibinfo  {journal} {Phys. Rev. B}\ }\textbf {\bibinfo {volume} {54}},\ \bibinfo {pages} {11169} (\bibinfo {year} {1996})}\BibitemShut {NoStop}%
\bibitem [{\citenamefont {Zeni}\ \emph {et~al.}(2025)\citenamefont {Zeni}, \citenamefont {Pinsler}, \citenamefont {Z{\"u}gner}, \citenamefont {Fowler}, \citenamefont {Horton}, \citenamefont {Fu}, \citenamefont {Wang}, \citenamefont {Shysheya}, \citenamefont {Crabb{\'e}}, \citenamefont {Ueda}, \citenamefont {Sordillo}, \citenamefont {Sun}, \citenamefont {Smith}, \citenamefont {Nguyen}, \citenamefont {Schulz}, \citenamefont {Lewis}, \citenamefont {Huang}, \citenamefont {Lu}, \citenamefont {Zhou}, \citenamefont {Yang}, \citenamefont {Hao}, \citenamefont {Li}, \citenamefont {Yang}, \citenamefont {Li}, \citenamefont {Tomioka},\ and\ \citenamefont {Xie}}]{zeniGenerativeModelInorganic2025}%
  \BibitemOpen
  \bibfield  {author} {\bibinfo {author} {\bibfnamefont {C.}~\bibnamefont {Zeni}}, \bibinfo {author} {\bibfnamefont {R.}~\bibnamefont {Pinsler}}, \bibinfo {author} {\bibfnamefont {D.}~\bibnamefont {Z{\"u}gner}}, \bibinfo {author} {\bibfnamefont {A.}~\bibnamefont {Fowler}}, \bibinfo {author} {\bibfnamefont {M.}~\bibnamefont {Horton}}, \bibinfo {author} {\bibfnamefont {X.}~\bibnamefont {Fu}}, \bibinfo {author} {\bibfnamefont {Z.}~\bibnamefont {Wang}}, \bibinfo {author} {\bibfnamefont {A.}~\bibnamefont {Shysheya}}, \bibinfo {author} {\bibfnamefont {J.}~\bibnamefont {Crabb{\'e}}}, \bibinfo {author} {\bibfnamefont {S.}~\bibnamefont {Ueda}}, \bibinfo {author} {\bibfnamefont {R.}~\bibnamefont {Sordillo}}, \bibinfo {author} {\bibfnamefont {L.}~\bibnamefont {Sun}}, \bibinfo {author} {\bibfnamefont {J.}~\bibnamefont {Smith}}, \bibinfo {author} {\bibfnamefont {B.}~\bibnamefont {Nguyen}}, \bibinfo {author} {\bibfnamefont {H.}~\bibnamefont {Schulz}}, \bibinfo {author} {\bibfnamefont {S.}~\bibnamefont {Lewis}}, \bibinfo {author} {\bibfnamefont {C.-W.}\ \bibnamefont {Huang}}, \bibinfo {author} {\bibfnamefont {Z.}~\bibnamefont {Lu}}, \bibinfo {author} {\bibfnamefont {Y.}~\bibnamefont {Zhou}}, \bibinfo {author} {\bibfnamefont {H.}~\bibnamefont {Yang}}, \bibinfo {author} {\bibfnamefont {H.}~\bibnamefont {Hao}}, \bibinfo {author} {\bibfnamefont {J.}~\bibnamefont {Li}}, \bibinfo {author} {\bibfnamefont {C.}~\bibnamefont {Yang}}, \bibinfo {author} {\bibfnamefont {W.}~\bibnamefont {Li}}, \bibinfo {author} {\bibfnamefont {R.}~\bibnamefont {Tomioka}},\ and\ \bibinfo {author} {\bibfnamefont {T.}~\bibnamefont {Xie}},\ }\bibfield  {title} {\bibinfo {title} {A generative model for inorganic materials design},\ }\href {https://doi.org/10.1038/s41586-025-08628-5} {\bibfield  {journal} {\bibinfo  {journal} {Nature}\ }\textbf {\bibinfo {volume} {639}},\ \bibinfo {pages} {624} (\bibinfo {year} {2025})}\BibitemShut {NoStop}%
\bibitem [{\citenamefont {Breuck}\ \emph {et~al.}(2025)\citenamefont {Breuck}, \citenamefont {Piracha}, \citenamefont {Rignanese},\ and\ \citenamefont {Marques}}]{breuckGenerativeMaterialTransformer2025}%
  \BibitemOpen
  \bibfield  {author} {\bibinfo {author} {\bibfnamefont {P.-P.~D.}\ \bibnamefont {Breuck}}, \bibinfo {author} {\bibfnamefont {H.~A.}\ \bibnamefont {Piracha}}, \bibinfo {author} {\bibfnamefont {G.-M.}\ \bibnamefont {Rignanese}},\ and\ \bibinfo {author} {\bibfnamefont {M.~A.~L.}\ \bibnamefont {Marques}},\ }\href {https://doi.org/10.48550/arXiv.2501.16051} {\bibinfo {title} {A generative material transformer using {{Wyckoff}} representation}} (\bibinfo {year} {2025}),\ \Eprint {https://arxiv.org/abs/2501.16051} {arXiv:2501.16051} \BibitemShut {NoStop}%
\bibitem [{\citenamefont {Choudhary}\ \emph {et~al.}(2020{\natexlab{b}})\citenamefont {Choudhary}, \citenamefont {Garrity}, \citenamefont {Reid}, \citenamefont {DeCost}, \citenamefont {Biacchi}, \citenamefont {Hight~Walker}, \citenamefont {Trautt}, \citenamefont {{Hattrick-Simpers}}, \citenamefont {Kusne}, \citenamefont {Centrone}, \citenamefont {Davydov}, \citenamefont {Jiang}, \citenamefont {Pachter}, \citenamefont {Cheon}, \citenamefont {Reed}, \citenamefont {Agrawal}, \citenamefont {Qian}, \citenamefont {Sharma}, \citenamefont {Zhuang}, \citenamefont {Kalinin}, \citenamefont {Sumpter}, \citenamefont {Pilania}, \citenamefont {Acar}, \citenamefont {Mandal}, \citenamefont {Haule}, \citenamefont {Vanderbilt}, \citenamefont {Rabe},\ and\ \citenamefont {Tavazza}}]{choudharyJointAutomatedRepository2020}%
  \BibitemOpen
  \bibfield  {author} {\bibinfo {author} {\bibfnamefont {K.}~\bibnamefont {Choudhary}}, \bibinfo {author} {\bibfnamefont {K.~F.}\ \bibnamefont {Garrity}}, \bibinfo {author} {\bibfnamefont {A.~C.~E.}\ \bibnamefont {Reid}}, \bibinfo {author} {\bibfnamefont {B.}~\bibnamefont {DeCost}}, \bibinfo {author} {\bibfnamefont {A.~J.}\ \bibnamefont {Biacchi}}, \bibinfo {author} {\bibfnamefont {A.~R.}\ \bibnamefont {Hight~Walker}}, \bibinfo {author} {\bibfnamefont {Z.}~\bibnamefont {Trautt}}, \bibinfo {author} {\bibfnamefont {J.}~\bibnamefont {{Hattrick-Simpers}}}, \bibinfo {author} {\bibfnamefont {A.~G.}\ \bibnamefont {Kusne}}, \bibinfo {author} {\bibfnamefont {A.}~\bibnamefont {Centrone}}, \bibinfo {author} {\bibfnamefont {A.}~\bibnamefont {Davydov}}, \bibinfo {author} {\bibfnamefont {J.}~\bibnamefont {Jiang}}, \bibinfo {author} {\bibfnamefont {R.}~\bibnamefont {Pachter}}, \bibinfo {author} {\bibfnamefont {G.}~\bibnamefont {Cheon}}, \bibinfo {author} {\bibfnamefont {E.}~\bibnamefont {Reed}}, \bibinfo {author} {\bibfnamefont {A.}~\bibnamefont {Agrawal}}, \bibinfo {author} {\bibfnamefont {X.}~\bibnamefont {Qian}}, \bibinfo {author} {\bibfnamefont {V.}~\bibnamefont {Sharma}}, \bibinfo {author} {\bibfnamefont {H.}~\bibnamefont {Zhuang}}, \bibinfo {author} {\bibfnamefont {S.~V.}\ \bibnamefont {Kalinin}}, \bibinfo {author} {\bibfnamefont {B.~G.}\ \bibnamefont {Sumpter}}, \bibinfo {author} {\bibfnamefont {G.}~\bibnamefont {Pilania}}, \bibinfo {author} {\bibfnamefont {P.}~\bibnamefont {Acar}}, \bibinfo {author} {\bibfnamefont {S.}~\bibnamefont {Mandal}}, \bibinfo {author} {\bibfnamefont {K.}~\bibnamefont {Haule}}, \bibinfo {author} {\bibfnamefont {D.}~\bibnamefont {Vanderbilt}}, \bibinfo {author} {\bibfnamefont {K.}~\bibnamefont {Rabe}},\ and\ \bibinfo {author} {\bibfnamefont {F.}~\bibnamefont {Tavazza}},\ }\bibfield  {title} {\bibinfo {title} {The joint automated repository for various integrated simulations ({{JARVIS}}) for data-driven materials design},\ }\href {https://doi.org/10.1038/s41524-020-00440-1} {\bibfield  {journal} {\bibinfo  {journal} {npj Comput. Mater.}\ }\textbf {\bibinfo {volume} {6}},\ \bibinfo {pages} {173} (\bibinfo {year} {2020}{\natexlab{b}})}\BibitemShut {NoStop}%
\bibitem [{\citenamefont {Andersen}\ \emph {et~al.}(2021)\citenamefont {Andersen}, \citenamefont {Armiento}, \citenamefont {Blokhin}, \citenamefont {Conduit}, \citenamefont {Dwaraknath}, \citenamefont {Evans}, \citenamefont {Fekete}, \citenamefont {Gopakumar}, \citenamefont {Gra\v{z}ulis}, \citenamefont {Merkys}, \citenamefont {Mohamed}, \citenamefont {Oses}, \citenamefont {Pizzi}, \citenamefont {Rignanese}, \citenamefont {Scheidgen}, \citenamefont {Talirz}, \citenamefont {Toher}, \citenamefont {Winston}, \citenamefont {Aversa}, \citenamefont {Choudhary}, \citenamefont {Colinet}, \citenamefont {Curtarolo}, \citenamefont {Di~Stefano}, \citenamefont {Draxl}, \citenamefont {Er}, \citenamefont {Esters}, \citenamefont {Fornari}, \citenamefont {Giantomassi}, \citenamefont {Govoni}, \citenamefont {Hautier}, \citenamefont {Hegde}, \citenamefont {Horton}, \citenamefont {Huck}, \citenamefont {Huhs}, \citenamefont {Hummelsh\o~j}, \citenamefont {Kariryaa}, \citenamefont {Kozinsky}, \citenamefont {Kumbhar}, \citenamefont {Liu}, \citenamefont {Marzari}, \citenamefont {Morris}, \citenamefont {Mostofi}, \citenamefont {Persson}, \citenamefont {Petretto}, \citenamefont {Purcell}, \citenamefont {Ricci}, \citenamefont {Rose}, \citenamefont {Scheffler}, \citenamefont {Speckhard}, \citenamefont {Uhrin}, \citenamefont {Vaitkus}, \citenamefont {Villars}, \citenamefont {Waroquiers}, \citenamefont {Wolverton}, \citenamefont {Wu},\ and\ \citenamefont {Yang}}]{Andersen2021}%
  \BibitemOpen
  \bibfield  {author} {\bibinfo {author} {\bibfnamefont {C.~W.}\ \bibnamefont {Andersen}}, \bibinfo {author} {\bibfnamefont {R.}~\bibnamefont {Armiento}}, \bibinfo {author} {\bibfnamefont {E.}~\bibnamefont {Blokhin}}, \bibinfo {author} {\bibfnamefont {G.~J.}\ \bibnamefont {Conduit}}, \bibinfo {author} {\bibfnamefont {S.}~\bibnamefont {Dwaraknath}}, \bibinfo {author} {\bibfnamefont {M.~L.}\ \bibnamefont {Evans}}, \bibinfo {author} {\bibfnamefont {A.}~\bibnamefont {Fekete}}, \bibinfo {author} {\bibfnamefont {A.}~\bibnamefont {Gopakumar}}, \bibinfo {author} {\bibfnamefont {S.}~\bibnamefont {Gra\v{z}ulis}}, \bibinfo {author} {\bibfnamefont {A.}~\bibnamefont {Merkys}}, \bibinfo {author} {\bibfnamefont {F.}~\bibnamefont {Mohamed}}, \bibinfo {author} {\bibfnamefont {C.}~\bibnamefont {Oses}}, \bibinfo {author} {\bibfnamefont {G.}~\bibnamefont {Pizzi}}, \bibinfo {author} {\bibfnamefont {G.-M.}\ \bibnamefont {Rignanese}}, \bibinfo {author} {\bibfnamefont {M.}~\bibnamefont {Scheidgen}}, \bibinfo {author} {\bibfnamefont {L.}~\bibnamefont {Talirz}}, \bibinfo {author} {\bibfnamefont {C.}~\bibnamefont {Toher}}, \bibinfo {author} {\bibfnamefont {D.}~\bibnamefont {Winston}}, \bibinfo {author} {\bibfnamefont {R.}~\bibnamefont {Aversa}}, \bibinfo {author} {\bibfnamefont {K.}~\bibnamefont {Choudhary}}, \bibinfo {author} {\bibfnamefont {P.}~\bibnamefont {Colinet}}, \bibinfo {author} {\bibfnamefont {S.}~\bibnamefont {Curtarolo}}, \bibinfo {author} {\bibfnamefont {D.}~\bibnamefont {Di~Stefano}}, \bibinfo {author} {\bibfnamefont {C.}~\bibnamefont {Draxl}}, \bibinfo {author} {\bibfnamefont {S.}~\bibnamefont {Er}}, \bibinfo {author} {\bibfnamefont {M.}~\bibnamefont {Esters}}, \bibinfo {author} {\bibfnamefont {M.}~\bibnamefont {Fornari}}, \bibinfo {author} {\bibfnamefont {M.}~\bibnamefont {Giantomassi}}, \bibinfo {author} {\bibfnamefont {M.}~\bibnamefont {Govoni}}, \bibinfo {author} {\bibfnamefont {G.}~\bibnamefont {Hautier}}, \bibinfo {author} {\bibfnamefont {V.}~\bibnamefont {Hegde}}, \bibinfo {author} {\bibfnamefont {M.~K.}\ \bibnamefont {Horton}}, \bibinfo {author} {\bibfnamefont {P.}~\bibnamefont {Huck}}, \bibinfo {author} {\bibfnamefont {G.}~\bibnamefont {Huhs}}, \bibinfo {author} {\bibfnamefont {J.}~\bibnamefont {Hummelsh\o~j}}, \bibinfo {author} {\bibfnamefont {A.}~\bibnamefont {Kariryaa}}, \bibinfo {author} {\bibfnamefont {B.}~\bibnamefont {Kozinsky}}, \bibinfo {author} {\bibfnamefont {S.}~\bibnamefont {Kumbhar}}, \bibinfo {author} {\bibfnamefont {M.}~\bibnamefont {Liu}}, \bibinfo {author} {\bibfnamefont {N.}~\bibnamefont {Marzari}}, \bibinfo {author} {\bibfnamefont {A.~J.}\ \bibnamefont {Morris}}, \bibinfo {author} {\bibfnamefont {A.~A.}\ \bibnamefont {Mostofi}}, \bibinfo {author} {\bibfnamefont {K.~A.}\ \bibnamefont {Persson}}, \bibinfo {author} {\bibfnamefont {G.}~\bibnamefont {Petretto}}, \bibinfo {author} {\bibfnamefont {T.}~\bibnamefont {Purcell}}, \bibinfo {author} {\bibfnamefont {F.}~\bibnamefont {Ricci}}, \bibinfo {author} {\bibfnamefont {F.}~\bibnamefont {Rose}}, \bibinfo {author} {\bibfnamefont {M.}~\bibnamefont {Scheffler}}, \bibinfo {author} {\bibfnamefont {D.}~\bibnamefont {Speckhard}}, \bibinfo {author} {\bibfnamefont {M.}~\bibnamefont {Uhrin}}, \bibinfo {author} {\bibfnamefont {A.}~\bibnamefont {Vaitkus}}, \bibinfo {author} {\bibfnamefont {P.}~\bibnamefont {Villars}}, \bibinfo {author} {\bibfnamefont {D.}~\bibnamefont {Waroquiers}}, \bibinfo {author} {\bibfnamefont {C.}~\bibnamefont {Wolverton}}, \bibinfo {author} {\bibfnamefont {M.}~\bibnamefont {Wu}},\ and\ \bibinfo {author} {\bibfnamefont {X.}~\bibnamefont {Yang}},\ }\bibfield  {title} {\bibinfo {title} {{OPTIMADE}, an {API} for exchanging materials data},\ }\bibfield  {journal} {\bibinfo  {journal} {Sci. Data}\ }\textbf {\bibinfo {volume} {8}},\ \href {https://doi.org/10.1038/s41597-021-00974-z} {10.1038/s41597-021-00974-z} (\bibinfo {year} {2021})\BibitemShut {NoStop}%
\bibitem [{\citenamefont {Evans}\ \emph {et~al.}(2024)\citenamefont {Evans}, \citenamefont {Bergsma}, \citenamefont {Merkys}, \citenamefont {Andersen}, \citenamefont {Andersson}, \citenamefont {Beltrán}, \citenamefont {Blokhin}, \citenamefont {Boland}, \citenamefont {Castañeda~Balderas}, \citenamefont {Choudhary}, \citenamefont {Díaz~Díaz}, \citenamefont {Domínguez~García}, \citenamefont {Eckert}, \citenamefont {Eimre}, \citenamefont {Fuentes~Montero}, \citenamefont {Krajewski}, \citenamefont {Mortensen}, \citenamefont {Nápoles~Duarte}, \citenamefont {Pietryga}, \citenamefont {Qi}, \citenamefont {Trejo~Carrillo}, \citenamefont {Vaitkus}, \citenamefont {Yu}, \citenamefont {Zettel}, \citenamefont {de~Castro}, \citenamefont {Carlsson}, \citenamefont {Cerqueira}, \citenamefont {Divilov}, \citenamefont {Hajiyani}, \citenamefont {Hanke}, \citenamefont {Jose}, \citenamefont {Oses}, \citenamefont {Riebesell}, \citenamefont {Schmidt}, \citenamefont {Winston}, \citenamefont {Xie}, \citenamefont {Yang}, \citenamefont {Bonella}, \citenamefont {Botti}, \citenamefont {Curtarolo}, \citenamefont {Draxl}, \citenamefont {Fuentes~Cobas}, \citenamefont {Hospital}, \citenamefont {Liu}, \citenamefont {Marques}, \citenamefont {Marzari}, \citenamefont {Morris}, \citenamefont {Ong}, \citenamefont {Orozco}, \citenamefont {Persson}, \citenamefont {Thygesen}, \citenamefont {Wolverton}, \citenamefont {Scheidgen}, \citenamefont {Toher}, \citenamefont {Conduit}, \citenamefont {Pizzi}, \citenamefont {Gražulis}, \citenamefont {Rignanese},\ and\ \citenamefont {Armiento}}]{Evans2024}%
  \BibitemOpen
  \bibfield  {author} {\bibinfo {author} {\bibfnamefont {M.~L.}\ \bibnamefont {Evans}}, \bibinfo {author} {\bibfnamefont {J.}~\bibnamefont {Bergsma}}, \bibinfo {author} {\bibfnamefont {A.}~\bibnamefont {Merkys}}, \bibinfo {author} {\bibfnamefont {C.~W.}\ \bibnamefont {Andersen}}, \bibinfo {author} {\bibfnamefont {O.~B.}\ \bibnamefont {Andersson}}, \bibinfo {author} {\bibfnamefont {D.}~\bibnamefont {Beltrán}}, \bibinfo {author} {\bibfnamefont {E.}~\bibnamefont {Blokhin}}, \bibinfo {author} {\bibfnamefont {T.~M.}\ \bibnamefont {Boland}}, \bibinfo {author} {\bibfnamefont {R.}~\bibnamefont {Castañeda~Balderas}}, \bibinfo {author} {\bibfnamefont {K.}~\bibnamefont {Choudhary}}, \bibinfo {author} {\bibfnamefont {A.}~\bibnamefont {Díaz~Díaz}}, \bibinfo {author} {\bibfnamefont {R.}~\bibnamefont {Domínguez~García}}, \bibinfo {author} {\bibfnamefont {H.}~\bibnamefont {Eckert}}, \bibinfo {author} {\bibfnamefont {K.}~\bibnamefont {Eimre}}, \bibinfo {author} {\bibfnamefont {M.~E.}\ \bibnamefont {Fuentes~Montero}}, \bibinfo {author} {\bibfnamefont {A.~M.}\ \bibnamefont {Krajewski}}, \bibinfo {author} {\bibfnamefont {J.~J.}\ \bibnamefont {Mortensen}}, \bibinfo {author} {\bibfnamefont {J.~M.}\ \bibnamefont {Nápoles~Duarte}}, \bibinfo {author} {\bibfnamefont {J.}~\bibnamefont {Pietryga}}, \bibinfo {author} {\bibfnamefont {J.}~\bibnamefont {Qi}}, \bibinfo {author} {\bibfnamefont {F.~d.~J.}\ \bibnamefont {Trejo~Carrillo}}, \bibinfo {author} {\bibfnamefont {A.}~\bibnamefont {Vaitkus}}, \bibinfo {author} {\bibfnamefont {J.}~\bibnamefont {Yu}}, \bibinfo {author} {\bibfnamefont {A.}~\bibnamefont {Zettel}}, \bibinfo {author} {\bibfnamefont {P.~B.}\ \bibnamefont {de~Castro}}, \bibinfo {author} {\bibfnamefont {J.}~\bibnamefont {Carlsson}}, \bibinfo {author} {\bibfnamefont {T.~F.~T.}\ \bibnamefont {Cerqueira}}, \bibinfo {author} {\bibfnamefont {S.}~\bibnamefont {Divilov}}, \bibinfo {author} {\bibfnamefont {H.}~\bibnamefont {Hajiyani}}, \bibinfo {author} {\bibfnamefont {F.}~\bibnamefont {Hanke}}, \bibinfo {author} {\bibfnamefont {K.}~\bibnamefont {Jose}}, \bibinfo {author} {\bibfnamefont {C.}~\bibnamefont {Oses}}, \bibinfo {author} {\bibfnamefont {J.}~\bibnamefont {Riebesell}}, \bibinfo {author} {\bibfnamefont {J.}~\bibnamefont {Schmidt}}, \bibinfo {author} {\bibfnamefont {D.}~\bibnamefont {Winston}}, \bibinfo {author} {\bibfnamefont {C.}~\bibnamefont {Xie}}, \bibinfo {author} {\bibfnamefont {X.}~\bibnamefont {Yang}}, \bibinfo {author} {\bibfnamefont {S.}~\bibnamefont {Bonella}}, \bibinfo {author} {\bibfnamefont {S.}~\bibnamefont {Botti}}, \bibinfo {author} {\bibfnamefont {S.}~\bibnamefont {Curtarolo}}, \bibinfo {author} {\bibfnamefont {C.}~\bibnamefont {Draxl}}, \bibinfo {author} {\bibfnamefont {L.~E.}\ \bibnamefont {Fuentes~Cobas}}, \bibinfo {author} {\bibfnamefont {A.}~\bibnamefont {Hospital}}, \bibinfo {author} {\bibfnamefont {Z.-K.}\ \bibnamefont {Liu}}, \bibinfo {author} {\bibfnamefont {M.~A.~L.}\ \bibnamefont {Marques}}, \bibinfo {author} {\bibfnamefont {N.}~\bibnamefont {Marzari}}, \bibinfo {author} {\bibfnamefont {A.~J.}\ \bibnamefont {Morris}}, \bibinfo {author} {\bibfnamefont {S.~P.}\ \bibnamefont {Ong}}, \bibinfo {author} {\bibfnamefont {M.}~\bibnamefont {Orozco}}, \bibinfo {author} {\bibfnamefont {K.~A.}\ \bibnamefont {Persson}}, \bibinfo {author} {\bibfnamefont {K.~S.}\ \bibnamefont {Thygesen}}, \bibinfo {author} {\bibfnamefont {C.}~\bibnamefont {Wolverton}}, \bibinfo {author} {\bibfnamefont {M.}~\bibnamefont {Scheidgen}}, \bibinfo {author} {\bibfnamefont {C.}~\bibnamefont {Toher}}, \bibinfo {author} {\bibfnamefont {G.~J.}\ \bibnamefont {Conduit}}, \bibinfo {author} {\bibfnamefont {G.}~\bibnamefont {Pizzi}}, \bibinfo {author} {\bibfnamefont {S.}~\bibnamefont {Gražulis}}, \bibinfo {author} {\bibfnamefont {G.-M.}\ \bibnamefont {Rignanese}},\ and\ \bibinfo {author} {\bibfnamefont {R.}~\bibnamefont {Armiento}},\ }\bibfield  {title} {\bibinfo {title} {Developments and applications of the {OPTIMADE API} for materials discovery, design, and data exchange},\ }\href {https://doi.org/10.1039/d4dd00039k} {\bibfield  {journal} {\bibinfo  {journal} {Digit. Discov.}\ }\textbf {\bibinfo {volume} {3}},\ \bibinfo {pages} {1509–1533} (\bibinfo {year} {2024})}\BibitemShut {NoStop}%
\bibitem [{\citenamefont {Xie}\ \emph {et~al.}(2022)\citenamefont {Xie}, \citenamefont {Fu}, \citenamefont {Ganea}, \citenamefont {Barzilay},\ and\ \citenamefont {Jaakkola}}]{xieCrystalDiffusionVariational2022}%
  \BibitemOpen
  \bibfield  {author} {\bibinfo {author} {\bibfnamefont {T.}~\bibnamefont {Xie}}, \bibinfo {author} {\bibfnamefont {X.}~\bibnamefont {Fu}}, \bibinfo {author} {\bibfnamefont {O.-E.}\ \bibnamefont {Ganea}}, \bibinfo {author} {\bibfnamefont {R.}~\bibnamefont {Barzilay}},\ and\ \bibinfo {author} {\bibfnamefont {T.}~\bibnamefont {Jaakkola}},\ }\href@noop {} {\bibinfo {title} {Crystal diffusion variational autoencoder for periodic material generation}} (\bibinfo {year} {2022}),\ \Eprint {https://arxiv.org/abs/2110.06197} {arXiv:2110.06197} \BibitemShut {NoStop}%
\bibitem [{\citenamefont {Castelli}\ \emph {et~al.}(2012{\natexlab{a}})\citenamefont {Castelli}, \citenamefont {Olsen}, \citenamefont {Datta}, \citenamefont {Landis}, \citenamefont {Dahl}, \citenamefont {Thygesen},\ and\ \citenamefont {Jacobsen}}]{castelliComputationalScreeningPerovskite2012}%
  \BibitemOpen
  \bibfield  {author} {\bibinfo {author} {\bibfnamefont {I.~E.}\ \bibnamefont {Castelli}}, \bibinfo {author} {\bibfnamefont {T.}~\bibnamefont {Olsen}}, \bibinfo {author} {\bibfnamefont {S.}~\bibnamefont {Datta}}, \bibinfo {author} {\bibfnamefont {D.~D.}\ \bibnamefont {Landis}}, \bibinfo {author} {\bibfnamefont {S.}~\bibnamefont {Dahl}}, \bibinfo {author} {\bibfnamefont {K.~S.}\ \bibnamefont {Thygesen}},\ and\ \bibinfo {author} {\bibfnamefont {K.~W.}\ \bibnamefont {Jacobsen}},\ }\bibfield  {title} {\bibinfo {title} {Computational screening of perovskite metal oxides for optimal solar light capture},\ }\href {https://doi.org/10.1039/C1EE02717D} {\bibfield  {journal} {\bibinfo  {journal} {Energy Environ. Sci.}\ }\textbf {\bibinfo {volume} {5}},\ \bibinfo {pages} {5814} (\bibinfo {year} {2012}{\natexlab{a}})}\BibitemShut {NoStop}%
\bibitem [{\citenamefont {Castelli}\ \emph {et~al.}(2012{\natexlab{b}})\citenamefont {Castelli}, \citenamefont {Landis}, \citenamefont {Thygesen}, \citenamefont {Dahl}, \citenamefont {Chorkendorff}, \citenamefont {Jaramillo},\ and\ \citenamefont {Jacobsen}}]{castelliNewCubicPerovskites2012}%
  \BibitemOpen
  \bibfield  {author} {\bibinfo {author} {\bibfnamefont {I.~E.}\ \bibnamefont {Castelli}}, \bibinfo {author} {\bibfnamefont {D.~D.}\ \bibnamefont {Landis}}, \bibinfo {author} {\bibfnamefont {K.~S.}\ \bibnamefont {Thygesen}}, \bibinfo {author} {\bibfnamefont {S.}~\bibnamefont {Dahl}}, \bibinfo {author} {\bibfnamefont {I.}~\bibnamefont {Chorkendorff}}, \bibinfo {author} {\bibfnamefont {T.~F.}\ \bibnamefont {Jaramillo}},\ and\ \bibinfo {author} {\bibfnamefont {K.~W.}\ \bibnamefont {Jacobsen}},\ }\bibfield  {title} {\bibinfo {title} {New cubic perovskites for one- and two-photon water splitting using the computational materials repository},\ }\href {https://doi.org/10.1039/C2EE22341D} {\bibfield  {journal} {\bibinfo  {journal} {Energy Environ. Sci.}\ }\textbf {\bibinfo {volume} {5}},\ \bibinfo {pages} {9034} (\bibinfo {year} {2012}{\natexlab{b}})}\BibitemShut {NoStop}%
\bibitem [{\citenamefont {Nouira}\ \emph {et~al.}(2019)\citenamefont {Nouira}, \citenamefont {Sokolovska},\ and\ \citenamefont {Crivello}}]{nouiraCrystalGANLearningDiscover2019}%
  \BibitemOpen
  \bibfield  {author} {\bibinfo {author} {\bibfnamefont {A.}~\bibnamefont {Nouira}}, \bibinfo {author} {\bibfnamefont {N.}~\bibnamefont {Sokolovska}},\ and\ \bibinfo {author} {\bibfnamefont {J.-C.}\ \bibnamefont {Crivello}},\ }\href {https://doi.org/10.48550/arXiv.1810.11203} {\bibinfo {title} {{{CrystalGAN}}: {L}earning to discover crystallographic structures with generative adversarial networks}} (\bibinfo {year} {2019}),\ \Eprint {https://arxiv.org/abs/1810.11203} {arXiv:1810.11203} \BibitemShut {NoStop}%
\bibitem [{\citenamefont {Hoffmann}\ \emph {et~al.}(2019)\citenamefont {Hoffmann}, \citenamefont {Maestrati}, \citenamefont {Sawada}, \citenamefont {Tang}, \citenamefont {Sellier},\ and\ \citenamefont {Bengio}}]{hoffmannDataDrivenApproachEncoding2019}%
  \BibitemOpen
  \bibfield  {author} {\bibinfo {author} {\bibfnamefont {J.}~\bibnamefont {Hoffmann}}, \bibinfo {author} {\bibfnamefont {L.}~\bibnamefont {Maestrati}}, \bibinfo {author} {\bibfnamefont {Y.}~\bibnamefont {Sawada}}, \bibinfo {author} {\bibfnamefont {J.}~\bibnamefont {Tang}}, \bibinfo {author} {\bibfnamefont {J.~M.}\ \bibnamefont {Sellier}},\ and\ \bibinfo {author} {\bibfnamefont {Y.}~\bibnamefont {Bengio}},\ }\href {https://doi.org/10.48550/arXiv.1909.00949} {\bibinfo {title} {Data-driven approach to encoding and decoding 3-{D} crystal structures}} (\bibinfo {year} {2019}),\ \Eprint {https://arxiv.org/abs/1909.00949} {arXiv:1909.00949} \BibitemShut {NoStop}%
\bibitem [{\citenamefont {Sawada}\ \emph {et~al.}(2019)\citenamefont {Sawada}, \citenamefont {Morikawa},\ and\ \citenamefont {Fujii}}]{sawadaStudyDeepGenerative2019}%
  \BibitemOpen
  \bibfield  {author} {\bibinfo {author} {\bibfnamefont {Y.}~\bibnamefont {Sawada}}, \bibinfo {author} {\bibfnamefont {K.}~\bibnamefont {Morikawa}},\ and\ \bibinfo {author} {\bibfnamefont {M.}~\bibnamefont {Fujii}},\ }\href {https://doi.org/10.48550/arXiv.1910.11499} {\bibinfo {title} {Study of deep generative models for inorganic chemical compositions}} (\bibinfo {year} {2019}),\ \Eprint {https://arxiv.org/abs/1910.11499} {arXiv:1910.11499} \BibitemShut {NoStop}%
\bibitem [{\citenamefont {Noh}\ \emph {et~al.}(2019)\citenamefont {Noh}, \citenamefont {Kim}, \citenamefont {Stein}, \citenamefont {{Sanchez-Lengeling}}, \citenamefont {Gregoire}, \citenamefont {{Aspuru-Guzik}},\ and\ \citenamefont {Jung}}]{nohInverseDesignSolidState2019}%
  \BibitemOpen
  \bibfield  {author} {\bibinfo {author} {\bibfnamefont {J.}~\bibnamefont {Noh}}, \bibinfo {author} {\bibfnamefont {J.}~\bibnamefont {Kim}}, \bibinfo {author} {\bibfnamefont {H.~S.}\ \bibnamefont {Stein}}, \bibinfo {author} {\bibfnamefont {B.}~\bibnamefont {{Sanchez-Lengeling}}}, \bibinfo {author} {\bibfnamefont {J.~M.}\ \bibnamefont {Gregoire}}, \bibinfo {author} {\bibfnamefont {A.}~\bibnamefont {{Aspuru-Guzik}}},\ and\ \bibinfo {author} {\bibfnamefont {Y.}~\bibnamefont {Jung}},\ }\bibfield  {title} {\bibinfo {title} {Inverse design of solid-state materials via a continuous representation},\ }\href {https://doi.org/10.1016/j.matt.2019.08.017} {\bibfield  {journal} {\bibinfo  {journal} {Matter}\ }\textbf {\bibinfo {volume} {1}},\ \bibinfo {pages} {1370} (\bibinfo {year} {2019})}\BibitemShut {NoStop}%
\bibitem [{\citenamefont {Dan}\ \emph {et~al.}(2020)\citenamefont {Dan}, \citenamefont {Zhao}, \citenamefont {Li}, \citenamefont {Li}, \citenamefont {Hu},\ and\ \citenamefont {Hu}}]{danGenerativeAdversarialNetworks2020}%
  \BibitemOpen
  \bibfield  {author} {\bibinfo {author} {\bibfnamefont {Y.}~\bibnamefont {Dan}}, \bibinfo {author} {\bibfnamefont {Y.}~\bibnamefont {Zhao}}, \bibinfo {author} {\bibfnamefont {X.}~\bibnamefont {Li}}, \bibinfo {author} {\bibfnamefont {S.}~\bibnamefont {Li}}, \bibinfo {author} {\bibfnamefont {M.}~\bibnamefont {Hu}},\ and\ \bibinfo {author} {\bibfnamefont {J.}~\bibnamefont {Hu}},\ }\bibfield  {title} {\bibinfo {title} {Generative adversarial networks ({{GAN}}) based efficient sampling of chemical composition space for inverse design of inorganic materials},\ }\href {https://doi.org/10.1038/s41524-020-00352-0} {\bibfield  {journal} {\bibinfo  {journal} {npj Comput. Mater.}\ }\textbf {\bibinfo {volume} {6}},\ \bibinfo {pages} {1} (\bibinfo {year} {2020})}\BibitemShut {NoStop}%
\bibitem [{\citenamefont {Kim}\ \emph {et~al.}(2020)\citenamefont {Kim}, \citenamefont {Noh}, \citenamefont {Gu}, \citenamefont {{Aspuru-Guzik}},\ and\ \citenamefont {Jung}}]{kimGenerativeAdversarialNetworks2020}%
  \BibitemOpen
  \bibfield  {author} {\bibinfo {author} {\bibfnamefont {S.}~\bibnamefont {Kim}}, \bibinfo {author} {\bibfnamefont {J.}~\bibnamefont {Noh}}, \bibinfo {author} {\bibfnamefont {G.~H.}\ \bibnamefont {Gu}}, \bibinfo {author} {\bibfnamefont {A.}~\bibnamefont {{Aspuru-Guzik}}},\ and\ \bibinfo {author} {\bibfnamefont {Y.}~\bibnamefont {Jung}},\ }\bibfield  {title} {\bibinfo {title} {Generative adversarial networks for crystal structure prediction},\ }\href {https://doi.org/10.1021/acscentsci.0c00426} {\bibfield  {journal} {\bibinfo  {journal} {ACS Cent. Sci.}\ }\textbf {\bibinfo {volume} {6}},\ \bibinfo {pages} {1412} (\bibinfo {year} {2020})}\BibitemShut {NoStop}%
\bibitem [{\citenamefont {Court}\ \emph {et~al.}(2020)\citenamefont {Court}, \citenamefont {Yildirim}, \citenamefont {Jain},\ and\ \citenamefont {Cole}}]{court3DInorganicCrystal2020}%
  \BibitemOpen
  \bibfield  {author} {\bibinfo {author} {\bibfnamefont {C.~J.}\ \bibnamefont {Court}}, \bibinfo {author} {\bibfnamefont {B.}~\bibnamefont {Yildirim}}, \bibinfo {author} {\bibfnamefont {A.}~\bibnamefont {Jain}},\ and\ \bibinfo {author} {\bibfnamefont {J.~M.}\ \bibnamefont {Cole}},\ }\bibfield  {title} {\bibinfo {title} {3-{D} inorganic crystal structure generation and property prediction via representation learning},\ }\href {https://doi.org/10.1021/acs.jcim.0c00464} {\bibfield  {journal} {\bibinfo  {journal} {J. Chem. Inf. Model.}\ }\textbf {\bibinfo {volume} {60}},\ \bibinfo {pages} {4518} (\bibinfo {year} {2020})}\BibitemShut {NoStop}%
\bibitem [{\citenamefont {Zhao}\ \emph {et~al.}(2021)\citenamefont {Zhao}, \citenamefont {{Al-Fahdi}}, \citenamefont {Hu}, \citenamefont {Siriwardane}, \citenamefont {Song}, \citenamefont {Nasiri},\ and\ \citenamefont {Hu}}]{zhaoHighThroughputDiscoveryNovel2021b}%
  \BibitemOpen
  \bibfield  {author} {\bibinfo {author} {\bibfnamefont {Y.}~\bibnamefont {Zhao}}, \bibinfo {author} {\bibfnamefont {M.}~\bibnamefont {{Al-Fahdi}}}, \bibinfo {author} {\bibfnamefont {M.}~\bibnamefont {Hu}}, \bibinfo {author} {\bibfnamefont {E.~M.~D.}\ \bibnamefont {Siriwardane}}, \bibinfo {author} {\bibfnamefont {Y.}~\bibnamefont {Song}}, \bibinfo {author} {\bibfnamefont {A.}~\bibnamefont {Nasiri}},\ and\ \bibinfo {author} {\bibfnamefont {J.}~\bibnamefont {Hu}},\ }\bibfield  {title} {\bibinfo {title} {High-throughput discovery of novel cubic crystal materials using deep generative neural networks},\ }\href {https://doi.org/10.1002/advs.202100566} {\bibfield  {journal} {\bibinfo  {journal} {Adv. Sci.}\ }\textbf {\bibinfo {volume} {8}},\ \bibinfo {pages} {2100566} (\bibinfo {year} {2021})}\BibitemShut {NoStop}%
\bibitem [{\citenamefont {Long}\ \emph {et~al.}(2021)\citenamefont {Long}, \citenamefont {Fortunato}, \citenamefont {Opahle}, \citenamefont {Zhang}, \citenamefont {Samathrakis}, \citenamefont {Shen}, \citenamefont {Gutfleisch},\ and\ \citenamefont {Zhang}}]{longConstrainedCrystalsDeep2021}%
  \BibitemOpen
  \bibfield  {author} {\bibinfo {author} {\bibfnamefont {T.}~\bibnamefont {Long}}, \bibinfo {author} {\bibfnamefont {N.~M.}\ \bibnamefont {Fortunato}}, \bibinfo {author} {\bibfnamefont {I.}~\bibnamefont {Opahle}}, \bibinfo {author} {\bibfnamefont {Y.}~\bibnamefont {Zhang}}, \bibinfo {author} {\bibfnamefont {I.}~\bibnamefont {Samathrakis}}, \bibinfo {author} {\bibfnamefont {C.}~\bibnamefont {Shen}}, \bibinfo {author} {\bibfnamefont {O.}~\bibnamefont {Gutfleisch}},\ and\ \bibinfo {author} {\bibfnamefont {H.}~\bibnamefont {Zhang}},\ }\bibfield  {title} {\bibinfo {title} {Constrained crystals deep convolutional generative adversarial network for the inverse design of crystal structures},\ }\href {https://doi.org/10.1038/s41524-021-00526-4} {\bibfield  {journal} {\bibinfo  {journal} {npj Comput. Mater.}\ }\textbf {\bibinfo {volume} {7}},\ \bibinfo {pages} {1} (\bibinfo {year} {2021})}\BibitemShut {NoStop}%
\bibitem [{\citenamefont {Ren}\ \emph {et~al.}(2022)\citenamefont {Ren}, \citenamefont {Tian}, \citenamefont {Noh}, \citenamefont {Oviedo}, \citenamefont {Xing}, \citenamefont {Li}, \citenamefont {Liang}, \citenamefont {Zhu}, \citenamefont {Aberle}, \citenamefont {Sun}, \citenamefont {Wang}, \citenamefont {Liu}, \citenamefont {Li}, \citenamefont {Jayavelu}, \citenamefont {Hippalgaonkar}, \citenamefont {Jung},\ and\ \citenamefont {Buonassisi}}]{renInvertibleCrystallographicRepresentation2022}%
  \BibitemOpen
  \bibfield  {author} {\bibinfo {author} {\bibfnamefont {Z.}~\bibnamefont {Ren}}, \bibinfo {author} {\bibfnamefont {S.~I.~P.}\ \bibnamefont {Tian}}, \bibinfo {author} {\bibfnamefont {J.}~\bibnamefont {Noh}}, \bibinfo {author} {\bibfnamefont {F.}~\bibnamefont {Oviedo}}, \bibinfo {author} {\bibfnamefont {G.}~\bibnamefont {Xing}}, \bibinfo {author} {\bibfnamefont {J.}~\bibnamefont {Li}}, \bibinfo {author} {\bibfnamefont {Q.}~\bibnamefont {Liang}}, \bibinfo {author} {\bibfnamefont {R.}~\bibnamefont {Zhu}}, \bibinfo {author} {\bibfnamefont {A.~G.}\ \bibnamefont {Aberle}}, \bibinfo {author} {\bibfnamefont {S.}~\bibnamefont {Sun}}, \bibinfo {author} {\bibfnamefont {X.}~\bibnamefont {Wang}}, \bibinfo {author} {\bibfnamefont {Y.}~\bibnamefont {Liu}}, \bibinfo {author} {\bibfnamefont {Q.}~\bibnamefont {Li}}, \bibinfo {author} {\bibfnamefont {S.}~\bibnamefont {Jayavelu}}, \bibinfo {author} {\bibfnamefont {K.}~\bibnamefont {Hippalgaonkar}}, \bibinfo {author} {\bibfnamefont {Y.}~\bibnamefont {Jung}},\ and\ \bibinfo {author} {\bibfnamefont {T.}~\bibnamefont {Buonassisi}},\ }\bibfield  {title} {\bibinfo {title} {An invertible crystallographic representation for general inverse design of inorganic crystals with targeted properties},\ }\href {https://doi.org/10.1016/j.matt.2021.11.032} {\bibfield  {journal} {\bibinfo  {journal} {Matter}\ }\textbf {\bibinfo {volume} {5}},\ \bibinfo {pages} {314} (\bibinfo {year} {2022})}\BibitemShut {NoStop}%
\bibitem [{\citenamefont {Long}\ \emph {et~al.}(2022)\citenamefont {Long}, \citenamefont {Zhang}, \citenamefont {Fortunato}, \citenamefont {Shen}, \citenamefont {Dai},\ and\ \citenamefont {Zhang}}]{longInverseDesignCrystal2022}%
  \BibitemOpen
  \bibfield  {author} {\bibinfo {author} {\bibfnamefont {T.}~\bibnamefont {Long}}, \bibinfo {author} {\bibfnamefont {Y.}~\bibnamefont {Zhang}}, \bibinfo {author} {\bibfnamefont {N.~M.}\ \bibnamefont {Fortunato}}, \bibinfo {author} {\bibfnamefont {C.}~\bibnamefont {Shen}}, \bibinfo {author} {\bibfnamefont {M.}~\bibnamefont {Dai}},\ and\ \bibinfo {author} {\bibfnamefont {H.}~\bibnamefont {Zhang}},\ }\bibfield  {title} {\bibinfo {title} {Inverse design of crystal structures for multicomponent systems},\ }\href {https://doi.org/10.1016/j.actamat.2022.117898} {\bibfield  {journal} {\bibinfo  {journal} {Acta Mater.}\ }\textbf {\bibinfo {volume} {231}},\ \bibinfo {pages} {117898} (\bibinfo {year} {2022})}\BibitemShut {NoStop}%
\bibitem [{\citenamefont {Zhao}\ \emph {et~al.}(2023)\citenamefont {Zhao}, \citenamefont {Siriwardane}, \citenamefont {Wu}, \citenamefont {Fu}, \citenamefont {{Al-Fahdi}}, \citenamefont {Hu},\ and\ \citenamefont {Hu}}]{zhaoPhysicsGuidedDeep2023}%
  \BibitemOpen
  \bibfield  {author} {\bibinfo {author} {\bibfnamefont {Y.}~\bibnamefont {Zhao}}, \bibinfo {author} {\bibfnamefont {E.~M.~D.}\ \bibnamefont {Siriwardane}}, \bibinfo {author} {\bibfnamefont {Z.}~\bibnamefont {Wu}}, \bibinfo {author} {\bibfnamefont {N.}~\bibnamefont {Fu}}, \bibinfo {author} {\bibfnamefont {M.}~\bibnamefont {{Al-Fahdi}}}, \bibinfo {author} {\bibfnamefont {M.}~\bibnamefont {Hu}},\ and\ \bibinfo {author} {\bibfnamefont {J.}~\bibnamefont {Hu}},\ }\bibfield  {title} {\bibinfo {title} {Physics guided deep learning for generative design of crystal materials with symmetry constraints},\ }\href {https://doi.org/10.1038/s41524-023-00987-9} {\bibfield  {journal} {\bibinfo  {journal} {npj Comput. Mater.}\ }\textbf {\bibinfo {volume} {9}},\ \bibinfo {pages} {1} (\bibinfo {year} {2023})}\BibitemShut {NoStop}%
\bibitem [{\citenamefont {{Flam-Shepherd}}\ and\ \citenamefont {{Aspuru-Guzik}}(2023)}]{flam-shepherdLanguageModelsCan2023}%
  \BibitemOpen
  \bibfield  {author} {\bibinfo {author} {\bibfnamefont {D.}~\bibnamefont {{Flam-Shepherd}}}\ and\ \bibinfo {author} {\bibfnamefont {A.}~\bibnamefont {{Aspuru-Guzik}}},\ }\href {https://doi.org/10.48550/arXiv.2305.05708} {\bibinfo {title} {Language models can generate molecules, materials, and protein binding sites directly in three dimensions as {{XYZ}}, {{CIF}}, and {{PDB}} files}} (\bibinfo {year} {2023}),\ \Eprint {https://arxiv.org/abs/2305.05708} {arXiv:2305.05708} \BibitemShut {NoStop}%
\bibitem [{\citenamefont {Liu}\ \emph {et~al.}(2023)\citenamefont {Liu}, \citenamefont {Gao}, \citenamefont {Yang},\ and\ \citenamefont {Han}}]{liuPCVAEPhysicsinformedNeural2023}%
  \BibitemOpen
  \bibfield  {author} {\bibinfo {author} {\bibfnamefont {K.}~\bibnamefont {Liu}}, \bibinfo {author} {\bibfnamefont {S.}~\bibnamefont {Gao}}, \bibinfo {author} {\bibfnamefont {K.}~\bibnamefont {Yang}},\ and\ \bibinfo {author} {\bibfnamefont {Y.}~\bibnamefont {Han}},\ }\bibfield  {title} {\bibinfo {title} {{{PCVAE}}: {A} physics-informed neural network for determining the symmetry and geometry of crystals},\ }in\ \href {https://doi.org/10.1109/IJCNN54540.2023.10191051} {\emph {\bibinfo {booktitle} {2023 {{International Joint Conference}} on {{Neural Networks}} ({{IJCNN}})}}}\ (\bibinfo {year} {2023})\ pp.\ \bibinfo {pages} {1--8}\BibitemShut {NoStop}%
\bibitem [{\citenamefont {Qi}\ \emph {et~al.}(2023)\citenamefont {Qi}, \citenamefont {Geng}, \citenamefont {Rando}, \citenamefont {Ohama}, \citenamefont {Kumar},\ and\ \citenamefont {Levine}}]{qiLatentConservativeObjective2023}%
  \BibitemOpen
  \bibfield  {author} {\bibinfo {author} {\bibfnamefont {H.}~\bibnamefont {Qi}}, \bibinfo {author} {\bibfnamefont {X.}~\bibnamefont {Geng}}, \bibinfo {author} {\bibfnamefont {S.}~\bibnamefont {Rando}}, \bibinfo {author} {\bibfnamefont {I.}~\bibnamefont {Ohama}}, \bibinfo {author} {\bibfnamefont {A.}~\bibnamefont {Kumar}},\ and\ \bibinfo {author} {\bibfnamefont {S.}~\bibnamefont {Levine}},\ }\href {https://doi.org/10.48550/arXiv.2310.10056} {\bibinfo {title} {Latent conservative objective models for data-driven crystal structure prediction}} (\bibinfo {year} {2023}),\ \Eprint {https://arxiv.org/abs/2310.10056} {arXiv:2310.10056} \BibitemShut {NoStop}%
\bibitem [{Neu(2023)}]{NeurIPSHierarchicalGFlowNet2023}%
  \BibitemOpen
  \href@noop {} {\bibinfo {title} {{NeurIPS} hierarchical {GFlowNet} for crystal structure generation}},\ \bibinfo {howpublished} {https://neurips.cc/virtual/2023/78549} (\bibinfo {year} {2023})\BibitemShut {NoStop}%
\bibitem [{\citenamefont {Xiao}\ \emph {et~al.}(2023)\citenamefont {Xiao}, \citenamefont {Li}, \citenamefont {Shi}, \citenamefont {Chen}, \citenamefont {Zhu}, \citenamefont {Chen},\ and\ \citenamefont {Wang}}]{xiaoInvertibleInvariantCrystal2023}%
  \BibitemOpen
  \bibfield  {author} {\bibinfo {author} {\bibfnamefont {H.}~\bibnamefont {Xiao}}, \bibinfo {author} {\bibfnamefont {R.}~\bibnamefont {Li}}, \bibinfo {author} {\bibfnamefont {X.}~\bibnamefont {Shi}}, \bibinfo {author} {\bibfnamefont {Y.}~\bibnamefont {Chen}}, \bibinfo {author} {\bibfnamefont {L.}~\bibnamefont {Zhu}}, \bibinfo {author} {\bibfnamefont {X.}~\bibnamefont {Chen}},\ and\ \bibinfo {author} {\bibfnamefont {L.}~\bibnamefont {Wang}},\ }\bibfield  {title} {\bibinfo {title} {An invertible, invariant crystal representation for inverse design of solid-state materials using generative deep learning},\ }\href {https://doi.org/10.1038/s41467-023-42870-7} {\bibfield  {journal} {\bibinfo  {journal} {Nat. Commun.}\ }\textbf {\bibinfo {volume} {14}},\ \bibinfo {pages} {7027} (\bibinfo {year} {2023})}\BibitemShut {NoStop}%
\bibitem [{\citenamefont {Luo}\ \emph {et~al.}(2023)\citenamefont {Luo}, \citenamefont {Liu},\ and\ \citenamefont {Ji}}]{luoSymmetryAwareGenerationPeriodic2023}%
  \BibitemOpen
  \bibfield  {author} {\bibinfo {author} {\bibfnamefont {Y.}~\bibnamefont {Luo}}, \bibinfo {author} {\bibfnamefont {C.}~\bibnamefont {Liu}},\ and\ \bibinfo {author} {\bibfnamefont {S.}~\bibnamefont {Ji}},\ }\href {https://doi.org/10.48550/arXiv.2307.02707} {\bibinfo {title} {Towards symmetry-aware generation of periodic materials}} (\bibinfo {year} {2023}),\ \Eprint {https://arxiv.org/abs/2307.02707} {arXiv:2307.02707} \BibitemShut {NoStop}%
\bibitem [{\citenamefont {AI4Science}\ \emph {et~al.}(2023)\citenamefont {AI4Science}, \citenamefont {{Hernandez-Garcia}}, \citenamefont {Duval}, \citenamefont {Volokhova}, \citenamefont {Bengio}, \citenamefont {Sharma}, \citenamefont {Carrier}, \citenamefont {Benabed}, \citenamefont {Koziarski},\ and\ \citenamefont {Schmidt}}]{ai4scienceCrystalGFNSamplingCrystals2023}%
  \BibitemOpen
  \bibfield  {author} {\bibinfo {author} {\bibfnamefont {M.}~\bibnamefont {AI4Science}}, \bibinfo {author} {\bibfnamefont {A.}~\bibnamefont {{Hernandez-Garcia}}}, \bibinfo {author} {\bibfnamefont {A.}~\bibnamefont {Duval}}, \bibinfo {author} {\bibfnamefont {A.}~\bibnamefont {Volokhova}}, \bibinfo {author} {\bibfnamefont {Y.}~\bibnamefont {Bengio}}, \bibinfo {author} {\bibfnamefont {D.}~\bibnamefont {Sharma}}, \bibinfo {author} {\bibfnamefont {P.~L.}\ \bibnamefont {Carrier}}, \bibinfo {author} {\bibfnamefont {Y.}~\bibnamefont {Benabed}}, \bibinfo {author} {\bibfnamefont {M.}~\bibnamefont {Koziarski}},\ and\ \bibinfo {author} {\bibfnamefont {V.}~\bibnamefont {Schmidt}},\ }\href {https://doi.org/10.48550/arXiv.2310.04925} {\bibinfo {title} {Crystal-{{GFN}}: Sampling crystals with desirable properties and constraints}} (\bibinfo {year} {2023}),\ \Eprint {https://arxiv.org/abs/2310.04925} {arXiv:2310.04925} \BibitemShut {NoStop}%
\bibitem [{\citenamefont {Klipfel}\ \emph {et~al.}(2023)\citenamefont {Klipfel}, \citenamefont {Fregier}, \citenamefont {Sayede},\ and\ \citenamefont {Bouraoui}}]{klipfelVectorFieldOriented2023}%
  \BibitemOpen
  \bibfield  {author} {\bibinfo {author} {\bibfnamefont {A.}~\bibnamefont {Klipfel}}, \bibinfo {author} {\bibfnamefont {Y.}~\bibnamefont {Fregier}}, \bibinfo {author} {\bibfnamefont {A.}~\bibnamefont {Sayede}},\ and\ \bibinfo {author} {\bibfnamefont {Z.}~\bibnamefont {Bouraoui}},\ }\href {https://doi.org/10.48550/arXiv.2401.05402} {\bibinfo {title} {Vector field oriented diffusion model for crystal material generation}} (\bibinfo {year} {2023}),\ \Eprint {https://arxiv.org/abs/2401.05402} {arXiv:2401.05402} \BibitemShut {NoStop}%
\bibitem [{\citenamefont {Novitskiy}\ \emph {et~al.}(2024)\citenamefont {Novitskiy}, \citenamefont {Lazarev}, \citenamefont {Tiutiulnikov}, \citenamefont {Vakhrameev}, \citenamefont {Eremin}, \citenamefont {Humonen}, \citenamefont {Kuznetsov}, \citenamefont {Dimitrov},\ and\ \citenamefont {Budennyy}}]{novitskiyUnleashingPowerNovel2024}%
  \BibitemOpen
  \bibfield  {author} {\bibinfo {author} {\bibfnamefont {L.}~\bibnamefont {Novitskiy}}, \bibinfo {author} {\bibfnamefont {V.}~\bibnamefont {Lazarev}}, \bibinfo {author} {\bibfnamefont {M.}~\bibnamefont {Tiutiulnikov}}, \bibinfo {author} {\bibfnamefont {N.}~\bibnamefont {Vakhrameev}}, \bibinfo {author} {\bibfnamefont {R.}~\bibnamefont {Eremin}}, \bibinfo {author} {\bibfnamefont {I.}~\bibnamefont {Humonen}}, \bibinfo {author} {\bibfnamefont {A.}~\bibnamefont {Kuznetsov}}, \bibinfo {author} {\bibfnamefont {D.}~\bibnamefont {Dimitrov}},\ and\ \bibinfo {author} {\bibfnamefont {S.}~\bibnamefont {Budennyy}},\ }\href {https://doi.org/10.48550/ARXIV.2411.03156} {\bibinfo {title} {Unleashing the power of novel conditional generative approaches for new materials discovery}} (\bibinfo {year} {2024})\BibitemShut {NoStop}%
\bibitem [{\citenamefont {Pakornchote}\ \emph {et~al.}(2024)\citenamefont {Pakornchote}, \citenamefont {{Choomphon-anomakhun}}, \citenamefont {Arrerut}, \citenamefont {Atthapak}, \citenamefont {Khamkaeo}, \citenamefont {Chotibut},\ and\ \citenamefont {Bovornratanaraks}}]{pakornchoteDiffusionProbabilisticModels2024}%
  \BibitemOpen
  \bibfield  {author} {\bibinfo {author} {\bibfnamefont {T.}~\bibnamefont {Pakornchote}}, \bibinfo {author} {\bibfnamefont {N.}~\bibnamefont {{Choomphon-anomakhun}}}, \bibinfo {author} {\bibfnamefont {S.}~\bibnamefont {Arrerut}}, \bibinfo {author} {\bibfnamefont {C.}~\bibnamefont {Atthapak}}, \bibinfo {author} {\bibfnamefont {S.}~\bibnamefont {Khamkaeo}}, \bibinfo {author} {\bibfnamefont {T.}~\bibnamefont {Chotibut}},\ and\ \bibinfo {author} {\bibfnamefont {T.}~\bibnamefont {Bovornratanaraks}},\ }\bibfield  {title} {\bibinfo {title} {Diffusion probabilistic models enhance variational autoencoder for crystal structure generative modeling},\ }\href {https://doi.org/10.1038/s41598-024-51400-4} {\bibfield  {journal} {\bibinfo  {journal} {Sci. Rep.}\ }\textbf {\bibinfo {volume} {14}},\ \bibinfo {pages} {1275} (\bibinfo {year} {2024})}\BibitemShut {NoStop}%
\bibitem [{\citenamefont {Alverson}\ \emph {et~al.}(2024)\citenamefont {Alverson}, \citenamefont {Baird}, \citenamefont {Murdock}, \citenamefont {Ho}, \citenamefont {Johnson},\ and\ \citenamefont {Sparks}}]{alversonGenerativeAdversarialNetworks2024}%
  \BibitemOpen
  \bibfield  {author} {\bibinfo {author} {\bibfnamefont {M.}~\bibnamefont {Alverson}}, \bibinfo {author} {\bibfnamefont {S.~G.}\ \bibnamefont {Baird}}, \bibinfo {author} {\bibfnamefont {R.}~\bibnamefont {Murdock}}, \bibinfo {author} {\bibfnamefont {E.~S.-H.}\ \bibnamefont {Ho}}, \bibinfo {author} {\bibfnamefont {J.}~\bibnamefont {Johnson}},\ and\ \bibinfo {author} {\bibfnamefont {T.~D.}\ \bibnamefont {Sparks}},\ }\bibfield  {title} {\bibinfo {title} {Generative adversarial networks and diffusion models in material discovery},\ }\href {https://doi.org/10.1039/D3DD00137G} {\bibfield  {journal} {\bibinfo  {journal} {Digit. Discov.}\ }\textbf {\bibinfo {volume} {3}},\ \bibinfo {pages} {62} (\bibinfo {year} {2024})}\BibitemShut {NoStop}%
\bibitem [{\citenamefont {Gruver}\ \emph {et~al.}(2024)\citenamefont {Gruver}, \citenamefont {Sriram}, \citenamefont {Madotto}, \citenamefont {Wilson}, \citenamefont {Zitnick},\ and\ \citenamefont {Ulissi}}]{gruverFineTunedLanguageModels2024}%
  \BibitemOpen
  \bibfield  {author} {\bibinfo {author} {\bibfnamefont {N.}~\bibnamefont {Gruver}}, \bibinfo {author} {\bibfnamefont {A.}~\bibnamefont {Sriram}}, \bibinfo {author} {\bibfnamefont {A.}~\bibnamefont {Madotto}}, \bibinfo {author} {\bibfnamefont {A.~G.}\ \bibnamefont {Wilson}}, \bibinfo {author} {\bibfnamefont {C.~L.}\ \bibnamefont {Zitnick}},\ and\ \bibinfo {author} {\bibfnamefont {Z.}~\bibnamefont {Ulissi}},\ }\href {https://doi.org/10.48550/arXiv.2402.04379} {\bibinfo {title} {Fine-tuned language models generate stable inorganic materials as text}} (\bibinfo {year} {2024}),\ \Eprint {https://arxiv.org/abs/2402.04379} {arXiv:2402.04379} \BibitemShut {NoStop}%
\bibitem [{\citenamefont {Jiao}\ \emph {et~al.}(2024{\natexlab{a}})\citenamefont {Jiao}, \citenamefont {Huang}, \citenamefont {Lin}, \citenamefont {Han}, \citenamefont {Chen}, \citenamefont {Lu},\ and\ \citenamefont {Liu}}]{jiaoCrystalStructurePrediction2024}%
  \BibitemOpen
  \bibfield  {author} {\bibinfo {author} {\bibfnamefont {R.}~\bibnamefont {Jiao}}, \bibinfo {author} {\bibfnamefont {W.}~\bibnamefont {Huang}}, \bibinfo {author} {\bibfnamefont {P.}~\bibnamefont {Lin}}, \bibinfo {author} {\bibfnamefont {J.}~\bibnamefont {Han}}, \bibinfo {author} {\bibfnamefont {P.}~\bibnamefont {Chen}}, \bibinfo {author} {\bibfnamefont {Y.}~\bibnamefont {Lu}},\ and\ \bibinfo {author} {\bibfnamefont {Y.}~\bibnamefont {Liu}},\ }\href {https://doi.org/10.48550/arXiv.2309.04475} {\bibinfo {title} {Crystal structure prediction by joint equivariant diffusion}} (\bibinfo {year} {2024}{\natexlab{a}}),\ \Eprint {https://arxiv.org/abs/2309.04475} {arXiv:2309.04475} \BibitemShut {NoStop}%
\bibitem [{\citenamefont {Jiao}\ \emph {et~al.}(2024{\natexlab{b}})\citenamefont {Jiao}, \citenamefont {Huang}, \citenamefont {Liu}, \citenamefont {Zhao},\ and\ \citenamefont {Liu}}]{jiaoSpaceGroupConstrained2024}%
  \BibitemOpen
  \bibfield  {author} {\bibinfo {author} {\bibfnamefont {R.}~\bibnamefont {Jiao}}, \bibinfo {author} {\bibfnamefont {W.}~\bibnamefont {Huang}}, \bibinfo {author} {\bibfnamefont {Y.}~\bibnamefont {Liu}}, \bibinfo {author} {\bibfnamefont {D.}~\bibnamefont {Zhao}},\ and\ \bibinfo {author} {\bibfnamefont {Y.}~\bibnamefont {Liu}},\ }\href {https://doi.org/10.48550/arXiv.2402.03992} {\bibinfo {title} {Space group constrained crystal generation}} (\bibinfo {year} {2024}{\natexlab{b}}),\ \Eprint {https://arxiv.org/abs/2402.03992} {arXiv:2402.03992} \BibitemShut {NoStop}%
\bibitem [{\citenamefont {Ye}\ \emph {et~al.}(2024)\citenamefont {Ye}, \citenamefont {Weng},\ and\ \citenamefont {Wu}}]{yeConCDVAEMethodConditional2024}%
  \BibitemOpen
  \bibfield  {author} {\bibinfo {author} {\bibfnamefont {C.-Y.}\ \bibnamefont {Ye}}, \bibinfo {author} {\bibfnamefont {H.-M.}\ \bibnamefont {Weng}},\ and\ \bibinfo {author} {\bibfnamefont {Q.-S.}\ \bibnamefont {Wu}},\ }\bibfield  {title} {\bibinfo {title} {Con-{{CDVAE}}: {{A}} method for the conditional generation of crystal structures},\ }\href {https://doi.org/10.1016/j.commt.2024.100003} {\bibfield  {journal} {\bibinfo  {journal} {Compt. Mater. Today}\ }\textbf {\bibinfo {volume} {1}},\ \bibinfo {pages} {100003} (\bibinfo {year} {2024})}\BibitemShut {NoStop}%
\bibitem [{\citenamefont {Li}\ and\ \citenamefont {Birbilis}(2024)}]{liNSGANNondominantSorting2024}%
  \BibitemOpen
  \bibfield  {author} {\bibinfo {author} {\bibfnamefont {Z.}~\bibnamefont {Li}}\ and\ \bibinfo {author} {\bibfnamefont {N.}~\bibnamefont {Birbilis}},\ }\bibfield  {title} {\bibinfo {title} {{{NSGAN}}: {A} non-dominant sorting optimisation-based generative adversarial design framework for alloy discovery},\ }\href {https://doi.org/10.1038/s41524-024-01294-7} {\bibfield  {journal} {\bibinfo  {journal} {npj Comput. Mater.}\ }\textbf {\bibinfo {volume} {10}},\ \bibinfo {pages} {1} (\bibinfo {year} {2024})}\BibitemShut {NoStop}%
\bibitem [{\citenamefont {Yang}\ \emph {et~al.}(2024{\natexlab{a}})\citenamefont {Yang}, \citenamefont {Cho}, \citenamefont {Merchant}, \citenamefont {Abbeel}, \citenamefont {Schuurmans}, \citenamefont {Mordatch},\ and\ \citenamefont {Cubuk}}]{yangScalableDiffusionMaterials2024}%
  \BibitemOpen
  \bibfield  {author} {\bibinfo {author} {\bibfnamefont {S.}~\bibnamefont {Yang}}, \bibinfo {author} {\bibfnamefont {K.}~\bibnamefont {Cho}}, \bibinfo {author} {\bibfnamefont {A.}~\bibnamefont {Merchant}}, \bibinfo {author} {\bibfnamefont {P.}~\bibnamefont {Abbeel}}, \bibinfo {author} {\bibfnamefont {D.}~\bibnamefont {Schuurmans}}, \bibinfo {author} {\bibfnamefont {I.}~\bibnamefont {Mordatch}},\ and\ \bibinfo {author} {\bibfnamefont {E.~D.}\ \bibnamefont {Cubuk}},\ }\href {https://doi.org/10.48550/arXiv.2311.09235} {\bibinfo {title} {Scalable diffusion for materials generation}} (\bibinfo {year} {2024}{\natexlab{a}}),\ \Eprint {https://arxiv.org/abs/2311.09235} {arXiv:2311.09235} \BibitemShut {NoStop}%
\bibitem [{\citenamefont {Miller}\ \emph {et~al.}(2024)\citenamefont {Miller}, \citenamefont {Chen}, \citenamefont {Sriram},\ and\ \citenamefont {Wood}}]{millerFlowMMGeneratingMaterials2024}%
  \BibitemOpen
  \bibfield  {author} {\bibinfo {author} {\bibfnamefont {B.~K.}\ \bibnamefont {Miller}}, \bibinfo {author} {\bibfnamefont {R.~T.~Q.}\ \bibnamefont {Chen}}, \bibinfo {author} {\bibfnamefont {A.}~\bibnamefont {Sriram}},\ and\ \bibinfo {author} {\bibfnamefont {B.~M.}\ \bibnamefont {Wood}},\ }\href {https://doi.org/10.48550/arXiv.2406.04713} {\bibinfo {title} {{{FlowMM}}: {{G}}enerating materials with riemannian flow matching}} (\bibinfo {year} {2024}),\ \Eprint {https://arxiv.org/abs/2406.04713} {arXiv:2406.04713} \BibitemShut {NoStop}%
\bibitem [{\citenamefont {Sinha}\ \emph {et~al.}(2024)\citenamefont {Sinha}, \citenamefont {Jia},\ and\ \citenamefont {Fung}}]{sinhaRepresentationspaceDiffusionModels2024}%
  \BibitemOpen
  \bibfield  {author} {\bibinfo {author} {\bibfnamefont {A.}~\bibnamefont {Sinha}}, \bibinfo {author} {\bibfnamefont {S.}~\bibnamefont {Jia}},\ and\ \bibinfo {author} {\bibfnamefont {V.}~\bibnamefont {Fung}},\ }\href {https://doi.org/10.48550/arXiv.2408.07213} {\bibinfo {title} {Representation-space diffusion models for generating periodic materials}} (\bibinfo {year} {2024}),\ \Eprint {https://arxiv.org/abs/2408.07213} {arXiv:2408.07213} \BibitemShut {NoStop}%
\bibitem [{\citenamefont {Cao}\ \emph {et~al.}(2024)\citenamefont {Cao}, \citenamefont {Luo}, \citenamefont {Lv},\ and\ \citenamefont {Wang}}]{caoSpaceGroupInformed2024}%
  \BibitemOpen
  \bibfield  {author} {\bibinfo {author} {\bibfnamefont {Z.}~\bibnamefont {Cao}}, \bibinfo {author} {\bibfnamefont {X.}~\bibnamefont {Luo}}, \bibinfo {author} {\bibfnamefont {J.}~\bibnamefont {Lv}},\ and\ \bibinfo {author} {\bibfnamefont {L.}~\bibnamefont {Wang}},\ }\href {https://doi.org/10.48550/arXiv.2403.15734} {\bibinfo {title} {Space group informed transformer for crystalline materials generation}} (\bibinfo {year} {2024}),\ \Eprint {https://arxiv.org/abs/2403.15734} {arXiv:2403.15734} \BibitemShut {NoStop}%
\bibitem [{\citenamefont {Qin}\ \emph {et~al.}(2024)\citenamefont {Qin}, \citenamefont {Liu}, \citenamefont {Ma}, \citenamefont {Du}, \citenamefont {Jiang},\ and\ \citenamefont {Zhao}}]{qinInverseDesignSemiconductor2024}%
  \BibitemOpen
  \bibfield  {author} {\bibinfo {author} {\bibfnamefont {C.}~\bibnamefont {Qin}}, \bibinfo {author} {\bibfnamefont {J.}~\bibnamefont {Liu}}, \bibinfo {author} {\bibfnamefont {S.}~\bibnamefont {Ma}}, \bibinfo {author} {\bibfnamefont {J.}~\bibnamefont {Du}}, \bibinfo {author} {\bibfnamefont {G.}~\bibnamefont {Jiang}},\ and\ \bibinfo {author} {\bibfnamefont {L.}~\bibnamefont {Zhao}},\ }\bibfield  {title} {\bibinfo {title} {Inverse design of semiconductor materials with deep generative models},\ }\href {https://doi.org/10.1039/D4TA02872D} {\bibfield  {journal} {\bibinfo  {journal} {J. Mater. Chem. A}\ }\textbf {\bibinfo {volume} {12}},\ \bibinfo {pages} {22689} (\bibinfo {year} {2024})}\BibitemShut {NoStop}%
\bibitem [{\citenamefont {Chenebuah}\ \emph {et~al.}(2024)\citenamefont {Chenebuah}, \citenamefont {Nganbe},\ and\ \citenamefont {Tchagang}}]{chenebuahDeepGenerativeModeling2024}%
  \BibitemOpen
  \bibfield  {author} {\bibinfo {author} {\bibfnamefont {E.~T.}\ \bibnamefont {Chenebuah}}, \bibinfo {author} {\bibfnamefont {M.}~\bibnamefont {Nganbe}},\ and\ \bibinfo {author} {\bibfnamefont {A.~B.}\ \bibnamefont {Tchagang}},\ }\bibfield  {title} {\bibinfo {title} {A deep generative modeling architecture for designing lattice-constrained perovskite materials},\ }\href {https://doi.org/10.1038/s41524-024-01381-9} {\bibfield  {journal} {\bibinfo  {journal} {npj Comput. Mater.}\ }\textbf {\bibinfo {volume} {10}},\ \bibinfo {pages} {1} (\bibinfo {year} {2024})}\BibitemShut {NoStop}%
\bibitem [{\citenamefont {Yang}\ \emph {et~al.}(2024{\natexlab{b}})\citenamefont {Yang}, \citenamefont {Batzner}, \citenamefont {Gao}, \citenamefont {Aykol}, \citenamefont {Gaunt}, \citenamefont {McMorrow}, \citenamefont {Rezende}, \citenamefont {Schuurmans}, \citenamefont {Mordatch},\ and\ \citenamefont {Cubuk}}]{yangGenerativeHierarchicalMaterials2024}%
  \BibitemOpen
  \bibfield  {author} {\bibinfo {author} {\bibfnamefont {S.}~\bibnamefont {Yang}}, \bibinfo {author} {\bibfnamefont {S.}~\bibnamefont {Batzner}}, \bibinfo {author} {\bibfnamefont {R.}~\bibnamefont {Gao}}, \bibinfo {author} {\bibfnamefont {M.}~\bibnamefont {Aykol}}, \bibinfo {author} {\bibfnamefont {A.~L.}\ \bibnamefont {Gaunt}}, \bibinfo {author} {\bibfnamefont {B.}~\bibnamefont {McMorrow}}, \bibinfo {author} {\bibfnamefont {D.~J.}\ \bibnamefont {Rezende}}, \bibinfo {author} {\bibfnamefont {D.}~\bibnamefont {Schuurmans}}, \bibinfo {author} {\bibfnamefont {I.}~\bibnamefont {Mordatch}},\ and\ \bibinfo {author} {\bibfnamefont {E.~D.}\ \bibnamefont {Cubuk}},\ }\href {https://doi.org/10.48550/arXiv.2409.06762} {\bibinfo {title} {Generative hierarchical materials search}} (\bibinfo {year} {2024}{\natexlab{b}}),\ \Eprint {https://arxiv.org/abs/2409.06762} {arXiv:2409.06762} \BibitemShut {NoStop}%
\bibitem [{\citenamefont {Zhu}\ \emph {et~al.}(2024)\citenamefont {Zhu}, \citenamefont {Nong}, \citenamefont {Yamazaki},\ and\ \citenamefont {Hippalgaonkar}}]{zhuWyCrystWyckoffInorganic2024}%
  \BibitemOpen
  \bibfield  {author} {\bibinfo {author} {\bibfnamefont {R.}~\bibnamefont {Zhu}}, \bibinfo {author} {\bibfnamefont {W.}~\bibnamefont {Nong}}, \bibinfo {author} {\bibfnamefont {S.}~\bibnamefont {Yamazaki}},\ and\ \bibinfo {author} {\bibfnamefont {K.}~\bibnamefont {Hippalgaonkar}},\ }\bibfield  {title} {\bibinfo {title} {{{WyCryst}}: {{Wyckoff}} inorganic crystal generator framework},\ }\href {https://doi.org/10.1016/j.matt.2024.05.042} {\bibfield  {journal} {\bibinfo  {journal} {Matter}\ }\textbf {\bibinfo {volume} {7}},\ \bibinfo {pages} {3469} (\bibinfo {year} {2024})}\BibitemShut {NoStop}%
\bibitem [{\citenamefont {Ding}\ \emph {et~al.}(2024)\citenamefont {Ding}, \citenamefont {Miret},\ and\ \citenamefont {Liu}}]{dingMatExpertDecomposingMaterials2024}%
  \BibitemOpen
  \bibfield  {author} {\bibinfo {author} {\bibfnamefont {Q.}~\bibnamefont {Ding}}, \bibinfo {author} {\bibfnamefont {S.}~\bibnamefont {Miret}},\ and\ \bibinfo {author} {\bibfnamefont {B.}~\bibnamefont {Liu}},\ }\href {https://doi.org/10.48550/arXiv.2410.21317} {\bibinfo {title} {{{MatExpert}}: {D}ecomposing materials discovery by mimicking human experts}} (\bibinfo {year} {2024}),\ \Eprint {https://arxiv.org/abs/2410.21317} {arXiv:2410.21317} \BibitemShut {NoStop}%
\bibitem [{\citenamefont {Sriram}\ \emph {et~al.}(2024)\citenamefont {Sriram}, \citenamefont {Miller}, \citenamefont {Chen},\ and\ \citenamefont {Wood}}]{sriramFlowLLMFlowMatching2024}%
  \BibitemOpen
  \bibfield  {author} {\bibinfo {author} {\bibfnamefont {A.}~\bibnamefont {Sriram}}, \bibinfo {author} {\bibfnamefont {B.~K.}\ \bibnamefont {Miller}}, \bibinfo {author} {\bibfnamefont {R.~T.~Q.}\ \bibnamefont {Chen}},\ and\ \bibinfo {author} {\bibfnamefont {B.~M.}\ \bibnamefont {Wood}},\ }\href {https://doi.org/10.48550/arXiv.2410.23405} {\bibinfo {title} {{{FlowLLM}}: {Flow} matching for material generation with large language models as base distributions}} (\bibinfo {year} {2024}),\ \Eprint {https://arxiv.org/abs/2410.23405} {arXiv:2410.23405} \BibitemShut {NoStop}%
\bibitem [{\citenamefont {Luo}\ \emph {et~al.}(2024)\citenamefont {Luo}, \citenamefont {Wang}, \citenamefont {Gao}, \citenamefont {Lv}, \citenamefont {Wang}, \citenamefont {Chen},\ and\ \citenamefont {Ma}}]{luoDeepLearningGenerative2024}%
  \BibitemOpen
  \bibfield  {author} {\bibinfo {author} {\bibfnamefont {X.}~\bibnamefont {Luo}}, \bibinfo {author} {\bibfnamefont {Z.}~\bibnamefont {Wang}}, \bibinfo {author} {\bibfnamefont {P.}~\bibnamefont {Gao}}, \bibinfo {author} {\bibfnamefont {J.}~\bibnamefont {Lv}}, \bibinfo {author} {\bibfnamefont {Y.}~\bibnamefont {Wang}}, \bibinfo {author} {\bibfnamefont {C.}~\bibnamefont {Chen}},\ and\ \bibinfo {author} {\bibfnamefont {Y.}~\bibnamefont {Ma}},\ }\bibfield  {title} {\bibinfo {title} {Deep learning generative model for crystal structure prediction},\ }\href {https://doi.org/10.1038/s41524-024-01443-y} {\bibfield  {journal} {\bibinfo  {journal} {npj Comput. Mater.}\ }\textbf {\bibinfo {volume} {10}},\ \bibinfo {pages} {1} (\bibinfo {year} {2024})}\BibitemShut {NoStop}%
\bibitem [{\citenamefont {Antunes}\ \emph {et~al.}(2024)\citenamefont {Antunes}, \citenamefont {Butler},\ and\ \citenamefont {{Grau-Crespo}}}]{antunesCrystalStructureGeneration2024}%
  \BibitemOpen
  \bibfield  {author} {\bibinfo {author} {\bibfnamefont {L.~M.}\ \bibnamefont {Antunes}}, \bibinfo {author} {\bibfnamefont {K.~T.}\ \bibnamefont {Butler}},\ and\ \bibinfo {author} {\bibfnamefont {R.}~\bibnamefont {{Grau-Crespo}}},\ }\bibfield  {title} {\bibinfo {title} {Crystal structure generation with autoregressive large language modeling},\ }\href {https://doi.org/10.1038/s41467-024-54639-7} {\bibfield  {journal} {\bibinfo  {journal} {Nat. Commun.}\ }\textbf {\bibinfo {volume} {15}},\ \bibinfo {pages} {10570} (\bibinfo {year} {2024})}\BibitemShut {NoStop}%
\bibitem [{\citenamefont {Mohanty}\ \emph {et~al.}(2024)\citenamefont {Mohanty}, \citenamefont {Mehta}, \citenamefont {Sayeed}, \citenamefont {Srikumar},\ and\ \citenamefont {Sparks}}]{mohantyCrysTextGenerativeAI2024}%
  \BibitemOpen
  \bibfield  {author} {\bibinfo {author} {\bibfnamefont {T.}~\bibnamefont {Mohanty}}, \bibinfo {author} {\bibfnamefont {M.}~\bibnamefont {Mehta}}, \bibinfo {author} {\bibfnamefont {H.~M.}\ \bibnamefont {Sayeed}}, \bibinfo {author} {\bibfnamefont {V.}~\bibnamefont {Srikumar}},\ and\ \bibinfo {author} {\bibfnamefont {T.~D.}\ \bibnamefont {Sparks}},\ }\href {https://doi.org/10.26434/chemrxiv-2024-gjhpq} {\bibinfo {title} {{{CrysText}}: {A} generative ai approach for text-conditioned crystal structure generation using {{LLM}}}} (\bibinfo {year} {2024})\BibitemShut {NoStop}%
\bibitem [{\citenamefont {Su}\ \emph {et~al.}(2024)\citenamefont {Su}, \citenamefont {Cao}, \citenamefont {Hu}, \citenamefont {Li},\ and\ \citenamefont {Zhang}}]{suCGWGANCrystalGenerative2024}%
  \BibitemOpen
  \bibfield  {author} {\bibinfo {author} {\bibfnamefont {T.}~\bibnamefont {Su}}, \bibinfo {author} {\bibfnamefont {B.}~\bibnamefont {Cao}}, \bibinfo {author} {\bibfnamefont {S.}~\bibnamefont {Hu}}, \bibinfo {author} {\bibfnamefont {M.}~\bibnamefont {Li}},\ and\ \bibinfo {author} {\bibfnamefont {T.-Y.}\ \bibnamefont {Zhang}},\ }\bibfield  {title} {\bibinfo {title} {{{CGWGAN}}: {C}rystal generative framework based on {{Wyckoff}} generative adversarial network},\ }\href {https://doi.org/10.20517/jmi.2024.24} {\bibfield  {journal} {\bibinfo  {journal} {J. Mater. Inf.}\ }\textbf {\bibinfo {volume} {4}},\ \bibinfo {pages} {N/A} (\bibinfo {year} {2024})}\BibitemShut {NoStop}%
\bibitem [{\citenamefont {Liu}\ \emph {et~al.}(2025)\citenamefont {Liu}, \citenamefont {Zhou}, \citenamefont {Zhang}, \citenamefont {Zhang}, \citenamefont {Lin},\ and\ \citenamefont {Pan}}]{liuEquivariantHypergraphDiffusion2025}%
  \BibitemOpen
  \bibfield  {author} {\bibinfo {author} {\bibfnamefont {Y.}~\bibnamefont {Liu}}, \bibinfo {author} {\bibfnamefont {C.}~\bibnamefont {Zhou}}, \bibinfo {author} {\bibfnamefont {S.}~\bibnamefont {Zhang}}, \bibinfo {author} {\bibfnamefont {P.}~\bibnamefont {Zhang}}, \bibinfo {author} {\bibfnamefont {X.}~\bibnamefont {Lin}},\ and\ \bibinfo {author} {\bibfnamefont {S.}~\bibnamefont {Pan}},\ }\href {https://doi.org/10.48550/arXiv.2501.18850} {\bibinfo {title} {Equivariant hypergraph diffusion for crystal structure prediction}} (\bibinfo {year} {2025}),\ \Eprint {https://arxiv.org/abs/2501.18850} {arXiv:2501.18850} \BibitemShut {NoStop}%
\bibitem [{\citenamefont {Wu}\ \emph {et~al.}(2025{\natexlab{a}})\citenamefont {Wu}, \citenamefont {Song}, \citenamefont {Gong}, \citenamefont {Cao}, \citenamefont {Ouyang}, \citenamefont {Zhang}, \citenamefont {Zhou}, \citenamefont {Ma},\ and\ \citenamefont {Liu}}]{wuPeriodicBayesianFlow2025}%
  \BibitemOpen
  \bibfield  {author} {\bibinfo {author} {\bibfnamefont {H.}~\bibnamefont {Wu}}, \bibinfo {author} {\bibfnamefont {Y.}~\bibnamefont {Song}}, \bibinfo {author} {\bibfnamefont {J.}~\bibnamefont {Gong}}, \bibinfo {author} {\bibfnamefont {Z.}~\bibnamefont {Cao}}, \bibinfo {author} {\bibfnamefont {Y.}~\bibnamefont {Ouyang}}, \bibinfo {author} {\bibfnamefont {J.}~\bibnamefont {Zhang}}, \bibinfo {author} {\bibfnamefont {H.}~\bibnamefont {Zhou}}, \bibinfo {author} {\bibfnamefont {W.-Y.}\ \bibnamefont {Ma}},\ and\ \bibinfo {author} {\bibfnamefont {J.}~\bibnamefont {Liu}},\ }\href {https://doi.org/10.48550/arXiv.2502.02016} {\bibinfo {title} {A periodic bayesian flow for material generation}} (\bibinfo {year} {2025}{\natexlab{a}}),\ \Eprint {https://arxiv.org/abs/2502.02016} {arXiv:2502.02016} \BibitemShut {NoStop}%
\bibitem [{\citenamefont {Chen}\ \emph {et~al.}(2025)\citenamefont {Chen}, \citenamefont {Yuan}, \citenamefont {Zheng}, \citenamefont {Guo}, \citenamefont {Liang}, \citenamefont {Wang},\ and\ \citenamefont {Wang}}]{chenTransformerEnhancedVariationalAutoencoder2025}%
  \BibitemOpen
  \bibfield  {author} {\bibinfo {author} {\bibfnamefont {Z.}~\bibnamefont {Chen}}, \bibinfo {author} {\bibfnamefont {Y.}~\bibnamefont {Yuan}}, \bibinfo {author} {\bibfnamefont {S.}~\bibnamefont {Zheng}}, \bibinfo {author} {\bibfnamefont {J.}~\bibnamefont {Guo}}, \bibinfo {author} {\bibfnamefont {S.}~\bibnamefont {Liang}}, \bibinfo {author} {\bibfnamefont {Y.}~\bibnamefont {Wang}},\ and\ \bibinfo {author} {\bibfnamefont {Z.}~\bibnamefont {Wang}},\ }\href {https://doi.org/10.48550/arXiv.2502.09423} {\bibinfo {title} {Transformer-enhanced variational autoencoder for crystal structure prediction}} (\bibinfo {year} {2025}),\ \Eprint {https://arxiv.org/abs/2502.09423} {arXiv:2502.09423} \BibitemShut {NoStop}%
\bibitem [{\citenamefont {Luo}\ \emph {et~al.}(2025)\citenamefont {Luo}, \citenamefont {Wang}, \citenamefont {Wang}, \citenamefont {Lv}, \citenamefont {Wang}, \citenamefont {Wang},\ and\ \citenamefont {Ma}}]{luoCrystalFlowFlowBasedGenerative2025}%
  \BibitemOpen
  \bibfield  {author} {\bibinfo {author} {\bibfnamefont {X.}~\bibnamefont {Luo}}, \bibinfo {author} {\bibfnamefont {Z.}~\bibnamefont {Wang}}, \bibinfo {author} {\bibfnamefont {Q.}~\bibnamefont {Wang}}, \bibinfo {author} {\bibfnamefont {J.}~\bibnamefont {Lv}}, \bibinfo {author} {\bibfnamefont {L.}~\bibnamefont {Wang}}, \bibinfo {author} {\bibfnamefont {Y.}~\bibnamefont {Wang}},\ and\ \bibinfo {author} {\bibfnamefont {Y.}~\bibnamefont {Ma}},\ }\href {https://doi.org/10.48550/arXiv.2412.11693} {\bibinfo {title} {{{CrystalFlow}}: {{A}} flow-based generative model for crystalline materials}} (\bibinfo {year} {2025}),\ \Eprint {https://arxiv.org/abs/2412.11693} {arXiv:2412.11693} \BibitemShut {NoStop}%
\bibitem [{\citenamefont {Gan}\ \emph {et~al.}(2025)\citenamefont {Gan}, \citenamefont {Zhong}, \citenamefont {Du}, \citenamefont {Zhu}, \citenamefont {Duan}, \citenamefont {Wang}, \citenamefont {Gomes}, \citenamefont {Persson}, \citenamefont {{Schwalbe-Koda}},\ and\ \citenamefont {Wang}}]{ganLargeLanguageModels2025}%
  \BibitemOpen
  \bibfield  {author} {\bibinfo {author} {\bibfnamefont {J.}~\bibnamefont {Gan}}, \bibinfo {author} {\bibfnamefont {P.}~\bibnamefont {Zhong}}, \bibinfo {author} {\bibfnamefont {Y.}~\bibnamefont {Du}}, \bibinfo {author} {\bibfnamefont {Y.}~\bibnamefont {Zhu}}, \bibinfo {author} {\bibfnamefont {C.}~\bibnamefont {Duan}}, \bibinfo {author} {\bibfnamefont {H.}~\bibnamefont {Wang}}, \bibinfo {author} {\bibfnamefont {C.~P.}\ \bibnamefont {Gomes}}, \bibinfo {author} {\bibfnamefont {K.~A.}\ \bibnamefont {Persson}}, \bibinfo {author} {\bibfnamefont {D.}~\bibnamefont {{Schwalbe-Koda}}},\ and\ \bibinfo {author} {\bibfnamefont {W.}~\bibnamefont {Wang}},\ }\href {https://doi.org/10.48550/arXiv.2502.20933} {\bibinfo {title} {Large language models are innate crystal structure generators}} (\bibinfo {year} {2025}),\ \Eprint {https://arxiv.org/abs/2502.20933} {arXiv:2502.20933} \BibitemShut {NoStop}%
\bibitem [{\citenamefont {Yan}\ \emph {et~al.}(2025)\citenamefont {Yan}, \citenamefont {Li}, \citenamefont {Ling}, \citenamefont {Ashen}, \citenamefont {Edwards}, \citenamefont {Arr{\'o}yave}, \citenamefont {Zitnik}, \citenamefont {Ji}, \citenamefont {Qian}, \citenamefont {Qian},\ and\ \citenamefont {Ji}}]{yanInvariantTokenizationCrystalline2025}%
  \BibitemOpen
  \bibfield  {author} {\bibinfo {author} {\bibfnamefont {K.}~\bibnamefont {Yan}}, \bibinfo {author} {\bibfnamefont {X.}~\bibnamefont {Li}}, \bibinfo {author} {\bibfnamefont {H.}~\bibnamefont {Ling}}, \bibinfo {author} {\bibfnamefont {K.}~\bibnamefont {Ashen}}, \bibinfo {author} {\bibfnamefont {C.}~\bibnamefont {Edwards}}, \bibinfo {author} {\bibfnamefont {R.}~\bibnamefont {Arr{\'o}yave}}, \bibinfo {author} {\bibfnamefont {M.}~\bibnamefont {Zitnik}}, \bibinfo {author} {\bibfnamefont {H.}~\bibnamefont {Ji}}, \bibinfo {author} {\bibfnamefont {X.}~\bibnamefont {Qian}}, \bibinfo {author} {\bibfnamefont {X.}~\bibnamefont {Qian}},\ and\ \bibinfo {author} {\bibfnamefont {S.}~\bibnamefont {Ji}},\ }\href {https://doi.org/10.48550/arXiv.2503.00152} {\bibinfo {title} {Invariant tokenization of crystalline materials for language model enabled generation}} (\bibinfo {year} {2025}),\ \Eprint {https://arxiv.org/abs/2503.00152} {arXiv:2503.00152} \BibitemShut {NoStop}%
\bibitem [{\citenamefont {Das}\ \emph {et~al.}(2025)\citenamefont {Das}, \citenamefont {Khastagir}, \citenamefont {Goyal}, \citenamefont {Lee}, \citenamefont {Bhattacharjee},\ and\ \citenamefont {Ganguly}}]{dasPeriodicMaterialsGeneration2025}%
  \BibitemOpen
  \bibfield  {author} {\bibinfo {author} {\bibfnamefont {K.}~\bibnamefont {Das}}, \bibinfo {author} {\bibfnamefont {S.}~\bibnamefont {Khastagir}}, \bibinfo {author} {\bibfnamefont {P.}~\bibnamefont {Goyal}}, \bibinfo {author} {\bibfnamefont {S.-C.}\ \bibnamefont {Lee}}, \bibinfo {author} {\bibfnamefont {S.}~\bibnamefont {Bhattacharjee}},\ and\ \bibinfo {author} {\bibfnamefont {N.}~\bibnamefont {Ganguly}},\ }\href {https://doi.org/10.48550/arXiv.2503.00522} {\bibinfo {title} {Periodic materials generation using text-guided joint diffusion model}} (\bibinfo {year} {2025}),\ \Eprint {https://arxiv.org/abs/2503.00522} {arXiv:2503.00522} \BibitemShut {NoStop}%
\bibitem [{\citenamefont {Xia}\ \emph {et~al.}(2025)\citenamefont {Xia}, \citenamefont {Jin}, \citenamefont {Xie}, \citenamefont {He}, \citenamefont {Cao}, \citenamefont {Luo}, \citenamefont {Liu}, \citenamefont {Wang}, \citenamefont {Liu}, \citenamefont {Chen}, \citenamefont {Guo}, \citenamefont {Bai}, \citenamefont {Deng}, \citenamefont {Min}, \citenamefont {Lu}, \citenamefont {Hao}, \citenamefont {Yang}, \citenamefont {Li}, \citenamefont {Liu}, \citenamefont {Zhang}, \citenamefont {Zhu}, \citenamefont {Bi}, \citenamefont {Wu}, \citenamefont {Zhang}, \citenamefont {Gao}, \citenamefont {Pei}, \citenamefont {Wang}, \citenamefont {Liu}, \citenamefont {Li}, \citenamefont {Zhu}, \citenamefont {Lu}, \citenamefont {Ma}, \citenamefont {Wang}, \citenamefont {Xie}, \citenamefont {Maziarz}, \citenamefont {Segler}, \citenamefont {Yang}, \citenamefont {Chen}, \citenamefont {Shi}, \citenamefont {Zheng}, \citenamefont {Wu}, \citenamefont {Hu}, \citenamefont {Dai}, \citenamefont {Liu}, \citenamefont {Liu},\ and\ \citenamefont {Qin}}]{xiaNatureLanguageModel2025}%
  \BibitemOpen
  \bibfield  {author} {\bibinfo {author} {\bibfnamefont {Y.}~\bibnamefont {Xia}}, \bibinfo {author} {\bibfnamefont {P.}~\bibnamefont {Jin}}, \bibinfo {author} {\bibfnamefont {S.}~\bibnamefont {Xie}}, \bibinfo {author} {\bibfnamefont {L.}~\bibnamefont {He}}, \bibinfo {author} {\bibfnamefont {C.}~\bibnamefont {Cao}}, \bibinfo {author} {\bibfnamefont {R.}~\bibnamefont {Luo}}, \bibinfo {author} {\bibfnamefont {G.}~\bibnamefont {Liu}}, \bibinfo {author} {\bibfnamefont {Y.}~\bibnamefont {Wang}}, \bibinfo {author} {\bibfnamefont {Z.}~\bibnamefont {Liu}}, \bibinfo {author} {\bibfnamefont {Y.-J.}\ \bibnamefont {Chen}}, \bibinfo {author} {\bibfnamefont {Z.}~\bibnamefont {Guo}}, \bibinfo {author} {\bibfnamefont {Y.}~\bibnamefont {Bai}}, \bibinfo {author} {\bibfnamefont {P.}~\bibnamefont {Deng}}, \bibinfo {author} {\bibfnamefont {Y.}~\bibnamefont {Min}}, \bibinfo {author} {\bibfnamefont {Z.}~\bibnamefont {Lu}}, \bibinfo {author} {\bibfnamefont {H.}~\bibnamefont {Hao}}, \bibinfo {author} {\bibfnamefont {H.}~\bibnamefont {Yang}}, \bibinfo {author} {\bibfnamefont {J.}~\bibnamefont {Li}}, \bibinfo {author} {\bibfnamefont {C.}~\bibnamefont {Liu}}, \bibinfo {author} {\bibfnamefont {J.}~\bibnamefont {Zhang}}, \bibinfo {author} {\bibfnamefont {J.}~\bibnamefont {Zhu}}, \bibinfo {author} {\bibfnamefont {R.}~\bibnamefont {Bi}}, \bibinfo {author} {\bibfnamefont {K.}~\bibnamefont {Wu}}, \bibinfo {author} {\bibfnamefont {W.}~\bibnamefont {Zhang}}, \bibinfo {author} {\bibfnamefont {K.}~\bibnamefont {Gao}}, \bibinfo {author} {\bibfnamefont {Q.}~\bibnamefont {Pei}}, \bibinfo {author} {\bibfnamefont {Q.}~\bibnamefont {Wang}}, \bibinfo {author} {\bibfnamefont {X.}~\bibnamefont {Liu}}, \bibinfo {author} {\bibfnamefont {Y.}~\bibnamefont {Li}}, \bibinfo {author} {\bibfnamefont {H.}~\bibnamefont {Zhu}}, \bibinfo {author} {\bibfnamefont {Y.}~\bibnamefont {Lu}}, \bibinfo {author} {\bibfnamefont {M.}~\bibnamefont {Ma}}, \bibinfo {author} {\bibfnamefont {Z.}~\bibnamefont {Wang}}, \bibinfo {author} {\bibfnamefont {T.}~\bibnamefont {Xie}}, \bibinfo {author} {\bibfnamefont {K.}~\bibnamefont {Maziarz}}, \bibinfo {author} {\bibfnamefont {M.}~\bibnamefont {Segler}}, \bibinfo {author} {\bibfnamefont {Z.}~\bibnamefont {Yang}}, \bibinfo {author} {\bibfnamefont {Z.}~\bibnamefont {Chen}}, \bibinfo {author} {\bibfnamefont {Y.}~\bibnamefont {Shi}}, \bibinfo {author} {\bibfnamefont {S.}~\bibnamefont {Zheng}}, \bibinfo {author} {\bibfnamefont {L.}~\bibnamefont {Wu}}, \bibinfo {author} {\bibfnamefont {C.}~\bibnamefont {Hu}}, \bibinfo {author} {\bibfnamefont {P.}~\bibnamefont {Dai}}, \bibinfo {author} {\bibfnamefont {T.-Y.}\ \bibnamefont {Liu}}, \bibinfo {author} {\bibfnamefont {H.}~\bibnamefont {Liu}},\ and\ \bibinfo {author} {\bibfnamefont {T.}~\bibnamefont {Qin}},\ }\href {https://doi.org/10.48550/arXiv.2502.07527} {\bibinfo {title} {Nature language model: {{Deciphering}} the language of nature for scientific discovery}} (\bibinfo {year} {2025}),\ \Eprint {https://arxiv.org/abs/2502.07527} {arXiv:2502.07527} \BibitemShut {NoStop}%
\bibitem [{\citenamefont {Tangsongcharoen}\ \emph {et~al.}(2025)\citenamefont {Tangsongcharoen}, \citenamefont {Pakornchote}, \citenamefont {Atthapak}, \citenamefont {{Choomphon-anomakhun}}, \citenamefont {Ektarawong}, \citenamefont {Alling}, \citenamefont {Sutton}, \citenamefont {Bovornratanaraks},\ and\ \citenamefont {Chotibut}}]{tangsongcharoenCrystalGRWGenerativeModeling2025}%
  \BibitemOpen
  \bibfield  {author} {\bibinfo {author} {\bibfnamefont {K.}~\bibnamefont {Tangsongcharoen}}, \bibinfo {author} {\bibfnamefont {T.}~\bibnamefont {Pakornchote}}, \bibinfo {author} {\bibfnamefont {C.}~\bibnamefont {Atthapak}}, \bibinfo {author} {\bibfnamefont {N.}~\bibnamefont {{Choomphon-anomakhun}}}, \bibinfo {author} {\bibfnamefont {A.}~\bibnamefont {Ektarawong}}, \bibinfo {author} {\bibfnamefont {B.}~\bibnamefont {Alling}}, \bibinfo {author} {\bibfnamefont {C.}~\bibnamefont {Sutton}}, \bibinfo {author} {\bibfnamefont {T.}~\bibnamefont {Bovornratanaraks}},\ and\ \bibinfo {author} {\bibfnamefont {T.}~\bibnamefont {Chotibut}},\ }\href {https://doi.org/10.48550/arXiv.2501.08998} {\bibinfo {title} {{{CrystalGRW}}: {G}enerative modeling of crystal structures with targeted properties via geodesic random walks}} (\bibinfo {year} {2025}),\ \Eprint {https://arxiv.org/abs/2501.08998} {arXiv:2501.08998} \BibitemShut {NoStop}%
\bibitem [{\citenamefont {Zhang}\ \emph {et~al.}(2025)\citenamefont {Zhang}, \citenamefont {Li}, \citenamefont {Luo}, \citenamefont {Hu}, \citenamefont {Zhao}, \citenamefont {Li}, \citenamefont {Liu}, \citenamefont {Wang}, \citenamefont {Bi}, \citenamefont {Gao}, \citenamefont {Guo}, \citenamefont {Xie}, \citenamefont {Liu}, \citenamefont {Zhang}, \citenamefont {Xie}, \citenamefont {Pinsler}, \citenamefont {Zeni}, \citenamefont {Lu}, \citenamefont {Xia}, \citenamefont {Segler}, \citenamefont {Riechert}, \citenamefont {Yuan}, \citenamefont {Chen}, \citenamefont {Liu},\ and\ \citenamefont {Qin}}]{zhangUniGenXUnifiedGeneration2025}%
  \BibitemOpen
  \bibfield  {author} {\bibinfo {author} {\bibfnamefont {G.}~\bibnamefont {Zhang}}, \bibinfo {author} {\bibfnamefont {Y.}~\bibnamefont {Li}}, \bibinfo {author} {\bibfnamefont {R.}~\bibnamefont {Luo}}, \bibinfo {author} {\bibfnamefont {P.}~\bibnamefont {Hu}}, \bibinfo {author} {\bibfnamefont {Z.}~\bibnamefont {Zhao}}, \bibinfo {author} {\bibfnamefont {L.}~\bibnamefont {Li}}, \bibinfo {author} {\bibfnamefont {G.}~\bibnamefont {Liu}}, \bibinfo {author} {\bibfnamefont {Z.}~\bibnamefont {Wang}}, \bibinfo {author} {\bibfnamefont {R.}~\bibnamefont {Bi}}, \bibinfo {author} {\bibfnamefont {K.}~\bibnamefont {Gao}}, \bibinfo {author} {\bibfnamefont {L.}~\bibnamefont {Guo}}, \bibinfo {author} {\bibfnamefont {Y.}~\bibnamefont {Xie}}, \bibinfo {author} {\bibfnamefont {C.}~\bibnamefont {Liu}}, \bibinfo {author} {\bibfnamefont {J.}~\bibnamefont {Zhang}}, \bibinfo {author} {\bibfnamefont {T.}~\bibnamefont {Xie}}, \bibinfo {author} {\bibfnamefont {R.}~\bibnamefont {Pinsler}}, \bibinfo {author} {\bibfnamefont {C.}~\bibnamefont {Zeni}}, \bibinfo {author} {\bibfnamefont {Z.}~\bibnamefont {Lu}}, \bibinfo {author} {\bibfnamefont {Y.}~\bibnamefont {Xia}}, \bibinfo {author} {\bibfnamefont {M.}~\bibnamefont {Segler}}, \bibinfo {author} {\bibfnamefont {M.}~\bibnamefont {Riechert}}, \bibinfo {author} {\bibfnamefont {L.}~\bibnamefont {Yuan}}, \bibinfo {author} {\bibfnamefont {L.}~\bibnamefont {Chen}}, \bibinfo {author} {\bibfnamefont {H.}~\bibnamefont {Liu}},\ and\ \bibinfo {author} {\bibfnamefont {T.}~\bibnamefont {Qin}},\ }\href {https://doi.org/10.48550/arXiv.2503.06687} {\bibinfo {title} {{{UniGenX}}: {{Unified}} generation of sequence and structure with autoregressive diffusion}} (\bibinfo {year} {2025}),\ \Eprint {https://arxiv.org/abs/2503.06687} {arXiv:2503.06687} \BibitemShut {NoStop}%
\bibitem [{\citenamefont {Wu}\ \emph {et~al.}(2025{\natexlab{b}})\citenamefont {Wu}, \citenamefont {Huang}, \citenamefont {Jiao}, \citenamefont {Huang}, \citenamefont {Liu}, \citenamefont {Zhou}, \citenamefont {Sun}, \citenamefont {Liu}, \citenamefont {Sun}, \citenamefont {Ren},\ and\ \citenamefont {Wen}}]{wuSiameseFoundationModels2025}%
  \BibitemOpen
  \bibfield  {author} {\bibinfo {author} {\bibfnamefont {L.}~\bibnamefont {Wu}}, \bibinfo {author} {\bibfnamefont {W.}~\bibnamefont {Huang}}, \bibinfo {author} {\bibfnamefont {R.}~\bibnamefont {Jiao}}, \bibinfo {author} {\bibfnamefont {J.}~\bibnamefont {Huang}}, \bibinfo {author} {\bibfnamefont {L.}~\bibnamefont {Liu}}, \bibinfo {author} {\bibfnamefont {Y.}~\bibnamefont {Zhou}}, \bibinfo {author} {\bibfnamefont {H.}~\bibnamefont {Sun}}, \bibinfo {author} {\bibfnamefont {Y.}~\bibnamefont {Liu}}, \bibinfo {author} {\bibfnamefont {F.}~\bibnamefont {Sun}}, \bibinfo {author} {\bibfnamefont {Y.}~\bibnamefont {Ren}},\ and\ \bibinfo {author} {\bibfnamefont {J.}~\bibnamefont {Wen}},\ }\href {https://doi.org/10.48550/arXiv.2503.10471} {\bibinfo {title} {Siamese foundation models for crystal structure prediction}} (\bibinfo {year} {2025}{\natexlab{b}}),\ \Eprint {https://arxiv.org/abs/2503.10471} {arXiv:2503.10471} \BibitemShut {NoStop}%
\bibitem [{\citenamefont {Lu}\ \emph {et~al.}(2025)\citenamefont {Lu}, \citenamefont {Lin}, \citenamefont {Yao}, \citenamefont {Gao}, \citenamefont {Ji}, \citenamefont {E}, \citenamefont {Zhang},\ and\ \citenamefont {Ke}}]{luUni3DARUnified3D2025}%
  \BibitemOpen
  \bibfield  {author} {\bibinfo {author} {\bibfnamefont {S.}~\bibnamefont {Lu}}, \bibinfo {author} {\bibfnamefont {H.}~\bibnamefont {Lin}}, \bibinfo {author} {\bibfnamefont {L.}~\bibnamefont {Yao}}, \bibinfo {author} {\bibfnamefont {Z.}~\bibnamefont {Gao}}, \bibinfo {author} {\bibfnamefont {X.}~\bibnamefont {Ji}}, \bibinfo {author} {\bibfnamefont {W.}~\bibnamefont {E}}, \bibinfo {author} {\bibfnamefont {L.}~\bibnamefont {Zhang}},\ and\ \bibinfo {author} {\bibfnamefont {G.}~\bibnamefont {Ke}},\ }\href {https://doi.org/10.48550/arXiv.2503.16278} {\bibinfo {title} {Uni-{{3DAR}}: {U}nified 3{D} generation and understanding via autoregression on compressed spatial tokens}} (\bibinfo {year} {2025}),\ \Eprint {https://arxiv.org/abs/2503.16278} {arXiv:2503.16278} \BibitemShut {NoStop}%
\bibitem [{\citenamefont {Park}\ \emph {et~al.}(2025)\citenamefont {Park}, \citenamefont {Onwuli},\ and\ \citenamefont {Walsh}}]{parkExplorationCrystalChemical2025}%
  \BibitemOpen
  \bibfield  {author} {\bibinfo {author} {\bibfnamefont {H.}~\bibnamefont {Park}}, \bibinfo {author} {\bibfnamefont {A.}~\bibnamefont {Onwuli}},\ and\ \bibinfo {author} {\bibfnamefont {A.}~\bibnamefont {Walsh}},\ }\bibfield  {title} {\bibinfo {title} {Exploration of crystal chemical space using text-guided generative artificial intelligence},\ }\href {https://doi.org/10.1038/s41467-025-59636-y} {\bibfield  {journal} {\bibinfo  {journal} {Nat. Commun.}\ }\textbf {\bibinfo {volume} {16}},\ \bibinfo {pages} {4379} (\bibinfo {year} {2025})}\BibitemShut {NoStop}%
\bibitem [{\citenamefont {Levy}\ \emph {et~al.}(2025)\citenamefont {Levy}, \citenamefont {Panigrahi}, \citenamefont {Kaba}, \citenamefont {Zhu}, \citenamefont {Lee}, \citenamefont {Galkin}, \citenamefont {Miret},\ and\ \citenamefont {Ravanbakhsh}}]{levySymmCDSymmetryPreservingCrystal2025}%
  \BibitemOpen
  \bibfield  {author} {\bibinfo {author} {\bibfnamefont {D.}~\bibnamefont {Levy}}, \bibinfo {author} {\bibfnamefont {S.~S.}\ \bibnamefont {Panigrahi}}, \bibinfo {author} {\bibfnamefont {S.-O.}\ \bibnamefont {Kaba}}, \bibinfo {author} {\bibfnamefont {Q.}~\bibnamefont {Zhu}}, \bibinfo {author} {\bibfnamefont {K.~L.~K.}\ \bibnamefont {Lee}}, \bibinfo {author} {\bibfnamefont {M.}~\bibnamefont {Galkin}}, \bibinfo {author} {\bibfnamefont {S.}~\bibnamefont {Miret}},\ and\ \bibinfo {author} {\bibfnamefont {S.}~\bibnamefont {Ravanbakhsh}},\ }\href {https://doi.org/10.48550/arXiv.2502.03638} {\bibinfo {title} {{{SymmCD}}: {S}ymmetry-preserving crystal generation with diffusion models}} (\bibinfo {year} {2025}),\ \Eprint {https://arxiv.org/abs/2502.03638} {arXiv:2502.03638} \BibitemShut {NoStop}%
\bibitem [{\citenamefont {Kelvinius}\ \emph {et~al.}(2025)\citenamefont {Kelvinius}, \citenamefont {Andersson}, \citenamefont {Parackal}, \citenamefont {Qian}, \citenamefont {Armiento},\ and\ \citenamefont {Lindsten}}]{kelviniusWyckoffDiffGenerativeDiffusion2025}%
  \BibitemOpen
  \bibfield  {author} {\bibinfo {author} {\bibfnamefont {F.~E.}\ \bibnamefont {Kelvinius}}, \bibinfo {author} {\bibfnamefont {O.~B.}\ \bibnamefont {Andersson}}, \bibinfo {author} {\bibfnamefont {A.~S.}\ \bibnamefont {Parackal}}, \bibinfo {author} {\bibfnamefont {D.}~\bibnamefont {Qian}}, \bibinfo {author} {\bibfnamefont {R.}~\bibnamefont {Armiento}},\ and\ \bibinfo {author} {\bibfnamefont {F.}~\bibnamefont {Lindsten}},\ }\href {https://doi.org/10.48550/arXiv.2502.06485} {\bibinfo {title} {{{WyckoffDiff}} -- {A} generative diffusion model for crystal symmetry}} (\bibinfo {year} {2025}),\ \Eprint {https://arxiv.org/abs/2502.06485} {arXiv:2502.06485} \BibitemShut {NoStop}%
\bibitem [{\citenamefont {Cornet}\ \emph {et~al.}(2025)\citenamefont {Cornet}, \citenamefont {Bergamin}, \citenamefont {Bhowmik}, \citenamefont {Lastra}, \citenamefont {Frellsen},\ and\ \citenamefont {Schmidt}}]{cornetKineticLangevinDiffusion2025}%
  \BibitemOpen
  \bibfield  {author} {\bibinfo {author} {\bibfnamefont {F.}~\bibnamefont {Cornet}}, \bibinfo {author} {\bibfnamefont {F.}~\bibnamefont {Bergamin}}, \bibinfo {author} {\bibfnamefont {A.}~\bibnamefont {Bhowmik}}, \bibinfo {author} {\bibfnamefont {J.~M.~G.}\ \bibnamefont {Lastra}}, \bibinfo {author} {\bibfnamefont {J.}~\bibnamefont {Frellsen}},\ and\ \bibinfo {author} {\bibfnamefont {M.~N.}\ \bibnamefont {Schmidt}},\ }\href {https://doi.org/10.48550/arXiv.2507.03602} {\bibinfo {title} {Kinetic langevin diffusion for crystalline materials generation}} (\bibinfo {year} {2025}),\ \Eprint {https://arxiv.org/abs/2507.03602} {arXiv:2507.03602} \BibitemShut {NoStop}%
\bibitem [{\citenamefont {Fredericks}\ \emph {et~al.}(2021)\citenamefont {Fredericks}, \citenamefont {Parrish}, \citenamefont {Sayre},\ and\ \citenamefont {Zhu}}]{fredericksPyXtalPythonLibrary2021}%
  \BibitemOpen
  \bibfield  {author} {\bibinfo {author} {\bibfnamefont {S.}~\bibnamefont {Fredericks}}, \bibinfo {author} {\bibfnamefont {K.}~\bibnamefont {Parrish}}, \bibinfo {author} {\bibfnamefont {D.}~\bibnamefont {Sayre}},\ and\ \bibinfo {author} {\bibfnamefont {Q.}~\bibnamefont {Zhu}},\ }\bibfield  {title} {\bibinfo {title} {{{PyXtal}}: {{A Python}} library for crystal structure generation and symmetry analysis},\ }\href {https://doi.org/10.1016/j.cpc.2020.107810} {\bibfield  {journal} {\bibinfo  {journal} {Comput. Phys. Commun.}\ }\textbf {\bibinfo {volume} {261}},\ \bibinfo {pages} {107810} (\bibinfo {year} {2021})}\BibitemShut {NoStop}%
\bibitem [{\citenamefont {Chen}\ and\ \citenamefont {Ong}(2022)}]{chenUniversalGraphDeep2022}%
  \BibitemOpen
  \bibfield  {author} {\bibinfo {author} {\bibfnamefont {C.}~\bibnamefont {Chen}}\ and\ \bibinfo {author} {\bibfnamefont {S.~P.}\ \bibnamefont {Ong}},\ }\bibfield  {title} {\bibinfo {title} {A universal graph deep learning interatomic potential for the periodic table},\ }\href {https://doi.org/10.1038/s43588-022-00349-3} {\bibfield  {journal} {\bibinfo  {journal} {Nat. Comput. Sci.}\ }\textbf {\bibinfo {volume} {2}},\ \bibinfo {pages} {718} (\bibinfo {year} {2022})}\BibitemShut {NoStop}%
\bibitem [{\citenamefont {Touvron}\ \emph {et~al.}(2023{\natexlab{a}})\citenamefont {Touvron}, \citenamefont {Martin}, \citenamefont {Stone}, \citenamefont {Albert}, \citenamefont {Almahairi}, \citenamefont {Babaei}, \citenamefont {Bashlykov}, \citenamefont {Batra}, \citenamefont {Bhargava}, \citenamefont {Bhosale}, \citenamefont {Bikel}, \citenamefont {Blecher}, \citenamefont {Ferrer}, \citenamefont {Chen}, \citenamefont {Cucurull}, \citenamefont {Esiobu}, \citenamefont {Fernandes}, \citenamefont {Fu}, \citenamefont {Fu}, \citenamefont {Fuller}, \citenamefont {Gao}, \citenamefont {Goswami}, \citenamefont {Goyal}, \citenamefont {Hartshorn}, \citenamefont {Hosseini}, \citenamefont {Hou}, \citenamefont {Inan}, \citenamefont {Kardas}, \citenamefont {Kerkez}, \citenamefont {Khabsa}, \citenamefont {Kloumann}, \citenamefont {Korenev}, \citenamefont {Koura}, \citenamefont {Lachaux}, \citenamefont {Lavril}, \citenamefont {Lee}, \citenamefont {Liskovich}, \citenamefont {Lu}, \citenamefont {Mao}, \citenamefont {Martinet}, \citenamefont {Mihaylov}, \citenamefont {Mishra}, \citenamefont {Molybog}, \citenamefont {Nie}, \citenamefont {Poulton}, \citenamefont {Reizenstein}, \citenamefont {Rungta}, \citenamefont {Saladi}, \citenamefont {Schelten}, \citenamefont {Silva}, \citenamefont {Smith}, \citenamefont {Subramanian}, \citenamefont {Tan}, \citenamefont {Tang}, \citenamefont {Taylor}, \citenamefont {Williams}, \citenamefont {Kuan}, \citenamefont {Xu}, \citenamefont {Yan}, \citenamefont {Zarov}, \citenamefont {Zhang}, \citenamefont {Fan}, \citenamefont {Kambadur}, \citenamefont {Narang}, \citenamefont {Rodriguez}, \citenamefont {Stojnic}, \citenamefont {Edunov},\ and\ \citenamefont {Scialom}}]{touvronLlama2Open2023}%
  \BibitemOpen
  \bibfield  {author} {\bibinfo {author} {\bibfnamefont {H.}~\bibnamefont {Touvron}}, \bibinfo {author} {\bibfnamefont {L.}~\bibnamefont {Martin}}, \bibinfo {author} {\bibfnamefont {K.}~\bibnamefont {Stone}}, \bibinfo {author} {\bibfnamefont {P.}~\bibnamefont {Albert}}, \bibinfo {author} {\bibfnamefont {A.}~\bibnamefont {Almahairi}}, \bibinfo {author} {\bibfnamefont {Y.}~\bibnamefont {Babaei}}, \bibinfo {author} {\bibfnamefont {N.}~\bibnamefont {Bashlykov}}, \bibinfo {author} {\bibfnamefont {S.}~\bibnamefont {Batra}}, \bibinfo {author} {\bibfnamefont {P.}~\bibnamefont {Bhargava}}, \bibinfo {author} {\bibfnamefont {S.}~\bibnamefont {Bhosale}}, \bibinfo {author} {\bibfnamefont {D.}~\bibnamefont {Bikel}}, \bibinfo {author} {\bibfnamefont {L.}~\bibnamefont {Blecher}}, \bibinfo {author} {\bibfnamefont {C.~C.}\ \bibnamefont {Ferrer}}, \bibinfo {author} {\bibfnamefont {M.}~\bibnamefont {Chen}}, \bibinfo {author} {\bibfnamefont {G.}~\bibnamefont {Cucurull}}, \bibinfo {author} {\bibfnamefont {D.}~\bibnamefont {Esiobu}}, \bibinfo {author} {\bibfnamefont {J.}~\bibnamefont {Fernandes}}, \bibinfo {author} {\bibfnamefont {J.}~\bibnamefont {Fu}}, \bibinfo {author} {\bibfnamefont {W.}~\bibnamefont {Fu}}, \bibinfo {author} {\bibfnamefont {B.}~\bibnamefont {Fuller}}, \bibinfo {author} {\bibfnamefont {C.}~\bibnamefont {Gao}}, \bibinfo {author} {\bibfnamefont {V.}~\bibnamefont {Goswami}}, \bibinfo {author} {\bibfnamefont {N.}~\bibnamefont {Goyal}}, \bibinfo {author} {\bibfnamefont {A.}~\bibnamefont {Hartshorn}}, \bibinfo {author} {\bibfnamefont {S.}~\bibnamefont {Hosseini}}, \bibinfo {author} {\bibfnamefont {R.}~\bibnamefont {Hou}}, \bibinfo {author} {\bibfnamefont {H.}~\bibnamefont {Inan}}, \bibinfo {author} {\bibfnamefont {M.}~\bibnamefont {Kardas}}, \bibinfo {author} {\bibfnamefont {V.}~\bibnamefont {Kerkez}}, \bibinfo {author} {\bibfnamefont {M.}~\bibnamefont {Khabsa}}, \bibinfo {author} {\bibfnamefont {I.}~\bibnamefont {Kloumann}}, \bibinfo {author} {\bibfnamefont {A.}~\bibnamefont {Korenev}}, \bibinfo {author} {\bibfnamefont {P.~S.}\ \bibnamefont {Koura}}, \bibinfo {author} {\bibfnamefont {M.-A.}\ \bibnamefont {Lachaux}}, \bibinfo {author} {\bibfnamefont {T.}~\bibnamefont {Lavril}}, \bibinfo {author} {\bibfnamefont {J.}~\bibnamefont {Lee}}, \bibinfo {author} {\bibfnamefont {D.}~\bibnamefont {Liskovich}}, \bibinfo {author} {\bibfnamefont {Y.}~\bibnamefont {Lu}}, \bibinfo {author} {\bibfnamefont {Y.}~\bibnamefont {Mao}}, \bibinfo {author} {\bibfnamefont {X.}~\bibnamefont {Martinet}}, \bibinfo {author} {\bibfnamefont {T.}~\bibnamefont {Mihaylov}}, \bibinfo {author} {\bibfnamefont {P.}~\bibnamefont {Mishra}}, \bibinfo {author} {\bibfnamefont {I.}~\bibnamefont {Molybog}}, \bibinfo {author} {\bibfnamefont {Y.}~\bibnamefont {Nie}}, \bibinfo {author} {\bibfnamefont {A.}~\bibnamefont {Poulton}}, \bibinfo {author} {\bibfnamefont {J.}~\bibnamefont {Reizenstein}}, \bibinfo {author} {\bibfnamefont {R.}~\bibnamefont {Rungta}}, \bibinfo {author} {\bibfnamefont {K.}~\bibnamefont {Saladi}}, \bibinfo {author} {\bibfnamefont {A.}~\bibnamefont {Schelten}}, \bibinfo {author} {\bibfnamefont {R.}~\bibnamefont {Silva}}, \bibinfo {author} {\bibfnamefont {E.~M.}\ \bibnamefont {Smith}}, \bibinfo {author} {\bibfnamefont {R.}~\bibnamefont {Subramanian}}, \bibinfo {author} {\bibfnamefont {X.~E.}\ \bibnamefont {Tan}}, \bibinfo {author} {\bibfnamefont {B.}~\bibnamefont {Tang}}, \bibinfo {author} {\bibfnamefont {R.}~\bibnamefont {Taylor}}, \bibinfo {author} {\bibfnamefont {A.}~\bibnamefont {Williams}}, \bibinfo {author} {\bibfnamefont {J.~X.}\ \bibnamefont {Kuan}}, \bibinfo {author} {\bibfnamefont {P.}~\bibnamefont {Xu}}, \bibinfo {author} {\bibfnamefont {Z.}~\bibnamefont {Yan}}, \bibinfo {author} {\bibfnamefont {I.}~\bibnamefont {Zarov}}, \bibinfo {author} {\bibfnamefont {Y.}~\bibnamefont {Zhang}}, \bibinfo {author} {\bibfnamefont {A.}~\bibnamefont {Fan}}, \bibinfo {author} {\bibfnamefont {M.}~\bibnamefont {Kambadur}}, \bibinfo {author} {\bibfnamefont {S.}~\bibnamefont {Narang}}, \bibinfo {author} {\bibfnamefont {A.}~\bibnamefont {Rodriguez}}, \bibinfo {author} {\bibfnamefont {R.}~\bibnamefont {Stojnic}}, \bibinfo {author} {\bibfnamefont {S.}~\bibnamefont {Edunov}},\ and\ \bibinfo {author} {\bibfnamefont {T.}~\bibnamefont {Scialom}},\ }\href {https://doi.org/10.48550/arXiv.2307.09288} {\bibinfo {title} {Llama 2: {Open} foundation and fine-tuned chat models}} (\bibinfo {year} {2023}{\natexlab{a}}),\ \Eprint {https://arxiv.org/abs/2307.09288} {arXiv:2307.09288 [cs]} \BibitemShut {NoStop}%
\bibitem [{\citenamefont {Touvron}\ \emph {et~al.}(2023{\natexlab{b}})\citenamefont {Touvron}, \citenamefont {Lavril}, \citenamefont {Izacard}, \citenamefont {Martinet}, \citenamefont {Lachaux}, \citenamefont {Lacroix}, \citenamefont {Rozi{\`e}re}, \citenamefont {Goyal}, \citenamefont {Hambro}, \citenamefont {Azhar}, \citenamefont {Rodriguez}, \citenamefont {Joulin}, \citenamefont {Grave},\ and\ \citenamefont {Lample}}]{touvronLLaMAOpenEfficient2023}%
  \BibitemOpen
  \bibfield  {author} {\bibinfo {author} {\bibfnamefont {H.}~\bibnamefont {Touvron}}, \bibinfo {author} {\bibfnamefont {T.}~\bibnamefont {Lavril}}, \bibinfo {author} {\bibfnamefont {G.}~\bibnamefont {Izacard}}, \bibinfo {author} {\bibfnamefont {X.}~\bibnamefont {Martinet}}, \bibinfo {author} {\bibfnamefont {M.-A.}\ \bibnamefont {Lachaux}}, \bibinfo {author} {\bibfnamefont {T.}~\bibnamefont {Lacroix}}, \bibinfo {author} {\bibfnamefont {B.}~\bibnamefont {Rozi{\`e}re}}, \bibinfo {author} {\bibfnamefont {N.}~\bibnamefont {Goyal}}, \bibinfo {author} {\bibfnamefont {E.}~\bibnamefont {Hambro}}, \bibinfo {author} {\bibfnamefont {F.}~\bibnamefont {Azhar}}, \bibinfo {author} {\bibfnamefont {A.}~\bibnamefont {Rodriguez}}, \bibinfo {author} {\bibfnamefont {A.}~\bibnamefont {Joulin}}, \bibinfo {author} {\bibfnamefont {E.}~\bibnamefont {Grave}},\ and\ \bibinfo {author} {\bibfnamefont {G.}~\bibnamefont {Lample}},\ }\href {https://doi.org/10.48550/arXiv.2302.13971} {\bibinfo {title} {{{LLaMA}}: {{Open}} and efficient foundation language models}} (\bibinfo {year} {2023}{\natexlab{b}}),\ \Eprint {https://arxiv.org/abs/2302.13971} {arXiv:2302.13971 [cs]} \BibitemShut {NoStop}%
\bibitem [{\citenamefont {Grattafiori}\ \emph {et~al.}(2024)\citenamefont {Grattafiori} \emph {et~al.}}]{grattafioriLlama3Herd2024}%
  \BibitemOpen
  \bibfield  {author} {\bibinfo {author} {\bibfnamefont {A.}~\bibnamefont {Grattafiori}} \emph {et~al.},\ }\href {https://doi.org/10.48550/arXiv.2407.21783} {\bibinfo {title} {The {{Llama}} 3 herd of models}} (\bibinfo {year} {2024}),\ \Eprint {https://arxiv.org/abs/2407.21783} {arXiv:2407.21783 [cs]} \BibitemShut {NoStop}%
\bibitem [{\citenamefont {Dettmers}\ \emph {et~al.}(2023)\citenamefont {Dettmers}, \citenamefont {Pagnoni}, \citenamefont {Holtzman},\ and\ \citenamefont {Zettlemoyer}}]{dettmersQLoRAEfficientFinetuning2023}%
  \BibitemOpen
  \bibfield  {author} {\bibinfo {author} {\bibfnamefont {T.}~\bibnamefont {Dettmers}}, \bibinfo {author} {\bibfnamefont {A.}~\bibnamefont {Pagnoni}}, \bibinfo {author} {\bibfnamefont {A.}~\bibnamefont {Holtzman}},\ and\ \bibinfo {author} {\bibfnamefont {L.}~\bibnamefont {Zettlemoyer}},\ }\href {https://doi.org/10.48550/arXiv.2305.14314} {\bibinfo {title} {{{QLoRA}}: {E}fficient finetuning of quantized {LLMs}}} (\bibinfo {year} {2023}),\ \Eprint {https://arxiv.org/abs/2305.14314} {arXiv:2305.14314 [cs]} \BibitemShut {NoStop}%
\bibitem [{\citenamefont {Bengio}\ \emph {et~al.}(2021{\natexlab{a}})\citenamefont {Bengio}, \citenamefont {Jain}, \citenamefont {Korablyov}, \citenamefont {Precup},\ and\ \citenamefont {Bengio}}]{bengioFlowNetworkBased}%
  \BibitemOpen
  \bibfield  {author} {\bibinfo {author} {\bibfnamefont {E.}~\bibnamefont {Bengio}}, \bibinfo {author} {\bibfnamefont {M.}~\bibnamefont {Jain}}, \bibinfo {author} {\bibfnamefont {M.}~\bibnamefont {Korablyov}}, \bibinfo {author} {\bibfnamefont {D.}~\bibnamefont {Precup}},\ and\ \bibinfo {author} {\bibfnamefont {Y.}~\bibnamefont {Bengio}},\ }\href {https://doi.org/10.48550/ARXIV.2106.04399} {\bibinfo {title} {Flow network based generative models for non-iterative diverse candidate generation}} (\bibinfo {year} {2021}{\natexlab{a}})\BibitemShut {NoStop}%
\bibitem [{\citenamefont {Zhang}\ \emph {et~al.}(2019)\citenamefont {Zhang}, \citenamefont {Hu},\ and\ \citenamefont {Jiang}}]{zhangEmbeddedAtomNeural2019}%
  \BibitemOpen
  \bibfield  {author} {\bibinfo {author} {\bibfnamefont {Y.}~\bibnamefont {Zhang}}, \bibinfo {author} {\bibfnamefont {C.}~\bibnamefont {Hu}},\ and\ \bibinfo {author} {\bibfnamefont {B.}~\bibnamefont {Jiang}},\ }\bibfield  {title} {\bibinfo {title} {Embedded atom neural network potentials: {{Efficient}} and accurate machine learning with a physically inspired representation},\ }\href {https://doi.org/10.1021/acs.jpclett.9b02037} {\bibfield  {journal} {\bibinfo  {journal} {J. Phys. Chem. Lett.}\ }\textbf {\bibinfo {volume} {10}},\ \bibinfo {pages} {4962} (\bibinfo {year} {2019})}\BibitemShut {NoStop}%
\bibitem [{\citenamefont {Riebesell}\ \emph {et~al.}(2025)\citenamefont {Riebesell}, \citenamefont {Goodall}, \citenamefont {Benner}, \citenamefont {Chiang}, \citenamefont {Deng}, \citenamefont {Ceder}, \citenamefont {Asta}, \citenamefont {Lee}, \citenamefont {Jain},\ and\ \citenamefont {Persson}}]{matterbench}%
  \BibitemOpen
  \bibfield  {author} {\bibinfo {author} {\bibfnamefont {J.}~\bibnamefont {Riebesell}}, \bibinfo {author} {\bibfnamefont {R.~E.~A.}\ \bibnamefont {Goodall}}, \bibinfo {author} {\bibfnamefont {P.}~\bibnamefont {Benner}}, \bibinfo {author} {\bibfnamefont {Y.}~\bibnamefont {Chiang}}, \bibinfo {author} {\bibfnamefont {B.}~\bibnamefont {Deng}}, \bibinfo {author} {\bibfnamefont {G.}~\bibnamefont {Ceder}}, \bibinfo {author} {\bibfnamefont {M.}~\bibnamefont {Asta}}, \bibinfo {author} {\bibfnamefont {A.~A.}\ \bibnamefont {Lee}}, \bibinfo {author} {\bibfnamefont {A.}~\bibnamefont {Jain}},\ and\ \bibinfo {author} {\bibfnamefont {K.~A.}\ \bibnamefont {Persson}},\ }\bibfield  {title} {\bibinfo {title} {A framework to evaluate machine learning crystal stability predictions},\ }\href {https://doi.org/10.1038/s42256-025-01055-1} {\bibfield  {journal} {\bibinfo  {journal} {Nat. Mach. Intell.}\ }\textbf {\bibinfo {volume} {7}},\ \bibinfo {pages} {836–847} (\bibinfo {year} {2025})}\BibitemShut {NoStop}%
\bibitem [{\citenamefont {Loew}\ \emph {et~al.}(2025)\citenamefont {Loew}, \citenamefont {Sun}, \citenamefont {Wang}, \citenamefont {Botti},\ and\ \citenamefont {Marques}}]{Loew2025}%
  \BibitemOpen
  \bibfield  {author} {\bibinfo {author} {\bibfnamefont {A.}~\bibnamefont {Loew}}, \bibinfo {author} {\bibfnamefont {D.}~\bibnamefont {Sun}}, \bibinfo {author} {\bibfnamefont {H.-C.}\ \bibnamefont {Wang}}, \bibinfo {author} {\bibfnamefont {S.}~\bibnamefont {Botti}},\ and\ \bibinfo {author} {\bibfnamefont {M.~A.~L.}\ \bibnamefont {Marques}},\ }\bibfield  {title} {\bibinfo {title} {Universal machine learning interatomic potentials are ready for phonons},\ }\bibfield  {journal} {\bibinfo  {journal} {npj Comput. Mater.}\ }\textbf {\bibinfo {volume} {11}},\ \href {https://doi.org/10.1038/s41524-025-01650-1} {10.1038/s41524-025-01650-1} (\bibinfo {year} {2025})\BibitemShut {NoStop}%
\bibitem [{\citenamefont {Cheetham}\ and\ \citenamefont {Seshadri}(2024)}]{Cheetham2024}%
  \BibitemOpen
  \bibfield  {author} {\bibinfo {author} {\bibfnamefont {A.~K.}\ \bibnamefont {Cheetham}}\ and\ \bibinfo {author} {\bibfnamefont {R.}~\bibnamefont {Seshadri}},\ }\bibfield  {title} {\bibinfo {title} {Artificial intelligence driving materials discovery? perspective on the article: Scaling deep learning for materials discovery},\ }\href {https://doi.org/10.1021/acs.chemmater.4c00643} {\bibfield  {journal} {\bibinfo  {journal} {Chem. Mater.}\ }\textbf {\bibinfo {volume} {36}},\ \bibinfo {pages} {3490–3495} (\bibinfo {year} {2024})}\BibitemShut {NoStop}%
\bibitem [{\citenamefont {Leeman}\ \emph {et~al.}(2024)\citenamefont {Leeman}, \citenamefont {Liu}, \citenamefont {Stiles}, \citenamefont {Lee}, \citenamefont {Bhatt}, \citenamefont {Schoop},\ and\ \citenamefont {Palgrave}}]{Leeman2024}%
  \BibitemOpen
  \bibfield  {author} {\bibinfo {author} {\bibfnamefont {J.}~\bibnamefont {Leeman}}, \bibinfo {author} {\bibfnamefont {Y.}~\bibnamefont {Liu}}, \bibinfo {author} {\bibfnamefont {J.}~\bibnamefont {Stiles}}, \bibinfo {author} {\bibfnamefont {S.~B.}\ \bibnamefont {Lee}}, \bibinfo {author} {\bibfnamefont {P.}~\bibnamefont {Bhatt}}, \bibinfo {author} {\bibfnamefont {L.~M.}\ \bibnamefont {Schoop}},\ and\ \bibinfo {author} {\bibfnamefont {R.~G.}\ \bibnamefont {Palgrave}},\ }\bibfield  {title} {\bibinfo {title} {Challenges in high-throughput inorganic materials prediction and autonomous synthesis},\ }\bibfield  {journal} {\bibinfo  {journal} {PRX Energy}\ }\textbf {\bibinfo {volume} {3}},\ \href {https://doi.org/10.1103/prxenergy.3.011002} {10.1103/prxenergy.3.011002} (\bibinfo {year} {2024})\BibitemShut {NoStop}%
\bibitem [{\citenamefont {Juelsholt}(2025)}]{Juelsholt2025}%
  \BibitemOpen
  \bibfield  {author} {\bibinfo {author} {\bibfnamefont {M.}~\bibnamefont {Juelsholt}},\ }\bibfield  {title} {\bibinfo {title} {Continued challenges in high-throughput materials predictions: Mattergen predicts compounds from the training dataset.},\ }\bibfield  {journal} {\bibinfo  {journal} {ChemRxiv}\ }\href {https://doi.org/10.26434/chemrxiv-2025-mkls8} {10.26434/chemrxiv-2025-mkls8} (\bibinfo {year} {2025})\BibitemShut {NoStop}%
\bibitem [{\citenamefont {Momma}\ and\ \citenamefont {Izumi}(2011)}]{VESTA}%
  \BibitemOpen
  \bibfield  {author} {\bibinfo {author} {\bibfnamefont {K.}~\bibnamefont {Momma}}\ and\ \bibinfo {author} {\bibfnamefont {F.}~\bibnamefont {Izumi}},\ }\bibfield  {title} {\bibinfo {title} {{VESTA} 3 for three-dimensional visualization of crystal, volumetric and morphology data},\ }\href {https://doi.org/10.1107/s0021889811038970} {\bibfield  {journal} {\bibinfo  {journal} {J. Appl. Crystallogr.}\ }\textbf {\bibinfo {volume} {44}},\ \bibinfo {pages} {1272–1276} (\bibinfo {year} {2011})}\BibitemShut {NoStop}%
\bibitem [{\citenamefont {Lyngby}\ and\ \citenamefont {Thygesen}(2022)}]{Lyngby2022}%
  \BibitemOpen
  \bibfield  {author} {\bibinfo {author} {\bibfnamefont {P.}~\bibnamefont {Lyngby}}\ and\ \bibinfo {author} {\bibfnamefont {K.~S.}\ \bibnamefont {Thygesen}},\ }\bibfield  {title} {\bibinfo {title} {Data-driven discovery of 2{D} materials by deep generative models},\ }\bibfield  {journal} {\bibinfo  {journal} {npj Comput. Mater.}\ }\textbf {\bibinfo {volume} {8}},\ \href {https://doi.org/10.1038/s41524-022-00923-3} {10.1038/s41524-022-00923-3} (\bibinfo {year} {2022})\BibitemShut {NoStop}%
\bibitem [{\citenamefont {Gjerding}\ \emph {et~al.}(2021)\citenamefont {Gjerding}, \citenamefont {Taghizadeh}, \citenamefont {Rasmussen}, \citenamefont {Ali}, \citenamefont {Bertoldo}, \citenamefont {Deilmann}, \citenamefont {Knøsgaard}, \citenamefont {Kruse}, \citenamefont {Larsen}, \citenamefont {Manti}, \citenamefont {Pedersen}, \citenamefont {Petralanda}, \citenamefont {Skovhus}, \citenamefont {Svendsen}, \citenamefont {Mortensen}, \citenamefont {Olsen},\ and\ \citenamefont {Thygesen}}]{Gjerding2021}%
  \BibitemOpen
  \bibfield  {author} {\bibinfo {author} {\bibfnamefont {M.~N.}\ \bibnamefont {Gjerding}}, \bibinfo {author} {\bibfnamefont {A.}~\bibnamefont {Taghizadeh}}, \bibinfo {author} {\bibfnamefont {A.}~\bibnamefont {Rasmussen}}, \bibinfo {author} {\bibfnamefont {S.}~\bibnamefont {Ali}}, \bibinfo {author} {\bibfnamefont {F.}~\bibnamefont {Bertoldo}}, \bibinfo {author} {\bibfnamefont {T.}~\bibnamefont {Deilmann}}, \bibinfo {author} {\bibfnamefont {N.~R.}\ \bibnamefont {Knøsgaard}}, \bibinfo {author} {\bibfnamefont {M.}~\bibnamefont {Kruse}}, \bibinfo {author} {\bibfnamefont {A.~H.}\ \bibnamefont {Larsen}}, \bibinfo {author} {\bibfnamefont {S.}~\bibnamefont {Manti}}, \bibinfo {author} {\bibfnamefont {T.~G.}\ \bibnamefont {Pedersen}}, \bibinfo {author} {\bibfnamefont {U.}~\bibnamefont {Petralanda}}, \bibinfo {author} {\bibfnamefont {T.}~\bibnamefont {Skovhus}}, \bibinfo {author} {\bibfnamefont {M.~K.}\ \bibnamefont {Svendsen}}, \bibinfo {author} {\bibfnamefont {J.~J.}\ \bibnamefont {Mortensen}}, \bibinfo {author} {\bibfnamefont {T.}~\bibnamefont {Olsen}},\ and\ \bibinfo {author} {\bibfnamefont {K.~S.}\ \bibnamefont {Thygesen}},\ }\bibfield  {title} {\bibinfo {title} {Recent progress of the computational 2{D} materials database {(C2DB)}},\ }\href {https://doi.org/10.1088/2053-1583/ac1059} {\bibfield  {journal} {\bibinfo  {journal} {2D Mater.}\ }\textbf {\bibinfo {volume} {8}},\ \bibinfo {pages} {044002} (\bibinfo {year} {2021})}\BibitemShut {NoStop}%
\bibitem [{\citenamefont {Moustafa}\ \emph {et~al.}(2023)\citenamefont {Moustafa}, \citenamefont {Lyngby}, \citenamefont {Mortensen}, \citenamefont {Thygesen},\ and\ \citenamefont {Jacobsen}}]{Moustafa2023}%
  \BibitemOpen
  \bibfield  {author} {\bibinfo {author} {\bibfnamefont {H.}~\bibnamefont {Moustafa}}, \bibinfo {author} {\bibfnamefont {P.~M.}\ \bibnamefont {Lyngby}}, \bibinfo {author} {\bibfnamefont {J.~J.}\ \bibnamefont {Mortensen}}, \bibinfo {author} {\bibfnamefont {K.~S.}\ \bibnamefont {Thygesen}},\ and\ \bibinfo {author} {\bibfnamefont {K.~W.}\ \bibnamefont {Jacobsen}},\ }\bibfield  {title} {\bibinfo {title} {Hundreds of new, stable, one-dimensional materials from a generative machine learning model},\ }\bibfield  {journal} {\bibinfo  {journal} {Phys. Rev. Mater.}\ }\textbf {\bibinfo {volume} {7}},\ \href {https://doi.org/10.1103/physrevmaterials.7.014007} {10.1103/physrevmaterials.7.014007} (\bibinfo {year} {2023})\BibitemShut {NoStop}%
\bibitem [{\citenamefont {Parida}\ \emph {et~al.}(2025)\citenamefont {Parida}, \citenamefont {Roy}, \citenamefont {Lastra},\ and\ \citenamefont {Bhowmik}}]{paridaMiningChemicalSpace2025}%
  \BibitemOpen
  \bibfield  {author} {\bibinfo {author} {\bibfnamefont {C.}~\bibnamefont {Parida}}, \bibinfo {author} {\bibfnamefont {D.}~\bibnamefont {Roy}}, \bibinfo {author} {\bibfnamefont {J.~M.~G.}\ \bibnamefont {Lastra}},\ and\ \bibinfo {author} {\bibfnamefont {A.}~\bibnamefont {Bhowmik}},\ }\href {https://doi.org/10.26434/chemrxiv-2025-q48jr} {\bibinfo {title} {Mining chemical space with generative models for battery materials}} (\bibinfo {year} {2025})\BibitemShut {NoStop}%
\bibitem [{\citenamefont {Yang}\ \emph {et~al.}(2024{\natexlab{c}})\citenamefont {Yang}, \citenamefont {Hu}, \citenamefont {Zhou}, \citenamefont {Liu}, \citenamefont {Shi}, \citenamefont {Li}, \citenamefont {Li}, \citenamefont {Chen}, \citenamefont {Chen}, \citenamefont {Zeni}, \citenamefont {Horton}, \citenamefont {Pinsler}, \citenamefont {Fowler}, \citenamefont {Z\"{u}gner}, \citenamefont {Xie}, \citenamefont {Smith}, \citenamefont {Sun}, \citenamefont {Wang}, \citenamefont {Kong}, \citenamefont {Liu}, \citenamefont {Hao},\ and\ \citenamefont {Lu}}]{MatterSim}%
  \BibitemOpen
  \bibfield  {author} {\bibinfo {author} {\bibfnamefont {H.}~\bibnamefont {Yang}}, \bibinfo {author} {\bibfnamefont {C.}~\bibnamefont {Hu}}, \bibinfo {author} {\bibfnamefont {Y.}~\bibnamefont {Zhou}}, \bibinfo {author} {\bibfnamefont {X.}~\bibnamefont {Liu}}, \bibinfo {author} {\bibfnamefont {Y.}~\bibnamefont {Shi}}, \bibinfo {author} {\bibfnamefont {J.}~\bibnamefont {Li}}, \bibinfo {author} {\bibfnamefont {G.}~\bibnamefont {Li}}, \bibinfo {author} {\bibfnamefont {Z.}~\bibnamefont {Chen}}, \bibinfo {author} {\bibfnamefont {S.}~\bibnamefont {Chen}}, \bibinfo {author} {\bibfnamefont {C.}~\bibnamefont {Zeni}}, \bibinfo {author} {\bibfnamefont {M.}~\bibnamefont {Horton}}, \bibinfo {author} {\bibfnamefont {R.}~\bibnamefont {Pinsler}}, \bibinfo {author} {\bibfnamefont {A.}~\bibnamefont {Fowler}}, \bibinfo {author} {\bibfnamefont {D.}~\bibnamefont {Z\"{u}gner}}, \bibinfo {author} {\bibfnamefont {T.}~\bibnamefont {Xie}}, \bibinfo {author} {\bibfnamefont {J.}~\bibnamefont {Smith}}, \bibinfo {author} {\bibfnamefont {L.}~\bibnamefont {Sun}}, \bibinfo {author} {\bibfnamefont {Q.}~\bibnamefont {Wang}}, \bibinfo {author} {\bibfnamefont {L.}~\bibnamefont {Kong}}, \bibinfo {author} {\bibfnamefont {C.}~\bibnamefont {Liu}}, \bibinfo {author} {\bibfnamefont {H.}~\bibnamefont {Hao}},\ and\ \bibinfo {author} {\bibfnamefont {Z.}~\bibnamefont {Lu}},\ }\href {https://doi.org/10.48550/ARXIV.2405.04967} {\bibinfo {title} {{MatterSim}: {A} deep learning atomistic model across elements, temperatures and pressures}} (\bibinfo {year} {2024}{\natexlab{c}})\BibitemShut {NoStop}%
\bibitem [{\citenamefont {Wines}\ \emph {et~al.}(2023)\citenamefont {Wines}, \citenamefont {Xie},\ and\ \citenamefont {Choudhary}}]{10.1021/acs.jpclett.3c01260}%
  \BibitemOpen
  \bibfield  {author} {\bibinfo {author} {\bibfnamefont {D.}~\bibnamefont {Wines}}, \bibinfo {author} {\bibfnamefont {T.}~\bibnamefont {Xie}},\ and\ \bibinfo {author} {\bibfnamefont {K.}~\bibnamefont {Choudhary}},\ }\bibfield  {title} {\bibinfo {title} {Inverse design of next-generation superconductors using data-driven deep generative models},\ }\href {https://doi.org/10.1021/acs.jpclett.3c01260} {\bibfield  {journal} {\bibinfo  {journal} {J. Phys. Chem. Lett.}\ }\textbf {\bibinfo {volume} {14}},\ \bibinfo {pages} {6630–6638} (\bibinfo {year} {2023})}\BibitemShut {NoStop}%
\bibitem [{\citenamefont {Choudhary}\ and\ \citenamefont {DeCost}(2021)}]{Choudhary2021}%
  \BibitemOpen
  \bibfield  {author} {\bibinfo {author} {\bibfnamefont {K.}~\bibnamefont {Choudhary}}\ and\ \bibinfo {author} {\bibfnamefont {B.}~\bibnamefont {DeCost}},\ }\bibfield  {title} {\bibinfo {title} {Atomistic line graph neural network for improved materials property predictions},\ }\bibfield  {journal} {\bibinfo  {journal} {npj Comput. Mater.}\ }\textbf {\bibinfo {volume} {7}},\ \href {https://doi.org/10.1038/s41524-021-00650-1} {10.1038/s41524-021-00650-1} (\bibinfo {year} {2021})\BibitemShut {NoStop}%
\bibitem [{\citenamefont {Li}\ \emph {et~al.}(2024)\citenamefont {Li}, \citenamefont {Okabe}, \citenamefont {Cheng}, \citenamefont {Chottratanapituk}, \citenamefont {Hung}, \citenamefont {Fu}, \citenamefont {Han}, \citenamefont {Wang}, \citenamefont {Xie}, \citenamefont {Cava}, \citenamefont {Jaakkola},\ and\ \citenamefont {Cheng}}]{Li2024SCIGEN}%
  \BibitemOpen
  \bibfield  {author} {\bibinfo {author} {\bibfnamefont {M.}~\bibnamefont {Li}}, \bibinfo {author} {\bibfnamefont {R.}~\bibnamefont {Okabe}}, \bibinfo {author} {\bibfnamefont {M.}~\bibnamefont {Cheng}}, \bibinfo {author} {\bibfnamefont {A.}~\bibnamefont {Chottratanapituk}}, \bibinfo {author} {\bibfnamefont {N.~T.}\ \bibnamefont {Hung}}, \bibinfo {author} {\bibfnamefont {X.}~\bibnamefont {Fu}}, \bibinfo {author} {\bibfnamefont {B.}~\bibnamefont {Han}}, \bibinfo {author} {\bibfnamefont {Y.}~\bibnamefont {Wang}}, \bibinfo {author} {\bibfnamefont {W.}~\bibnamefont {Xie}}, \bibinfo {author} {\bibfnamefont {R.}~\bibnamefont {Cava}}, \bibinfo {author} {\bibfnamefont {T.}~\bibnamefont {Jaakkola}},\ and\ \bibinfo {author} {\bibfnamefont {Y.}~\bibnamefont {Cheng}},\ }\bibfield  {title} {\bibinfo {title} {Structural constraint integration in generative model for discovery of quantum material candidates},\ }\bibfield  {journal} {\bibinfo  {journal} {Preprint}\ }\href {https://doi.org/10.21203/rs.3.rs-4765336/v1} {10.21203/rs.3.rs-4765336/v1} (\bibinfo {year} {2024})\BibitemShut {NoStop}%
\bibitem [{\citenamefont {Gao}\ \emph {et~al.}(2025)\citenamefont {Gao}, \citenamefont {Huang}, \citenamefont {Huang}, \citenamefont {Li}, \citenamefont {Liu}, \citenamefont {Sa}, \citenamefont {Yu}, \citenamefont {Xue}, \citenamefont {Liu},\ and\ \citenamefont {Dai}}]{Gao2025Deep}%
  \BibitemOpen
  \bibfield  {author} {\bibinfo {author} {\bibfnamefont {S.}~\bibnamefont {Gao}}, \bibinfo {author} {\bibfnamefont {Q.}~\bibnamefont {Huang}}, \bibinfo {author} {\bibfnamefont {C.}~\bibnamefont {Huang}}, \bibinfo {author} {\bibfnamefont {C.}~\bibnamefont {Li}}, \bibinfo {author} {\bibfnamefont {K.}~\bibnamefont {Liu}}, \bibinfo {author} {\bibfnamefont {B.}~\bibnamefont {Sa}}, \bibinfo {author} {\bibfnamefont {Y.}~\bibnamefont {Yu}}, \bibinfo {author} {\bibfnamefont {D.}~\bibnamefont {Xue}}, \bibinfo {author} {\bibfnamefont {Z.}~\bibnamefont {Liu}},\ and\ \bibinfo {author} {\bibfnamefont {M.}~\bibnamefont {Dai}},\ }\bibfield  {title} {\bibinfo {title} {Deep generative model for the inverse design of van der waals heterostructures},\ }\bibfield  {journal} {\bibinfo  {journal} {Sci. Rep.}\ }\textbf {\bibinfo {volume} {15}},\ \href {https://doi.org/10.1038/s41598-025-06432-9} {10.1038/s41598-025-06432-9} (\bibinfo {year} {2025})\BibitemShut {NoStop}%
\bibitem [{\citenamefont {Lecun}\ \emph {et~al.}(2006)\citenamefont {Lecun}, \citenamefont {Chopra}, \citenamefont {Hadsell}, \citenamefont {Ranzato},\ and\ \citenamefont {Huang}}]{lecunTutorialEnergybasedLearning2006}%
  \BibitemOpen
  \bibfield  {author} {\bibinfo {author} {\bibfnamefont {Y.}~\bibnamefont {Lecun}}, \bibinfo {author} {\bibfnamefont {S.}~\bibnamefont {Chopra}}, \bibinfo {author} {\bibfnamefont {R.}~\bibnamefont {Hadsell}}, \bibinfo {author} {\bibfnamefont {M.~A.}\ \bibnamefont {Ranzato}},\ and\ \bibinfo {author} {\bibfnamefont {F.~J.}\ \bibnamefont {Huang}},\ }\bibfield  {title} {\bibinfo {title} {{A tutorial on energy-based learning}},\ }in\ \href@noop {} {\emph {\bibinfo {booktitle} {{Predicting structured data}}}},\ \bibinfo {editor} {edited by\ \bibinfo {editor} {\bibfnamefont {G.}~\bibnamefont {Bakir}}, \bibinfo {editor} {\bibfnamefont {T.}~\bibnamefont {Hofman}}, \bibinfo {editor} {\bibfnamefont {B.}~\bibnamefont {Scholkopt}}, \bibinfo {editor} {\bibfnamefont {A.}~\bibnamefont {Smola}},\ and\ \bibinfo {editor} {\bibfnamefont {B.}~\bibnamefont {Taskar}}}\ (\bibinfo  {publisher} {MIT Press},\ \bibinfo {year} {2006})\BibitemShut {NoStop}%
\bibitem [{\citenamefont {Du}\ and\ \citenamefont {Mordatch}(2019)}]{duImplicitGenerationModeling2019}%
  \BibitemOpen
  \bibfield  {author} {\bibinfo {author} {\bibfnamefont {Y.}~\bibnamefont {Du}}\ and\ \bibinfo {author} {\bibfnamefont {I.}~\bibnamefont {Mordatch}},\ }\bibfield  {title} {\bibinfo {title} {Implicit generation and modeling with energy based models},\ }in\ \href@noop {} {\emph {\bibinfo {booktitle} {Advances in Neural Information Processing Systems}}},\ Vol.~\bibinfo {volume} {32},\ \bibinfo {editor} {edited by\ \bibinfo {editor} {\bibfnamefont {H.}~\bibnamefont {Wallach}}, \bibinfo {editor} {\bibfnamefont {H.}~\bibnamefont {Larochelle}}, \bibinfo {editor} {\bibfnamefont {A.}~\bibnamefont {Beygelzimer}}, \bibinfo {editor} {\bibfnamefont {F.}~\bibnamefont {{dAlch{\'e}-Buc}}}, \bibinfo {editor} {\bibfnamefont {E.}~\bibnamefont {Fox}},\ and\ \bibinfo {editor} {\bibfnamefont {R.}~\bibnamefont {Garnett}}}\ (\bibinfo  {publisher} {Curran Associates, Inc.},\ \bibinfo {year} {2019})\BibitemShut {NoStop}%
\bibitem [{\citenamefont {Bengio}\ \emph {et~al.}(2021{\natexlab{b}})\citenamefont {Bengio}, \citenamefont {Jain}, \citenamefont {Korablyov}, \citenamefont {Precup},\ and\ \citenamefont {Bengio}}]{bengioFlowNetworkBased2021}%
  \BibitemOpen
  \bibfield  {author} {\bibinfo {author} {\bibfnamefont {E.}~\bibnamefont {Bengio}}, \bibinfo {author} {\bibfnamefont {M.}~\bibnamefont {Jain}}, \bibinfo {author} {\bibfnamefont {M.}~\bibnamefont {Korablyov}}, \bibinfo {author} {\bibfnamefont {D.}~\bibnamefont {Precup}},\ and\ \bibinfo {author} {\bibfnamefont {Y.}~\bibnamefont {Bengio}},\ }\bibfield  {title} {\bibinfo {title} {Flow {{Network}} based {{Generative Models}} for {{Non-Iterative Diverse Candidate Generation}}},\ }in\ \href@noop {} {\emph {\bibinfo {booktitle} {Advances in {{Neural Information Processing Systems}}}}},\ Vol.~\bibinfo {volume} {34}\ (\bibinfo  {publisher} {Curran Associates, Inc.},\ \bibinfo {year} {2021})\ pp.\ \bibinfo {pages} {27381--27394}\BibitemShut {NoStop}%
\bibitem [{\citenamefont {Kirkpatrick}\ \emph {et~al.}(1983)\citenamefont {Kirkpatrick}, \citenamefont {Gelatt},\ and\ \citenamefont {Vecchi}}]{kirkpatrickOptimizationSimulatedAnnealing1983}%
  \BibitemOpen
  \bibfield  {author} {\bibinfo {author} {\bibfnamefont {S.}~\bibnamefont {Kirkpatrick}}, \bibinfo {author} {\bibfnamefont {C.~D.}\ \bibnamefont {Gelatt}},\ and\ \bibinfo {author} {\bibfnamefont {M.~P.}\ \bibnamefont {Vecchi}},\ }\bibfield  {title} {\bibinfo {title} {Optimization by simulated annealing},\ }\href {https://doi.org/10.1126/science.220.4598.671} {\bibfield  {journal} {\bibinfo  {journal} {Science}\ }\textbf {\bibinfo {volume} {220}},\ \bibinfo {pages} {671} (\bibinfo {year} {1983})}\BibitemShut {NoStop}%
\bibitem [{\citenamefont {De~Breuck}\ \emph {et~al.}(2021)\citenamefont {De~Breuck}, \citenamefont {Evans},\ and\ \citenamefont {Rignanese}}]{debreuckRobustModelBenchmarking2021}%
  \BibitemOpen
  \bibfield  {author} {\bibinfo {author} {\bibfnamefont {P.-P.}\ \bibnamefont {De~Breuck}}, \bibinfo {author} {\bibfnamefont {M.~L.}\ \bibnamefont {Evans}},\ and\ \bibinfo {author} {\bibfnamefont {G.-M.}\ \bibnamefont {Rignanese}},\ }\bibfield  {title} {\bibinfo {title} {Robust model benchmarking and bias-imbalance in data-driven materials science: A case study on {{MODNet}}},\ }\href {https://doi.org/10.1088/1361-648X/ac1280} {\bibfield  {journal} {\bibinfo  {journal} {J. Phys.: Condens. Matter}\ }\textbf {\bibinfo {volume} {33}},\ \bibinfo {pages} {404002} (\bibinfo {year} {2021})}\BibitemShut {NoStop}%
\end{thebibliography}%






\end{document}